\PassOptionsToPackage{unicode}{hyperref}
\PassOptionsToPackage{hyphens}{url}
\documentclass[10 pt]{article}
\usepackage{setspace}
\usepackage[margin=0.8 in]{geometry}    
\usepackage[title,titletoc,toc]{appendix}
\usepackage{hyperref}     
 \usepackage{color}
 \usepackage[most]{tcolorbox}
 \tcbset{colback=yellow!10!white, colframe=red!50!black, 
        highlight math style= {enhanced, 
            colframe=red,colback=red!10!white,boxsep=0pt}
        }        
\usepackage[compat=1.1.0]{tikz-feynman}
\usepackage{graphicx}
\usepackage{amssymb}
\usepackage{amsmath}
\usepackage{amsfonts}
\usepackage{epstopdf}
\usepackage{empheq}

\newcommand{\p}{\partial}
\newcommand{\pdp}{p\frac{d }{dp}}
\newcommand{\qdq}{q\frac{d }{dq}}

\newcommand{\lo}{\Lambda_0}
\newcommand{\lm}{\Lambda}
\newcommand{\hf}{{1\over 2}}
\newcommand{\be}{\begin{equation}}
\newcommand{\br}{\begin{eqnarray}}
\newcommand{\er}{\end{eqnarray}}
\newcommand{\ee}{\end{equation}}
\newcommand{\bt}{\begin{tabular}}
\newcommand{\et}{\end{tabular}}
\newcommand{\bp}{\bar p}

\newcommand{\bphi}{\bar \phi}

\newcommand{\dd}{\delta}
\newcommand{\DD}{\Delta}
\newcommand{\Dt}{\frac{D}{2}}

\newcommand{\Dp}{\frac{d^D p}{(2\pi)^D}}

\newcommand{\eps}{\epsilon}
\newcommand{\la}{\lambda}

\newcommand{\kp}{ K'(p^2)}
\newcommand{\kpi}{ K'(p_i^2)}

\title{Finite Cutoff CFT's and Composite Operators.}
\author{S.~Dutta, B.~Sathiapalan\\
Institute of
  Mathematical Sciences\\CIT Campus, Tharamani\\ 
  Chennai 600113, India\\and\\Homi Bhabha National Institute\\Training
  School Complex, Anushakti Nagar\\Mumbai 400085, India\\}

\begin{document}

\hspace{12cm} IMSc/2021/05/01\\

{\let\newpage\relax\maketitle}

\maketitle
\begin{abstract}
Recently a conformally invariant action describing the Wilson-Fisher fixed point in $D=4-\eps$ dimensions in the presence of a {\em finite} UV cutoff was constructed \cite{Dutta}. In the present paper we construct two composite operator perturbations of this action with definite scaling dimension also in the presence of a  finite cutoff.  Thus the operator (as well as the fixed point action) is well defined at all momenta $0\leq p\leq \infty$ and at low energies they reduce to $\int_x \phi^2$ and $\int _x \phi^4$ respectively.
The construction includes terms up to $O(\eps^2)$.  In the presence of a finite cutoff they mix with higher order $\int_x \phi^n$ operators. The dimensions are also calculated to this order and agree with known results.
  
\end{abstract}
\newpage
\tableofcontents
\newpage

\section{Introduction}

Conformal field theories have been the subject of much study over the last many decades. A very important motivation comes from condensed matter physics in the study of critical phenomena. Critical phenomena are characterized by a large (infinitely large at the critical point) correlation length. It was argued long ago that such systems are conformally invariant at the critical point.
\cite
{Polyakov1970}. The idea of
bootstrap was also introduced soon after, which allowed further non perturbative constraints to be placed on the system \cite{Polyakov1974}. Particularly in two dimensions these
ideas have been very fruitful \cite{Polyakov1984} and have applications in the world sheet description of string theory. Reviews of later
developments and references are given in
\cite{DiFrancesco1997,Rychkov2016}. 
The  AdS/CFT correpondence
\cite{Maldacena,Polyakov,Witten1,Witten2} or ``holography'' between a
boundary CFT and a bulk gravity theory gives another motivation for studying CFT's. \footnote{It also opens up the amazing possibility of
  rewriting quantum gravity as a quantum field theory in flat space.}
There is a large amount of literature on this.  See, for example,
\cite{Penedones2016} for a review.

In condensed matter physics there is an underlying lattice structure that provides a natural ultraviolet cutoff. At the critical point the correlation length being much larger than the lattice spacing, one can for many purposes treat it as a continuum theory, much as is done in high energy physics. Nevertheless conceptually it is important to understand field theories with a finite cutoff. In particular it is interesting to study {\em conformal} field theories in the presence of a finite cutoff.\footnote{If one speculates as for instance in \cite{BSLV} that space time itself in string theory is dicrete, then that is additional motivation for studying such theories.}   

Naively, a cutoff would violate scale invariance. But in fact scale invariance merely imposes the restriction, that the action(or hamiltonian) expressed in dimensionless variables - i.e. dimensionful quantities are expresssed in units of the lattice spacing - is unchanged when one performs a coarse graining of the lattice\footnote{Unchanged up to a wave function renormalization}. For this to happen, an infinite number of dimensionless parameters that characterize the action, which are  coefficients of all higher dimension operators, must be tuned to  specific values. This gives the ``fixed point " action. Scale invariance is then just the statement that ``there is no other scale in the problem".  

Expressed in dimensionful variables, the higher dimensional operators in the fixed point action are down by powers of the cutoff and are not important for energy scales much lower than the cutoff. They are technically ``irrelevant".  Thus in the continuum limit (lattice spacing going to zero, or momentum cutoff taken to infinity \footnote{It is important to realize that the precise value of the cutoff has no significance, since in a CFT  there is no other scale to compare it with. What matters is whether it is finite or infinite.}) we are left with the only a finite number of lower dimension ``relevant" and ``marginal" operators in the action. This is the situation that is usually conidered in renormalizable quantum field theories in the continuum limit.
\footnote{A caveat needs to be made here. We have used the words ``lattice spacing" and "momentum cutoff" interchangeably. But it is important for the momentum space ERG method followed in this paper that the cutoff be analytic in momentum space. Thus a geometric lattice interpretation should not be taken literally here. It may be possible to do these calculations decribed in this paper using real space RG techniques. This is an open question.}

Besides applications in critical phenomena, such fixed point actions arise in 
the world sheet description of string theory in the presence of background fields. Each of the coupling constants in the two dimensional theory corresponds to background values of the space time fields corresponding to the (infinite tower of) string modes. The world sheet RG equations can be interpreted as equations of motion for these fields. 
In continuum description one imposes vanishing $\beta$- function conditions for the marginal and relevant couplings. In string theory these correspond to equations for the massless modes and tachyon (in the bosonic string) respectively. In theories with a finite cutoff one imposes the fixed point condition as an Exact RG (ERG) equation. This gives equations of motion for all the massless and massive modes of the string. This technique can be used to obtain background and gauge invariant interacting equations for all modes. 
This can  be interpreted as equations of some ``string field theory". \footnote{See \cite{Sathiapalan} and references therein.}

Recently a role has been proposed for the Exact
Renormalization Group (ERG) equation
\cite{Wilson,Wegner,Wilson2,Polchinski,MorrisERG,Bagnuls1,Bagnuls2,Rosten} in the AdS-CFT correspondence.
  In the AdS/CFT correspondence the radial direction can be interpreted
as the scale of the boundary field theory. Thus, a radial evolution
can be thought of as an RG evolution and has been dubbed ``holgraphic
RG''
\cite{Akhmedov,Akhmedov1,Akhmedov2,Alvarez,Kraus,Warner,Verlinde,Boer,Faulkner,Klebanov:1999tb,Heemskerk,Morris,Bzowski:2015pba,deHaro:2000vlm}.
  If one starts with a conformally invariant fixed point action in $D$ dimensions  and perturbs it, then an ERG describes the evolution of these perturbations.
It was shown in
\cite{Sathiapalan1,Sathiapalan2,Sathiapalan3} that the evolution operator of this ERG can be written as a functional integral of a field theory in $AdS_{D+1}$ space.  
The boundary values of these fields are typically sources for the perturbing operators, though other interpretations are also possible.

Motivated by the ideas describe above, fixed point Wilson action for the $O(N)$ model in $4-\eps$ dimension was constructed in \cite{Dutta} to $O(\eps^2)$. 
An important operator in this theory is the energy momentum tensor which was also constructed to this order using techniques in ERG \cite{Sonoda-emt,Rosten2}. \footnote{In AdS-CFT correpondence, the bulk graviton is the field corresponding to this perturbation.} The energy momentum tensor was shown to be traceless. This implies that the fixed point theory is also conformally invariant, as expected on general grounds for most field theories. Indeed the tracelessness of the energy-momentum tensor
defines what we mean by a CFT \cite{Callan:1970ze, Coleman2,Brown,  Polchinski2}.
The Wilson action is usually understood as a low energy effective action valid only at energies well below the cutoff. But when it is obtained as a solution to the ERG it is valid at all energies. It thus has a lot of information about the high energy physics as well.

In this paper we take a logical next step in understanding these fixed point theories (with finite cutoff). The leading perturbations to the fixed point action involving operators with definite scaling dimension are constructed. (The energy momentum tensor is one such operator and was constructed as mentioned in the last paragraph.) The lowest dimension operator is the mass perturbation $\int_x \phi^2$. In the presence of interactions and a finite cutoff, one can expect this to mix with higher dimension operators such as $\int_x \phi^4, \int_x \phi^6,...$. Similarly, in an interacting theory, the continuum operator $\phi^4$ also mixes with higher dimensional operators - again in the presence of a finite cutoff. 

These operators are generically referred to as ``composite" operators.   In continuum field theory these have to be renormalized so that Green functions involving these are finite. This is an interesting problem in its own right.  This is described in many textbooks such as \cite{CollinsI}. The renormalization of these operators in $\phi^4$ theory in four dimensions is described in detail in \cite{Brown, Collins}. Analogous study of $\phi^3$ theory in six dimensions has also been done \cite{Pavan}. In contrast to the situation with finite cutoff, in the continuum theory an operator mixes with other operators of the same dimension or less. 

 In conformal field theory these operators are particularly interesting \footnote{See for instance \cite{Rychkov2016} and references therein}. In CFT's the Hamiltonian does not have a mass gap. There is a continuum of energy eigenstates starting with the vacuum ground state. It is useful then to use as Hamiltonian the dilatation generator that has a discrete spectrum as the Hamiltonian. These eigenvalues are the dimensions of the operators. The dilatation generator  is related by a finite conformal transformation to the usual Hamiltonian. This gives rise to the idea of radial quantization. Thus identifying the composite operators is equivalent to understanding the eigenstates and eigenvalues of this new ``Hamiltonian".

In ERG composite operators are solutions of the linearized equations \cite{Wilson}.   See \cite{Igarashi} for a review. They show up in many situations such as in realization of symmetry of the Wilson action such as to define either Ward-Takhashi identity in continuum limit for continuous symmetry or Quantum Master equation in Antifield formalism. Many aspects of composite operators have been studied in \cite{SonodaI,SonodaII,Pagani}. Lowest order eigenoperators of definite scaling dimension have also been constructed \cite{PaganiI}.

  One simple method to construct composite operators is to perturb the bare action and follows its linearized evolution. A general perturbation will mix with other operators. We are interested in eigen-operators that maintain their form as they evolve.  These operators should obey the usual properties of  operators with definite scaling dimension in a CFT. The added complication is the presence of a finite cutoff. The eigenvector equation, which is the ERG equation, can be solved perturbatively in powers of $\la$ the coupling constant. This is also related to $\eps$ since $\la \approx O(\eps)$. It involves making a fairly general (momentum dependent) ansatz for the eigen operators and solving for the momentum dependence order by order. We do this up to $O(\la^2)$ which is already quite tedious algebraically. The composite operator so constructed has a property in common with the fixed point action constructed in \cite{Dutta}, and quite unlike the compsite operators of continuum field theory, that there is no restriction that momenta should be small relative to the UV cutoff. The form of a typical term is illustrated by the quartic term  $\int_{p_i}B(p_1,p_2,p_3,p_4)\phi(p_1)\phi(p_2)\phi(p_3)\phi(p_4)$.
 These descriptions are valid at all momenta $p_i$ including energies comparable or even larger than the UV cutoff\footnote{Since a smooth analytic cutoff is used, the range of momentum extends all the way to infinity}.
 For the simplest case which is the leading order relevant operator, we  construct the local operator i.e. $\phi^2(x)$ or in momentum space $\phi^2(q)$ with $q\neq 0$. In all other cases, for reasons of computational simplicity, especially at second order, in this paper we have focused on the integrated operators  $\int_x \phi^2 (x)$ and $\int _x \phi^4 (x)$. This amounts to imposing $\sum _i p_i=q=0$. The unintegrated operator can be extracted from this modulo total derivative terms. The scaling dimensions are also calculated and agree with the literature to this order. 
 
Construction of local operators  enables us to do other analysis. Important one among them is to find whether the composite operators are primary or not. This has to be done by checking whether the correponding correlation function  satisfy the Conformal Ward Identity. However, we did not pursue that in this paper.
 
This paper is organized as follow. Section 2 contains some background material about ERG and composite operators and an application of these ideas to the Gaussian case. Section 3 describes the $O(\la)$ construction of the two operators. Only the relevant operator $\phi^2(q)$ is constructed for non zero $q$ to illustrate the procedure. In Section 4 the $O(\la^2)$ construction is given. Section 5 contains some conclusions. The appendices contain some background material and most of the details of the calculation. For convenience of readers, we have summarized the results obtained in this paper in \eqref{dim_lead}, \eqref{relevant_lead}, \eqref{irrelevant_lead}, \eqref{dim_sublead}, \eqref{irrelevant_sublead} and \eqref{relevant_sublead}.

\section{Background}
\label{secA}
\subsection{Composite Operators in Field Theory}
\label{secA1}

CFT's are defined by the spectrum of primary operators and their three point correlators, say in position space or equivalently by their Operator Product Expansion (OPE). Higher point correlations can be obtained once  
this data is given. This description does not require a Lagrangian description of the theory. It is a group theoretic approach where the symmetry group is the conformal group.

One can ask whether there exists a  description for any given CFT as a fixed point of some quantum field theory (QFT) with a known Lagrangian. If so one can hope to construct these primary operators as composites of the fundamental fields of the QFT. This may be useful in a physical situation where one may also be interested in the physics far away from the fixed point.

A good example of the above is the usual 
$\phi^4$ scalar field theory. In $D=4-\eps$ this is known to have a non trivial fixed point - Wilson-Fisher (WF). This is in addition to the fixed point corresponding to the free theory also known as the Gaussian fixed point. If $\eps$ is small the WF theory can be treated perturbatively. 

In free field theory in $D=4$ the scalar field $\phi$ has engineering (or canonical) dimension one.  The composite $\phi^n$ thus has dimension $n$. Thus we consider a term in the action $\DD S_2= \hf \int m^2\phi^2$. Let the UV cutoff be $\lm$. 
We write this action in terms of dimensionless fields and coordinates. Define 
\[
\phi = \lm \bphi, ~~x=\frac{\bar x }{\lm}
\]
Then
\[
\DD S_2 =\hf\int d^4\bar x~ \frac{m^2}{\lm^2}\bphi^2 =\hf\int d^4\bar x~ r\bphi^2
\]
Here $r$ is dimensionless. On coarse graining, $\lm$ decreases, so for fixed $m^2$, $r$ increases. Thus if we write $\lm = \lo e^{-t}$
we see that
\[
\frac{d \DD S_2}{dt}\equiv d_m \DD S_2=2\DD S_2
\]
and we call it relevant. $d_m$ is the overall length scaling dimension of $\DD S_2$ (not counting the parameter $m^2$, which is included to make the whole thing have dimension zero). 

If we add a term 
\[
\DD S_4=\int d^4x~  u\phi^4
\]
one immediately sees that $u$ is already dimensionless and
\[
\frac{d \DD S_4}{dt}\equiv d_m \DD S_4=0
\]
and we call it marginal.

But this is not the whole story even in a free theory. The operation $d/dt$ refers not to just changing $\lm$ that was introduced here to make things dimensionless, but it refers to the whole process of {\em integrating out} modes between $\lm$ and $\lm (1-dt)$. This physical coarse graining process fixes the $\lm$ dependence of the action. It introduces an extra $\lm$ dependence over and above what is required for writing everything in terms of dimensionless variables.

We illustrate this with a simple calculation. Write $\phi = \phi_h+\phi_l$. We assume that $\phi_h$ are modes between $\lm, \lo$ and are integrated out. Thus
\[
\phi^4 = \phi_l^4 + 6 \phi_l^2\phi_h^2 +\phi_h^4
\]
Integrating out $\phi_h$ in the second term gives
\[
\int_x 6 \phi_l^2 \frac{1}{(4\pi)^2} \int_{\lm^2}^{\lo^2} dp^2 p^2\frac{1}{p^2}=\frac{6}{(4\pi)^2}[\lo^2-\lm^2]\int_x \phi_l^2
\]
If we take $\lm$ and $\lm (1-dt)$ instead of $\lo$ and $\lm$ we get
\be   \label{2.1}
\frac{d \DD S_4}{dt}= \frac{6u}{(4\pi)^2}[2\lm^2 ]\int_x \phi_l^2
\ee

Thus we see that $\dot \DD S_4 \neq 0$ even in a free theory. One must add $\DD S_2$ with
$r_0=-\frac{6u}{(4\pi)^2}$. So in dimensionless variables
\be   \label{2.2}
\DD S =    \frac{1}{4!}\int d^4\bar x~ u\bphi^4+\hf\int d^4\bar x~ r_0\bphi^2 
\ee
satisfies $\dot \DD S=0$ and has $d_m=0$. This is the usual "quadratic" divergence in scalar field theory in another guise.

The simple calculation above is in the spirit of the Wilsonian RG and is described further in the next section below.
The above simple calculation also indicates the need to renormalize the operators when taking the continuum limit.
In the interacting case the $\lm$ dependence will be more complicated. There will in general be mixing among all operators of a given dimension. 

In the usual continuum field theory approach these operators have to be renormalized as one takes $\lo\to \infty$. The renormalization of these operators is complicated by this mixing and  the RG flow is described as a matrix equation. 
Renormalization of composite operators are described in many field theory text books (for eg \cite{CollinsI}). A careful analysis of the composite operators is described in \cite{Brown,Collins} for $\phi^4$ theory in four dimensions  using dimensional regularization , in \cite{CollinsI} for $\phi^3$ theory in six dimension. In particular the composite energy momentum tensor operator is constructed there. A similar analysis has been done recently for the $\phi^3$ theory in six dimensions \cite{Pavan}.

In contrast, in the Wilsonian RG one studies the evolution of an operator as longer and longer wavelength modes are integrated out. 
This is done by requiring that $\DD S$ obey the Wilsonian RG equation linearized about a fixed point. This leads to the definition of a composite operator in ERG given below.

\subsection{Composite operator in ERG}

Composite Operators of definite scaling dimension using the ERG were discused in \cite{Wilson}. A good discussion of composite operators is given in \cite{Igarashi} and some of it is summarized in this section below. Many other aspects of composite operators in $\phi^4$ field theory are discussed in \cite{SonodaI,SonodaII, Pagani,PaganiI}. In particular, few works on energy moemntum tensor and corresponding correlators have been done \cite{Sonoda-emt,Rosten2}.

A Composite operator in ERG is defined as the operator obtained by evolution of a bare operator under ERG flow.  Consider an operator $O_B$ in the bare theory.
Define the low energy propagator as
\[
\DD_l=\frac{K(p)}{p^2}
\]
where $K(p)$ is a smooth momentum cutoff function. For eg.
\[
K(p)=e^{-\frac{p^2}{\lm^2}}
\]
We also define 
\[
K_0(p)=e^{-\frac{p^2}{\lo^2}}
\]where $\lo$ is the UV cutoff in the bare theory that we can take to $\infty$ in the continuum limit. Then define 
\[
\DD_h(p)=\frac{K_0(p)-K(p)}{p^2}
\]
 the high energy propagator. It propagates modes mainly between $\lo,\lm$. The full propagator of the bare theory is $\DD=\DD_l+\DD_h$.
 
Define the Wilson Action  $S_\lm$ and  the interacting part of the Wilson Action $S_{\lm,I}$ by
\begin{align}\label{wilson}
\int \mathcal{D}\phi_h e ^{- S_{B}[\phi_l+\phi_h]}=\int \mathcal{D}\phi_h e ^{-\hf\int \phi_l\DD_l^{-1}\phi_l-\frac{1}{2}\int \phi_h \Delta_h^{-1} \phi_h- S_{B,I}[\phi_l+\phi_h]}= e^{\hf\int \phi_l\DD_l^{-1}\phi_l-S_{\Lambda,I}[\phi_l]}=e^{-S_{\lm}}
\end{align}
where $S_{B,I}$ is the interacting part of the bare action. The first equality in this can be proved \cite{Igarashi}. The rest are definitions.
This defines an ERG flow from $\lo$ to $\lm$.

$S_\lm$ is a theory where $\lm$ is a UV cutoff. It may be obtained as above  by integrating out modes in a bare theory defined at a higher scale. From the point of view of this bare theory, $\lm$ is an IR cutoff during the integration process. Nevertheless a fixed point Wilson action $S_\lm$  defined as a stationary solution of the ERG equation, has an existence in its own right without reference to a bare theory from which it is derived. In this viewpoint $\lm$ is indeed a UV cutoff. We take this viewpoint in this paper.

We give below some equivalent ways of defining a composite operator in ERG:

{\bf Definition I}

The composite operator of this operator at scale $\Lambda$ , $O_\Lambda $ is defined as:

\begin{align*}
\int \mathcal{D} \phi_h O_B[\phi_l+\phi_h]e ^{-\frac{1}{2}\int \phi_h \Delta_h^{-1} \phi_h- S_{B,I}[\phi_l+\phi_h]}= O_\Lambda[\phi_l] e^{-S_{I,\Lambda}[\phi_l]}
\end{align*}
The composite operator defined as above has the useful property: \cite{Igarashi}
\[
\langle O_B(x) \phi(x_1) \phi(x_2)..\phi(x_n)\rangle_{\lo} =\prod_{i=1}^n\frac{K_0(p_i)}{K(p_i)}
\langle [O]_\lm(x) \phi(x_1) \phi(x_2)...\phi(x_n)\rangle_\lm
\]

{\bf Definition II}

A useful way to think about composite operators in ERG is in terms of evolution operators. Define an ERG evolution operator $U$ by
\[
e^{-S_\lm[\phi_f]}=U(f,i)e^{-S_B[\phi_i]}
\] Then
\[
O_\lm [\phi_f]U(f,i) e^{-S_B[\phi_i]}=U(f,i)O_B[\phi_i] e^{-S_B[\phi_i]}
\] Thus formally one can write this as
\be    \label{2.3}
O_\lm[\phi_f]= U(f,i)O_B[\phi_i][U(f,i)]^{-1}
\ee

{\bf Definition III}

We can also think of perturbing $S_B$ by a term of order $\epsilon$ and calculate the change in $S_\Lambda$ to order $\epsilon$:

\begin{align}\label{ERG_comp}
\int \mathcal{D}\phi_h e ^{-\frac{1}{2}\int \phi_h \Delta_h^{-1} \phi_h- S_{B,I}[\phi_l+\phi_h]+\epsilon O_B[\phi_l+\phi_h]}= e^{-S_{I,\Lambda}[\phi_l]+ \epsilon O_\Lambda[\phi_l]}
\end{align}

This definition leads to a functional differential equation
and is also a convenient way of defining $O_\lm$. In this paper we use this approach.
This equation is in fact the linearized ERG equation for a perturbation $\Delta S$ obtained from \eqref{2.3}\footnote{See Appendix \ref{A} for a review of the ERG equation}:

\begin{align}\label{main}
\nonumber &\frac{\partial \Delta S}{\partial t}= \int_p  \bigg \lbrace (-K'(p^2)) [\underbrace{\frac{\delta^2 \Delta S}{\delta \phi(p)\delta \phi(-p)}}_{1}\underbrace{-2\frac{\delta S}{\delta \phi(p)}\frac{\delta \Delta S}{\delta \phi(-p)}}_{2}] \underbrace{-2\frac{p^2K'}{K}\phi(p)\frac{\delta \Delta S}{\delta \phi(p)}}_{3}+\\
 \nonumber &+ \frac{-\eta}{2} \frac{K(p^2)(1-K(p^2)}{p^2} [\frac{\delta^2 \Delta S}{\delta \phi(p)\delta \phi(-p)} -2\frac{\delta S}{\delta \phi(p)}\frac{\delta \Delta S}{\delta \phi(-p)}]+ \frac{-\eta}{2} \phi(p) \frac{\delta \Delta S}{\delta \phi(p)} \bigg \rbrace\\
&+[(1-\frac{D}{2})N_\phi+D-N_p]\Delta S
\end{align}

All the variables are  dimensionless. This equation defines the $\lm$ or $t$ dependence, given some starting operator at the initial time. Eigen-operators are defined by the property that
\be   \label{eigen}
\frac{\p \DD S}{\p t} =d_m \DD S+ \beta(\lambda) \frac{\p \Delta S}{\p \la}
\ee
i.e. under RG evolution they just scale as $e^{d_mt}$ where $d_m$ is the (length) scaling dimension.The second term $\beta(\lambda)$ is zero at the fixed point. Actually this is true for operators integrated over all space.
In most places in this paper $\DD S$ is chosen to be of the form $g_i\int _x O^i(x)$, i.e. integrated over space and thus correspond to some coupling constant in the action. From the integrated form one can determine $O(x)$ up to {\em total derivatives}. Thus $O(x)$ and $O(x)+\p_\mu O^\mu(x)$ will give the same $\DD S$.
To determine $O(x)$ unambiguously one would have to make $g_i(x)$ space dependent. This complicates the (already involved) algebra especially at two loops and is not attempted here. \footnote{Just as an illustration, the leading order result for the relevant unintegrated operator $\phi^2$ is given in Section 3.1} .

\vspace{0.1 in}

{\bf Boundary Conditions on Composite Operators:}

In the first two definitions it is evident that there is a boundary condition for $O_\lm[\phi]$, namely that at $\lm=\lo$ it becomes equal to $O_B[\phi]$.  Similarly, while solving the eigenvalue equation at Wisher-Fisher fixed point in this paper we  put initial condition that at $\lm=\lo$ it reduces to $O_B[\phi]$. 
We choose $O_B$ as in the Gaussian theory, namely 
\[
{\cal O}_2 =\phi^2~~~at~~\lm=\lo
\]
and
\[
{\cal O}_4 = \phi^4 ~~~at~~\lm=\lo
\] 
Correction to this will be evaluated in a perturbation series as powers of $\la$. Thus
$\Delta S(\lambda=0)$ will be equal to correponding operator in Gaussian theory ( which is given in the next subsection). The corrections will be chosen to be in terms of $h(p)$, the high energy propagator, that vanishes when $\lm=\lo$. All the correction terms thus vanish at $\lm=\lo$.  This implements the required boundary condition. 

Many aspects of these local operators are discussed in  \cite{SonodaI,SonodaII,Pagani,PaganiI}. Some scaling properties  are described in Appendix A.

It is also important to point out that the concept of scaling dimension makes sense only if the theory has scale invariance. Thus $S$ must correspond to a fixed point action that obeys 
\[
\frac{\p  S}{\p t} =0
\]

But in general one can solve a more general equation by putting 

\[
\frac{\p S}{\p t}= \beta(\lambda) \frac{ \p S}{\p \la}
\]

As we will calculate the anomalous dimension of the composite operators for the Wilson-Fisher fixed point, let us do some simple calcualtion to understand what to expect as the anomalous dimensions. Consider a bare action at scale $\Lambda_0$ and evolve to $\Lambda$ which is close to $\Lambda_0$.

\begin{align}
S_{\Lambda_0}= \int_x \left[ \hf \partial_\mu \phi \partial^\mu \phi + \hf m_0^2 \phi^2+ \la_0 \frac{\phi^4}{4!}\right]
\end{align}

The operator $\frac{\phi^4}{4!}$ is the relevant operator of the bare theory and can be seen as

\begin{align}
\frac{\partial S_{\Lambda_0}}{\partial \la_0}= \int_x \frac{\partial \mathcal{L}_{\Lambda_0}}{\partial \la_0}= \int_x \frac{\phi^4}{4!}
\end{align}

And the relavant operator $\hf \phi^2$ as

\begin{align}
\frac{\partial S_{\Lambda_0}}{\partial m_0^2}= \int_x \frac{\partial \mathcal{L}_{\la_0}}{\partial \la_0}= \int_x \frac{\phi^2}{2}
\end{align}

$S_\lm$ is obtained by evolving down from $\lo$ to $\lm$ i.e. by integrating modes  $\lm<p<\lo$.  

If we apply Definition III given above for a composite operator,  $\frac{\p { S}_{\lm}}{\p \la_0}$ is a composite operator and defines in fact what we call $[\phi^4]/4!$.  
\be \label{co}
\frac{\p S_\lm}{\p \la_0} \equiv \int _x \frac{[\phi^4]_\lm}{4!}
\ee
We can expect $S_\lm$ to look like the following:
\be
S_\lm = \int _x [(1-\dd Z(t))\hf \p_\mu\phi \p^\mu \phi + \hf  (m_0^2+ \delta m_0(t)^2)\phi^2 + (\la_0+\dd \la_0(t)) \frac{\phi^4}{4!} + O(1/\lm)]
\ee
Here $\dd Z$ is the correction to the kinetic term coming from the two loop diagram at $O(\la^2)$, $\dd m_0^2\approx O(\la)$ and $\dd \la_0 \approx O(\la^2)$ are the corrections starting at one loop.


Adding and subtracting terms we can write $S_\lm$ as:

\[=
\int _x [\hf \p_\mu\phi \p^\mu \phi + \hf  (m_0^2+ \delta m_0(t)^2 + \dd Z m_0^2)\phi^2 + (\la_0+\underbrace{\dd \la_0(t)+2\dd Z \la_0}_{\bar \dd \la_0(t)}) \frac{\phi^4}{4!} + O(1/\lm)]
\]
\[-\dd Z [\hf \p_\mu\phi \p^\mu \phi + \hf  m_0^2\phi^2+2 \la_0 \frac{\phi^4}{4!}]
\]
The beta function is defined by
\be
\bar \dd \la_0  \approx\beta(\la_0)t
\ee
and
\[
\dd Z \approx -\frac{\eta}{2} t
\]
with
\be
\frac{\eta}{2} = \frac{\la_0^2}{(16\pi^2)^2}\frac{1}{12}
\ee
The mass anomalous dimension is defined by,

\be
\delta m_0(t) \approx \gamma_m t
\ee

We write $\hf \p_\mu\phi \p^\mu \phi=-\hf \phi \Box \phi$ and then use
\[-\dd Z [\hf \p_\mu\phi \p^\mu \phi + \hf  m_0^2\phi^2+2 \la_0 \frac{\phi^4}{4!}]=-\dd Z \hf \phi \frac{\dd S}{\dd \phi}
\]

$ \phi \frac{\dd S}{\dd \phi}$ is called as the equation of motion operator. \footnote{ More correctly at higher orders it should include the change in measure and becomes
the ``number operator". Here $[\phi]$ is the ``composite operator" correponding to $\phi$ and is defined by \cite{Igarashi}
\[
[\phi]_\lm(p)=\frac{K_0}{K} \phi(p) + \frac{K_0-K}{p^2}\frac{\dd S_\lm}{\dd \phi(-p)}
\]}
\be \label{n}
{\cal N}= -\int _p Ke^{+S_\lm}\frac{\dd }{\dd \phi}([\phi]_\lm  e^{-S_\lm}) \approx \int _p \phi \frac{\dd  S}{\dd \phi} 
\ee

\[S_\lm=
\int _x [\hf \p_\mu\phi \p^\mu \phi + \hf  (m_0^2+m_0^2\gamma_m t)\phi^2 + (\la_0+\beta(\la_0)t) \frac{\phi^4}{4!} + O(1/\lm)]
\]
\be  \label{2.17}
+\frac{\eta}{2} t\int_x \hf \phi \frac{\dd S}{\dd \phi(x)}
\ee
According to \eqref{co}
\be \label{co1}
\frac{\p S_\lm}{\p \la_0} \equiv \int _x \frac{[\phi^4]_\lm}{4!}= (1+ \frac{\p \beta(\la_0)}{\p \la_0}t)\frac{\phi^4}{4!}+ m_0^2  \frac{\p \gamma_m(\la_0)}{\p \la_0}t\hf \phi^2 +\hf \frac{\p \eta(\la_0)}{\p \la_0} t {\cal N} +O(1/\lm)]
\ee
So,
\be
\boxed{
\frac{\p}{\p t} \int _x \frac{[\phi^4]_{\lm}}{4!}= (\frac{\p \beta(\la_0)}{\p \la_0})\frac{\phi^4}{4!}+ m_0^2  \frac{\p \gamma_m(\la_0)}{\p \la_0}\hf \phi^2+ \hf\frac{\p \eta(\la_0)}{\p \la_0}  {\cal N}+O(1/\lm)]}
\ee

From chapter 11 of \cite{Kleinert,zinn}\footnote{The coupling constants in the relevant equations in these two books differ by a factor of 2}, we get $\beta(\la_0)$ in our convention as,

\begin{align}
\beta(\la_0)= \la_0(\eps-\frac{1}{16\pi^2}3\la_0+\frac{1}{(16\pi^2)^2}\frac{17}{3}\la_0^2)
\end{align}

In the critical theory we can set $m_0^2=0$. 
So if we collect the coefficient of $\phi^4$ we get what we have defined above as $d_m$ in the ERG evolution (we denote anomalous dimension of irrelevant operator as $d_4$ and that of relevant operator as $d_2$):

\begin{align}\label{expect_irr}
\nonumber d_4= &~\eps-\frac{1}{16\pi^2}6\lambda_0+\frac{1}{(16\pi^2)^2}17 \la_0^2+\frac{4}{(16\pi^2)^2}\frac{2\la_0^2}{12}\\
=&~ \eps -\frac{1}{16\pi^2}6\lambda_0+\frac{1}{(16\pi^2)^2}\frac{53\la_0^2}{3}
\end{align}

%
%
%

For the relevant operator $\phi^2$, analogously one can define it as 
\be
\frac{\p S_\lm}{\p m_0^2} =\int _x \frac{[\phi^2]_\lm}{2}
\ee

So applying this to \eqref{2.17}
\be
\frac{\p}{\p t} \int _x \frac{[\phi^2]_\lm}{2}
= \gamma_m \int _x \frac{[\phi^2]_\lm}{2}
\ee
From \cite{Kleinert,zinn}\footnote{There is a factor of two in the definition of $d_m$} we get for the two loop anomalous dimension

\be
\gamma_m =\frac{\la_0}{16\pi^2}-\frac{1}{(16\pi^2)^2}\frac{5}{6}\la_0^2
\ee

So length scaling dimension $d_2$ (in our notation) of the relevant operator $\int_x \phi^2$ is given by,

\be\label{expect_rel}
d_2= 2-\frac{\la_0}{16\pi^2}+\frac{1}{(16\pi^2)^2}\frac{5}{6}\la_0^2
\ee

Note that the results of \cite{Kleinert,zinn} are obtained using the mass independent dimensional renormalization scheme or ``minimal subtraction". The scheme used in this paper is also mass independent. In mass independent schemes the first two orders in the power series expansion of the beta function are well known to be scheme independent. 
(Proof: Let
\[
\beta (\la)=\frac{d\la}{dt}=b_2 \la^2 +b_3 \la^3
\] 
Let 
\[
\la'=\la + a \la^2
\]
and 
\[
\beta'(\la')=\frac{d\la'}{dt}=b'_2\la'^2 +b'_3 \la'^3=b'_2(\la + a \la^2)^2 + b_3'(\la + a \la^2)^3= b'_2\la^2 +(b'_3 +2ab'_2)\la^3 +... 
\]
But also
\[\frac{d\la'}{dt}=\beta (\la) + a 2\la \beta (\la)
=b_2 \la^2 +b_3 \la^3 + 2a\la (b_2 \la^2 +b_3 \la^3) =b_2\la^2+ (b_3+2ab_2)\la^3\]
Comparing, we see that $b_2=b'_2$ and $b_3=b'_3$.)
Thus, upto and including $O(\la^3)$, the beta functions in the ERG calculation and in dimensional regularization MS scheme are identical. This also means that at the fixed point (given by vanishing of beta function) the expresions relating $\eps$ and $\la$ are scheme independent to the same order. 
Now, at the fixed point, the dimensions of operators expressed in terms of $\eps$  are eigenvalues of the dilatation operator of the CFT and thus universal (to any order in $\eps$).
These universal expressions in powers of $\eps$, when re-expressed in terms of $\la$, will thus have to match to the lowest two orders in any mass independent scheme. Thus the expressions obtained for $d_2,d_4$ in the ERG scheme must agree with the expressions given above. These expectations will be confirmed in Sections 3 and 4. 

\subsection{Gaussian Theory ERG}
As mentioned above, one  fixed point action is the free scalar field theory in four (or any other) dimension. As a warm up exercise let us solve the eigenvalue equation \eqref{eigen} for the two operators, $\phi^2,\phi^4$ discussed above. 

The composite operators found here will be the $\la\to 0$ limit of the composite operators at the Wilson-Fisher fixed point in next section.

The action we take to be

\be S=\hf \int \Dp \phi(p)\frac{p^2}{K(p)}\phi(-p)\ee

It obeys Polchinski equation: 

\be \frac{\p S}{\p t}= \int_p \lbrace -K'(p^2) \rbrace [\frac{\dd^2S}{\dd \phi(p)\dd \phi(-p)}-\frac{\dd S}{\delta \phi(p)}\frac{\dd S}{\delta \phi(-p)}] -2\frac{p^2K'}{K}\phi(p)\frac{\dd S}{\delta \phi(p)}+\nonumber\ee

\be +[(1-\Dt)N_\phi+D-N_p]S\ee
and is also a fixed point solution, i.e.
\be
\frac{\p S}{\p t}=0
\ee

(Anomalous dimension $\frac{\eta}{2}$, beta function $\beta(\lambda)$ has been set to zero since it is a
Gaussian fixed point.)

Let \(\DD S (q)\) be a {\em local} composite operator of momentum $q$ with definite dimension -
 added to the action. So as a composite
operator it obeys the linearized equation

\be \frac{\p \DD S(q)}{\p t}= \int_p \lbrace -K'(p^2)\rbrace [\frac{\dd^2 \DD S(q)}{\dd \phi(p)\dd \phi(-p)}-2\frac{\dd S}{\delta \phi(p)}\frac{\dd \DD S(q)}{\delta \phi(-p)}] -2\frac{p^2K'}{K}\phi(p)\frac{\dd \DD S(q)}{\delta \phi(p)}+\nonumber\ee

\be +\underbrace{[(1-\Dt)N_\phi-N_p]}_{{\cal G}_{dil}^c}\DD S(q)= (d_m +\qdq) \DD S(q) \label{3}\ee

Here \(d_m\) is the length dimension.

The expression \(N_p\) in \({\cal G}_{dil}^c\) in \(\eqref{3}\) stands
for \(\sum _i p_i \frac{\p}{\p  p_i}\). 

Take

\be \DD S(q) = \hf \int_{p_1} \int _{p_2}A(p_1,p_2,q)\phi(p_1)\phi(p_2)\ee

The second and third term in \(\eqref{3}\) cancel (and the first term is
field independent), so we get

(set \(D=4-\eps\))

\be (d_m +\qdq)A(p_1,p_2,q)=(2-D-\sum_{i=1}^2 p_i\frac{\p}{\p p_i} )A(p_1,p_2,q)\ee

\begin{enumerate}
\def\labelenumi{\arabic{enumi}.}
\item
From $\eqref{gdil}$ we see that 
\[
A(p_1,p_2,q)=\dd(p_1+p_2-q)~~~
\]
satisfies this equation. Note that $d_\phi^x=\Dt-1$ so $d_m= -2d_\phi^x+D=2$.
This is the (length) dimension of $\int_p \phi(p)\phi(q-p)$ as mentioned earlier.

\item
  Take \(A(p_1,p_2,q)=p_1.p_2\dd (p_1+p_2-q)\). We get

  \be (d_m +\qdq)A(p_1,p_2,q)=(2-D-\sum_{i=1}^2 p_i\frac{\p}{\p p_i} )A(p_1,p_2,q)\ee
  From \eqref{a68} and the subsequent discussion we see that $d_m=0$.
%

\item  Now we consider higher dimensional operators:

  Take

  \be \DD S(q) = \frac{1}{4!}\int_{p_1,p_2,p_3,p_4}B(p_1,p_2,p_3,p_4,q)\phi(p_1)\phi(p_2)\phi(p_3)\phi(p_4)\nonumber\ee

  \be +\hf \int_{p_1} \int _{p_2}A(p_1,p_2,q)\phi(p_1)\phi(p_2)\ee

Assume once again that this operator has definite momentum $q$. We get

\be (d_m+\qdq) \left \lbrace \frac{1}{4!}\int_{p_1,p_2,p_3,p_4}B(p_1,p_2,p_3,p_4,q)\phi(p_1)\phi(p_2)\phi(p_3)\phi(p_4) +\hf \int_{p_1} \int _{p_2} A(p_1,p_2,q) \phi(p_1) \phi(p_2) \right \rbrace \nonumber\ee

\be = -\hf\int_pK'(p^2)\int_{p_1,p_2}B(p_1,p_2,p,-p,q)\phi(p_1)\phi(p_2) +[(1-\Dt)2 -\sum _i p_i\frac{\p}{\p p_i} ]\hf \int_{p_1} \int _{p_2} A(p_1,p_2,q) \phi(p_1) \phi(p_2)) \nonumber\ee

\be [(1-\Dt)4 -\sum _i p_i\frac{\p}{\p p_i} ]\frac{1}{4!}\int_{p_1,p_2,p_3,p_4}B(p_1,p_2,p_3,p_4,q)\phi(p_1)\phi(p_2)\phi(p_3)\phi(p_4)\ee

We see that a quartic term cannot be an eigen-operator by itself - need
a quadratic piece.

For simplicity if we take 

\[B(p_1,p_2,p_3,p_4,q)=  \dd(p_1+p_2+p_3+p_4-q)\]
 \[ A(p_1,p_2,q)=A \dd(p_1+p_2-q)\]

we find ($D=4-\eps$) using \eqref{dfn} and its  generalization:
\[
\sum_{i=1}^4p_i\frac{\p}{\p p_i} \dd(\sum_{j=1}^4 p_j-q)=-D\dd(\sum_{j=1}^4 p_j-q)+\qdq\dd(\sum_{j=1}^4 p_j-q)
\]

from the $\phi^4$ term:
\be d_m - (4-2D)+D =0 \implies d_m=\eps ~~~\ee
This operator is relevant in the Gaussian theory in $D<4$ as is also obvious from simple dimensional analysis.

From the quadratic term we get an equation for \(A\) 

\be 
\dd(\sum_{j=1}^2 p_j-q)[\hf F + (2-D)\frac{A}{2} +D \frac{A}{2}]+ \qdq \frac{A}{2}\dd(\sum_{j=1}^2 p_j-q) = (\eps+\qdq) \frac{A}{2}\dd(\sum_{j=1}^2 p_j-q)
\ee
where 
\[
F=\int_p(-K'(p^2))=\frac{1}{16\pi^2}
\]
Since \(d_m=\eps\), \(A= -\frac{F}{2-\eps}\). Thus our operator is

\be \DD S=\frac{1}{4!}\int _{p_1,p_2,p_3}\phi(p_1)\phi(p_2)\phi(p_3)\phi(-p_1-p_2-p_3+q)-\frac{F}{2-\eps}\hf  \int_p \phi(p)\phi(q-p)\ee

which agrees with \eqref{2.2} if we take $u=\frac{1}{4!}$  for $q=0$ and $\eps =0$.
\end{enumerate}

\section{ Wilson-Fisher Composite operator at the leading order}
\label{secB}

In this section we will construct, for the Wilson-Fisher fixed point theory, the two lowest dimension composite operators that were studied in the last section for the Gausian fixed point theory namely $\phi^2$ and $\phi^4$. We know the Wilson-Fisher fixed point action from \cite{Dutta}. $\phi^2$ is a relevant operator at both fixed points. $\phi^4$, which was relevant at the Gaussian fixed point in $D=4-\eps$ (and marginal in $D=4$) turns out to be irrelevant at the W-F fixed point. We use perturbation theory in $\la$. In principle one can also do perturbation in $\eps$. At the W-F fixed point $\la \approx \eps$ and there is not much difference. However even in the Gaussian theory in $D=4-\eps$, we have seen that $\eps$ shows up in the dimension so it is clear that the two expansions are in principle different. The relevant and irrelevant operator for WF fixed point is denoted by $\mathcal{O}_2(q)$ and $\mathcal{O}_4(q)$. Though for simplicity we have taken external momentum $q=0$ for all the calculation except while finding $\mathcal{O}_2^{(1)}(q)$.
In the calculation both in this section and the next one, in principle one can put the fixed point condition right in the begining itself to interpret $O(\lambda^n)$ terms as $O(\eps^n)$, but there is a subtlety  there - \hspace{0.01 in} ideally all the momentum integrations are to be done in $D=4-\eps$ dimensions. So there are implicit factors of $\epsilon$ hidden in there. It therefore makes sense to keep track of $\eps$ and $\lambda$ separately and to take $\la
=O(\eps)$ in the end at the fixed point. Our expressions are in general true for general $D=4-\eps$, but while calculating the anomalous dimension, in order to compare with known
results for $\phi^4$ in $D=4$ \cite{Kleinert,zinn} that have been
obtained using dimensional regularization, we have performed the final integrals in four dimensions. 

\vspace{0.1 in}

 We make the following general ansatz for both $\mathcal{O}_2(q)$ and $\mathcal{O}_4(q)$ as :
 \[ \DD S(q) =\hf \int_{p_1} \int _{p_2}A(p_1,p_2)\phi(p_1)\phi(p_2)\]

\[ + \frac{1}{4!}\int_{p_1,p_2,p_3,p_4}B(p_1,p_2,p_3,p_4)\phi(p_1)\phi(p_2)\phi(p_3)\phi(p_4)\]
\be \label{DeltaS}
+ \frac{1}{6!}\int_{p_1,...p_5,p_6}D(p_1,...,p_6)\phi(p_1)...\phi(p_6)~~~+~~~~O(\phi^8)~~~+...\ee


We will assume an ansatz of the form:
\br
A(p_1,p_2)&=& \dd(p_1+p_2-q)[A^{(0)} + A^{(1)}(p_1,p_2,q)+...] \nonumber \\
B(p_1,p_2,p_3,p_4)&=& \dd(p_1+p_2+p_3+p_4-q)[B^{(0)} +  B^{(1)}(p_1,p_2,p_3,p_4,q)+...]\nonumber \\
D(p_1,p_2,p_3,p_4,p_5,p_6)&=& \dd(p_1+p_2+p_3+p_4+p_5+p_6-q)[ D^{(1)}(p_1,p_2,p_3,p_4,p_5,p_6,q)+...] \\
\er

Further we will write each term as a sum of several terms with different momentum structures. For instance $B^{(1)}$  will turn out to be:
\be
B^{(1)}(p_1,p_2,p_3,p_4,q)=\la \sum _{i=1}^4 B_I(p_i,q)+ \la \hf \underbrace{\sum_{i,j=1,2,3,4}}_{6~perm}B_{II}(p_i+p_j,q)+...
\ee

For the irrelevant operator, $\mathcal{O}_4(q)$, our starting approximation will be to take $B^{(0)}=1$. Thus
 \be
 B(p_1,..,p_4)=\dd(p_1+p_2+p_3+p_4-q)[1+O(\la)] 
 \ee
 Since even in the Gaussian theory this is accompanied by a $\phi^2$ term it is clear that $A^{(0)}$ also starts at $O(1)$.
 Thus
 \be 
A(p_1,p_2)=\dd (p_1+p_2-q)[\frac{F}{\eps-2}+O(\la)]
\ee
 Everything else is $O(\la)$ or higher.

\vspace{0.02 in}

On the other hand for the relevant operator , $\mathcal{O}_2(q)$ 
we start with 
\be 
A(p_1,p_2)=\dd(p_1+p_2-q)[1+O(\la)]
\ee 
and everything else is higher order in $\la$. 

The strategy will be to take these as the starting inputs and solve the linearized ERG equation \eqref{main} order by order in $\la$. Typically at each order the coefficient of a new higher dimensional irrelevant operator enters the equation.

We write the WF fixed point action at the first order of $\la$.

\[S= \hf \int_p \left(\frac{p^2}{K}+U_2(p)\right)\phi(p)\phi(-p) + \frac{1}{4!}\int_{p_1,p_2,p_3}(\la +U_4)\phi(p_1)\phi(p_2)\phi(p_3)\phi(p_4)~~\]\[p_4=-p_1-p_2-p_3 \]\[+\frac{1}{6!} \int_{p_1,...p_5}V_6 \phi(p_1)...\phi(p_6)~~\]\[~p_6=-p_1-...-p_5\]

\[U_2=-\la \frac{1}{2-\eps}\underbrace{\int_p (-K'(p^2))}_{F} +O(\la^2)\]\[
V_6=-\la^2 \sum_{10~perm~i,j,k}h(p_i+p_j+p_k)\]\[
h(p)\equiv\frac{1-K(p^2)}{p^2}\]

\[ U_4 \approx O(\la^2)\]

We number the Polchinski's equation (part with the anomalous dimension is not required at this order since $\eta \approx O(\la^2)$) in the following way:

\begin{align}\label{pol_lead}
\nonumber \frac{\p \DD S(q)}{\p t}= &\int_p -K'(p^2) [\underbrace{\frac{\dd^2 \DD S}{\dd \phi(p)\dd \phi(-p)}}_{\textbf{(1)}}\underbrace{-2\frac{\dd S}{\phi(p)}\frac{\dd \DD S}{\phi(-p)}}_{\textbf{(2)}}] \underbrace{-2\frac{p^2K'}{K}\phi(p)\frac{\dd \DD S}{\phi(p)}}_{\textbf{(3)}}++\underbrace{[(1-\Dt)N_\phi+D-N_p]}_{{\cal G}_{dil}^c=\textbf{(4a)}}\DD S\\
&= (d_m +\qdq)\DD S(q)
\end{align}
The second equality is the requirement that $\DD (q)$ be a scaling operator of length dimension $d_m$. Note that we donot have to include the term $\beta(\lambda) \frac{\p \Delta S}{\p \la}$ in this order.
We have calculated different parts of \eqref{pol_lead} in Appendix \ref{appB}. we collect all the results we have found at leading order below. In next two subsections we have derived them.

 The anomalous dimension at the leading order we get as,

\begin{subequations}\label{dim_lead}
\begin{empheq}[box=\fbox]{align}
 d_2= &2-\la F \\ 
d_4= & \eps-6F\la 
\end{empheq}
\end{subequations}

The corresponding eigenvectors are given by,

\begin{subequations}\label{relevant_lead}
\begin{empheq}[box=\fbox]{align}
 \nonumber &\mathcal{O}_2(q)=\hf \int_{p_1,p_2}\dd(p_1+p_2-q)[1+\la \frac{F}{2}(h(p_1-q)+h(p_2-q))+ \la{\cal F}(q)]~~\phi(p_1)\phi(p_2)\\
 \nonumber &-\frac{1}{4!}\int_{p_1,p_2,p_3}\dd(p_1+p_2+p_3+p_4-q)\la \sum_{i=1}^4 h(p_i-q)\phi(p_1)\phi(p_2)\phi(p_3)\phi(p_4)\\
 \end{empheq}
 \end{subequations}
 
\begin{subequations}\label{irrelevant_lead}
\begin{empheq}[box=\fbox]{align}
\nonumber &\mathcal{O}_4(0)= -\frac{1}{6!}\int_{p_1,p_2,p_3,p_4,p_5}\sum_{10~perm~(i,j,k)}2 \la h(p_i+p_j+p_k)\phi(p_1)\phi(p_2)\phi(p_3)\phi(p_4)\phi(p_5)\phi(p_6)\\
\nonumber & +\frac{1}{4!}\int_{p_1,p_2,p_3} \left[\sum_{i=1}^4 F \la h(p_i) -\sum_{3~perm~(i,j)} 2\la \mathcal{F}(p_i+p_j)\right] \phi(p_1)\phi(p_2)\phi(p_3)\phi(p_4)\\
& +\hf \int_{p_1} \frac{F}{d_4-2} \phi(p_1)\phi(p_2)
\end{empheq}
\end{subequations}

Where all the functions have been defined in Appendix \eqref{appF}. Note that here we have not found 2-pt vertex of first order, because that is not required to find the anomlous dimension at first order. We have calculated that in the next section where it is needed to find second order anomalous dimension

\subsection{The Relevant Operator}

We start with $A=1$ and $d_2\approx 2$.
\br
A(p_1,p_2)&=&\dd(p_1+p_2-q)[1+ A^{(1)}(p_1,p_2,q) +...\nonumber\\
B(p_1,p_2,p_3,p_4)&=&\dd(p_1+p_2+p_3+p_4-q)[ \sum_{i=1}^4 B_I^{(1)}(p_i,q)+ \hf \sum_{6~perm~(i,j)}B_{II}^{(1)}(p_i+p_j,q)...]\nonumber \\
D(p_1,p_2,p_3,p_4,p_5,p_6)&=&\dd(p_1+p_2+p_3+p_4+p_5+p_6-q)[ D^{(1)}(p_1,p_2,p_3,p_4,p_5,p_6,q)+...]\nonumber \\
d_2&=&2+ d_2^{(1)}+...\nonumber\\
&&  \label{relans}
\er
It turns out that at leading order we can set $B_{II}^{(1)}(p_I+p_j,q)=D^{(1)}(p_1,p_2,p_3,p_4,p_5,p_6,q)=0$. 

\subsubsection{$O(\la)$ Equation for $\phi^4$}

\[
-\frac{2\la}{4!} \sum_{i=1}^4 -K'((p_i-q)^2)\dd(\sum p_i-q)
+\frac{1}{4!}((4-D) +\qdq)\dd(\sum p_i-q)(\sum _{i=1}^4 B_I^{(1)}(p_i,q))
\]
\[
-\frac{1}{4!}\dd(\sum p_i-q)(\sum_i p_i\frac{\p}{\p p_i} +\qdq)\sum_{i=1}^4 B_I^{(1)}(p_i,q)= (d_2 +\qdq)\dd(\sum p_i-q)(\sum _{i=1}^4 B_I^{(1)}(p_i,q))
\]
Canceling terms and dropping $O(\eps \la)$ or $O(\la^2)$ terms we get
\[
-2\la (-K'((p_i-q)^2))-(\sum_i p_i\frac{\p}{\p p_i} +\qdq)B_I^{(1)}(p_i,q)= 2 B_I^{(1)}(p_i,q)
\]
This is solved by 
\be \label{B1}
B_I^{(1)}(p_i,q)=-\la h(p_i-q)
\ee
The $\phi^2$ terms in the equation are:

\[
\frac{F}{2}\dd(p_1+p_2-q)(B_I(p_1,q)+B_I(p_2,q)) + \hf \int_p (-K'(p^2))(B_I(p,q)+B_I(-p,q))\dd(p_1+p_2-q)
\]
\[
+\frac{\la F}{2-\eps} \sum _i -K'(p_i^2) \dd(p_1+p_2-q)
\]
\[+
\hf(2-D+D+\qdq)\dd(p_1+p_2-q)(1+ A^{(1)}(p_1,p_2,q)) - \hf\dd(p_1+p_2-q)(\sum_i p_i\frac{\p}{\p p_i} +\qdq)A^{(1)}(p_1,p_2,q) 
\]
\[=
(d_2 +\qdq)\dd(p_1+p_2-q)\hf(1+ A^{(1)}(p_1,p_2,q))
\]

$\mathbf {O(1)}$

The $O(1)$ part of this equation (after canceling terms) gives 
\be
d_2^{(0)}=2
\ee

$\mathbf {O(\la)}$

We substitute \eqref{B1} in the $O(\la )$ part to get

\[
\frac{\la F}{2}[-h(p_1-q)-h(p_2-q)] + \la \hf \int_p (-K'(p^2))[-h(p-q)-h(p+q)]
\]
\[+
\frac{\la F}{2-\eps} [-K'((p_1-q)^2)-K'((p_2-q)^2)]
\]
\[+
A^{(1)}(p_1,p_2,q) -\hf (\sum_i p_i\frac{\p}{\p p_i} +\qdq)A^{(1)}(p_1,p_2,q)\]
\[= d_2^{(1)} \hf + A^{(1)}(p_1,p_2,q)
\]

The second term of the first line can be rewritten as
\be \label{31}
\hf \la \int_p (-K'(p^2))[(h(p)-h(p-q))+(h(p)-h(p+q))]
+\hf \la \int_p (-K'(p^2))[-2h(p)]
\ee

The $q$ independent term evaluates to  $-F$ and we thus get
\be
\tcboxmath{d_2^{(1)}=-\la F \approx -\frac{\eps}{3}}
\ee

The first term in \eqref{31} which is independent of $p_i$ can be canceled by choosing 
\be 
A^{(1)}(p_1,p_2,q)= \hf  \la({\cal F}(q) +{\cal F}(-q))=\la {\cal F}(q)
\ee
which is defined in Appendix F. Note that ${\cal F}(0)=0$.

The remaining equation is satisfied by setting
\be\label{S2lead_rel}
A^{(1)}(p_1,p_2,q)= \frac{\la F}{2} (h(p_1-q) +h(p_2-q))
\ee

%
%
%
%
%

 $A^{(1)}(p)$ looks like first diagram in Fig\ref{relevant_lead}.

\begin{figure}[b]
   \centering
  \begin{tikzpicture}
  \begin{feynman}
    	\vertex (a) at (0,0);
    	\vertex (b) at (2,0);
    	\vertex (c) at (4,0);
    	\vertex (d) at (6,0);
    	\vertex (x) at (2,2);
      \diagram* {
        (a) --[edge label=$\phi(p)$,insertion=1.0](b)--[edge label=$h(p-q)$,insertion=1.0](c)--[edge label=$\phi(q-p)$](d),
        (x) --[photon, edge label= $\phi^2(q)$](b), 
         };
    \end{feynman}
  \end{tikzpicture}
  \hspace{0.2 in}
  \begin{tikzpicture}
  \begin{feynman}
    	\vertex (a) at (0,0);
    	\vertex (b) at (2,0);
    	\vertex (c) at (4,0);
    	\vertex (x) at (2,3);
    	\vertex (y) at (2,2);
      \diagram* {
        (a) --[near start, edge label=$\phi(p)$](b)--[ near end,edge label=$\phi(q-p)$](c),
        (b) --[half left, looseness=1.2, edge label= $h(q-p)$](y)--[half left, looseness=1.2, edge label=$\left \lbrace-K^\prime(p)\right \rbrace$](b),
        (x) --[photon, insertion=1.0, edge label= $\phi^2(q)$](y),
         };
    \end{feynman}
  \end{tikzpicture}
  \caption{The left diagram is for the the relevant operator $A^{(1)}(p)$. The right one is the diagrammatic representation of the term contributing to the anomalous dimension $d_2^{(1)}$. Note that the right diagram is a loagarithmically divergent diagram made finite by replacing the propagator $h(p)$ by $-K^\prime(p^2)$. It is the $q$ independent part that gives $d_2^{(1)}$.}
  \label{relevant_lead}
\end{figure}
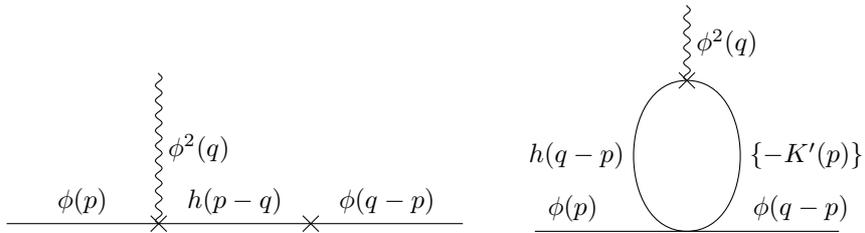

 This gives $d_2^{(1)}=-\la F = -\frac{\eps}{3}$. The value of $d_2^{(1)}$ is coming from the second diagram in Fig. \ref{relevant_lead}. As expected the origin of the anomalous dimension is the logarithmically divergent diagram.

Thus the relevant eigen-operator  and its dimension is given as:

\[
\tcboxmath{
 d_2=2-\la F
 }
\]
\[
\tcboxmath{
\mathcal{O}_2(q)= \hf \int_{p_1,p_2}\dd(p_1+p_2-q)[1+\la \frac{F}{2}(h(p_1-q)+h(p_2-q))+ \la{\cal F}(q)]~~\phi(p_1)\phi(p_2)}
\]
\[
\tcboxmath{
-\frac{1}{4!}\int_{p_1,p_2,p_3}\dd(p_1+p_2+p_3+p_4-q)\la \sum_{i=1}^4 h(p_i-q)\phi(p_1)\phi(p_2)\phi(p_3)\phi(p_4)}
\]

Note that the value of the anomalous dimension agrees with \eqref{expect_rel} to this order.

\subsection{The Irrelevant Operator}

For simplicity we set $q=0$. The ansatz simplifies to (momentum conservation is implicit, i.e. $\sum_i p_i=0$):

\[A(p)=A^{(0)}+ A^{(1)}(p) +...\]\[
B(p_1,p_2,p_3,p_4)=B^{(0)}(p_1,p_2,p_3,p_4)+B_I^{(1)}(p_1,p_2,p_3,p_4)+B_{II}^{(2)}(p_1,p_2,p_3,p_4)\]
\[= \sum_{i=1}^4 B^{(0)}(p_i)+ \sum_{i=1}^4 B_I^{(1)} (p_i)+  \sum_{3~perm~(i,j)} B_{II}^{(1)}(p_i+p_j)\]\[
D(p_1,p_2,p_3,p_4,p_5,p_6)= D^{(1)}(p_1,p_2,p_3,p_4,p_5,p_6)\]
\[ d_4= d_4^{(1)}+...\]

Below we are writing $\phi^2$, $\phi^4$ and $\phi^6$ terms separtely to obtain different quantities.

\subsubsection{\texorpdfstring{Equation for \(\phi^2 \) -
\(O(1)\):}{Equation for \textbackslash phi \textbackslash phi - O(1):}}

Different parts of \eqref{pol_lead} gives,

\textbf{(1)}

\[\int_q \left \lbrace -K'(q^2)\right \rbrace \lbrace B^{(0)}(q)+B_I^{(1)}(q) \rbrace + F \lbrace B^{(0)}(p)+ B^{(1)}(p) \rbrace  +\hf F\la B_{II}(0) -\hf \int_q K'(q^2)\la[B_{II}(p+q)+B_{II}(p-q)]\]

\textbf{(2)+(3)}

\[-2(-\kp )U_2(p) A^{(0)}(p)\]

\textbf{(4a)}

\[A^{(0)}-p^2\frac{d}{dp^2}A^{(0)}\]

Collecting terms of $O(1)$:

\[A^{(0)}(p)-p^2 \frac{d}{dp^2}A^{(0)}(p)+ \hf F (4B=1)=d_4 \frac{A^{(0)}(p)}{2}\]

Assuming that \(A^{(0)}(p)\) is a constant and \(O(1)\) we obtain (neglecting
\(O(\la)\) terms)

\begin{align}\label{A0}
\tcboxmath{A^{(0)}=\frac{F}{d_4-2}\approx -\hf F}
\end{align}

\(d_4\) is expected to be of \(O(\eps)\) since \(\phi^4\) is marginal in
\(D=4\).

\subsubsection{\texorpdfstring{Equation for
\(\phi^6\):}{Equation for \textbackslash phi\^{}6:}}

Now we turn to the \(\phi^6\) equation:

\[-\frac{4}{6!}\int_{p_1,..p_5} \sum_{10~perm}-K'((p_i+p_j+p_k)^2)\la (4B=1)\phi(p_1)...\phi(p_6)\]\[+\frac{1}{6!}\int_{p_1...p_5}(6-2D-2\sum_i2p_i^2\frac{d}{dp_i^2})\sum_{10~perm} D^{(1)}(p_i+p_j+p_k)\phi(p_1)...\phi(p_6)\]\[= \frac{d_m}{6!}\int_{p_1,..p_5}\sum_{10~perm} D^{(1)}(p_i+p_j+p_k)\phi(p_1)...\phi(p_6)\]

At order $\la$ the equation is

\[(1+p^2\frac{d}{dp^2}) D^{(1)}(p)=2\la \kp\]

considering \eqref{h} we see that

\begin{align}\label{S6lead_irr}
\tcboxmath{D^{(1)}(p)=-2 \la h(p)}
\end{align}

is the solution to this order.

\subsubsection{\texorpdfstring{Equation for
\(\phi^4\):}{Equation for \textbackslash phi\^{}4:}}

Now we turn to the \(\phi^4\) equation:

\textbf{(1)}

\[\frac{1}{4!} \int_p \lbrace -\kp \rbrace \int_{p_1,p_2,p_3}\left[ D^{(1)}(p_1)+D^{(1)}(p_2)+D^{(1)}(p_3)+D^{(1)}(p_4)\right]\phi(p_1)...\phi(p_4) ~~(~``type~1")\]\[+
\frac{1}{4!} \int_p \lbrace -\kp \rbrace\int_{p_1,p_2,p_3}[D^{(1)}(p+p_1+p_2)+D^{(1)}(p+p_1+p_3)+D^{(1)}(p+p_1+p_4)\]\[+D^{(1)}(p-p_1-p_2)+D^{(1)}(p-p_1-p_3)+D^{(1)}(p-p_1-p_4)]\phi(p_1)...\phi(p_4)~~~(``type~2")\]

We have written the expression in the first line as type 1, because we will see below that quadratically divergent 4-pt vertex will be obtained from these expresssions, while from type 2 expressions logarithically divergent 4-pt vertex  will be obtained. We will see the contribution from type 1 diagram will be cancelled and those from the type 2 diagram will contribute to the anomalous dimension.

\textbf{(2)+(3)}

\[(-2)\left[\frac{1}{4!}\int_{p_1,p_2,p_3}[\sum_i\lbrace -\kpi)A(p_i \rbrace]\la +\frac{1}{4!}\int_{p_1,p_2,p_3}[\sum_i\lbrace-\kpi\rbrace U_2(p_i)][1+...]\right]\phi(p_1)...\phi(p_4)\]

\textbf{(4a)}

\[\frac{1}{4!}\int_{p_1,p_2,p_3}(4-D-2\sum_i p_i^2\frac{d}{dp_i^2})[1+\sum_{i=1}^4 B_I^{(1)}(p_i)+ \sum_{3~perm(i,j)}B_{II}^{(1)}(p_i+p_j)]\phi(p_1)..\phi(p_4)\]

Collect type 1 terms and (2)+(3) part above, we get,

\[\int_p\lbrace -\kp \rbrace[D^{(1)}(p_1)+D^{(1)}(p_2)+D^{(1)}(p_3)+D^{(1)}(p_4)]+2[\sum_i(\kpi)A(p_i)]\la +2[\sum_i(\kpi)U_2(p_i)]\]\[
+(4-D-2\sum_i p_i^2\frac{d}{dp_i^2})[1+\sum_{i=1}^4  B_I^{(1)}(p_i)]= d_4 \sum_{i=1}^4B_I^{(1)}(p_i)\]

%
%
%
%
%

Ignoring $\mathcal{O}(\eps \la)$ or $O(\la^2)$ terms from \eqref{h} we get,

\begin{align}\label{S4lead1_irr}
\tcboxmath{B_I^{(1)}(p)=\la Fh(p)}
\end{align}

\begin{figure}[b]
   \centering
  \begin{tikzpicture}
  \begin{feynman}
    	\vertex (e) at (1,0);
      \vertex (m) at ( 0, 0);
      \vertex (n) at ( 3,0);
      \vertex (x) at (-2,0){\(\phi\)};
      \vertex (y) at (5,0);
      \vertex (a) at (-1,-2){\(\phi\)};
      \vertex (b) at ( 5,0){\(\phi\)};
      \vertex (c) at (-1, 2){\(\phi\)};
      \vertex (d) at (3,2);
      \diagram* {
        (m) -- [edge label=$h(p_4)$](n),
      	(m) -- [edge label =$p_2$](x),
        (a) -- [edge label'=$p_3$](m),
        (c) -- [edge label = $p_1$](m),
        (b) -- [near start,edge label = $p_4$,insertion=1.0] (n),
        (d) -- [photon,edge label= $\phi^4(0)$] (n),
         };
    \end{feynman}
   \end{tikzpicture}
   \hspace{0.2 in}
   \begin{tikzpicture}
  \begin{feynman}
    	\vertex (e) at (1,0);
      \vertex (m) at ( 0, 0);
      \vertex (n) at ( 4,0);
      \vertex (x) at (-2,0);
      \vertex (a) at (-1,-2){\(\phi\)};
      \vertex (c) at (-1, 2){\(\phi\)};
      \vertex (d) at (3,2);
      \vertex (e) at (5,2);
      \vertex (f) at (5,-2);
      \diagram* {
        (m) -- [half left, looseness=1.0, edge label=$h(p_i+p_j+q)$](n) -- [half left, looseness=1.0, edge label=$h(q)$](m),
      	(m) -- [photon,edge label =$\phi^4(0)$, insertion=0.0](x),
        (a) -- [edge label= $p_j$](m),
        (c) -- [edge label = $p_i$](m),
        (e) -- (n),
        (f) -- (n),
         };
    \end{feynman}
  \end{tikzpicture}
  \caption{The left diagram represents Type-I diagram corresponding to $B_I(p)$, while the right one represents type-II diagram representing $B_{II}(p_i+p_j)$. Anomalous dimension is coming from the process of making the latter diagram zero at zero external momenta. Note that the $B_{II}(p_i+p_j)$ is nothing but the usual logarithmic divergent  diagram made finite by adding a counterterm.}
  \label{irrelevant_lead}
\end{figure}
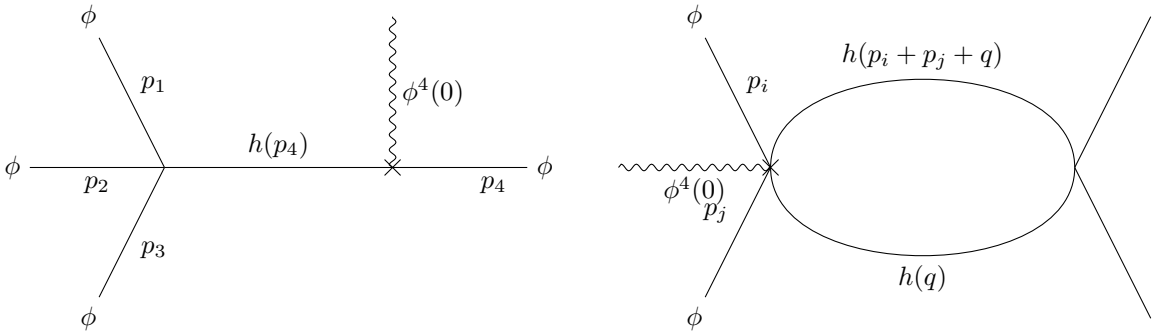

Where $h(p)=\frac{1-K(p^2)}{p^2}$. It looks like first diagram in Fig.\ref{irrelevant_lead}.

\vspace{0.2 in}

The leftover terms on LHS is

\[(4-D)- d_4 \sum_{i=1}^4 B_I^{(1)}(p_i)\]

We will keep record of all leftover terms in LHS as we need those in sub-leading order calculation.

Now we collect type 2 terms and the rest of the equation, we have put
\(D(p)=-2\lambda h(p)\).

\[4\lambda\int_p\kp[h(p+p_1+p_2)+h(p+p_1+p_3)+h(p+p_1+p_4)]\]\[
 +(4-D-\sum_{l=1}^4 p_l.\frac{d}{dp_l}) [B_{II}^{(1)}(p_1+p_2)+B_{II}^{(1)}(p_1+p_3)+B_{II}^{(1)}(p_1+p_4)]=  d_4 \]

Considering \eqref{F}, if we add and subtract $6F\la$ as momentum independent term we get,

\begin{align}\label{S4lead2_irr}
 \tcboxmath{B_{II}^{(1)}(p)=-2 \la \mathcal{F}(p)}
\end{align}

which looks like second diagram in Fig.\ref{irrelevant_lead}. Here $\mathcal{F}(p)= \frac{1}{2}\int_q \left \lbrace h(p+q)h(q)-h(q)h(q) \right\rbrace $, it is defined in \eqref{F} in Appendix \ref{appF}.

While the Leftover terms in the L.H.S are:

\[(4-D) - d_4\sum_{i=1}^4 B^{(1)}(p_i)-6F\lambda\]

Keeping only \(\lambda^1\) and $\eps^1$ terms and equating with R.H.S we get,

\[4-D-6F\lambda= d_4 \sum_{i=1}^4 B^{(0)}(p_i)\]

so we get the anomalous dimension at the leading order as,

\be
d_4^{(1)}=\epsilon -6F\la
\ee

in agreement with \eqref{expect_irr} at this order.

At \(F\lambda=\frac{\epsilon}{3}\) we get,

\[\boxed{d_4^{(1)}=-\epsilon}\]

It is to be noted the origin of the anomalous dimension is the Type-II diagram (second diagram in Fig.\ref{irrelevant_lead}). It is expected as anomalous dimension should come from the process of logarithmic divergent digram finite as it happens in the continuum field theory.

%
%
%
%
%

So the irrelevant eigen-operator and its anomalous diemsnion is given as:

\vspace{0.2 in}

\[
\tcboxmath{d_4=\epsilon-6F\la}
\]
\[
\tcboxmath{
\mathcal{O}_4(0)= -\frac{1}{6!}\int_{p_1,p_2,p_3,p_4,p_5}\sum_{10~perm~(i,j,k)}2 \la h(p_i+p_j+p_k)\phi(p_1)\phi(p_2)\phi(p_3)\phi(p_4)\phi(p_5)\phi(p_6)}
\]
\[
\tcboxmath{
+\frac{1}{4!}\int_{p_1,p_2,p_3} \left[\sum_{i=1}^4 F \la h(p_i) -\sum_{3~perm~(i,j)} 2\la \mathcal{F}(p_i+p_j)\right] \phi(p_1)\phi(p_2)\phi(p_3)\phi(p_4)
}
\]
\[
\tcboxmath{
+\hf \int_{p_1} \frac{F}{d_4-2} \phi(p_1)\phi(p_2)
}
\]

Where $F=\frac{1}{16\pi^2}$. Thus at the fixed point, we get a composite operator with a dimension
\(-\eps\) which is (just a little) irrelevant in contrast with the Gaussian case. \footnote{This also agrees with Kogut and
Wilson (page 109)\cite{Wilson} to this order.}

\section{Wilson-Fisher Composite operator at the subleading order}
\label{secC}

Now we turn to find the irrelevant and relevant operators with their respective anomalous dimensions at the order $\epsilon^2$. We set $q=0$ for simplicity. At this order we have to include anomalous dimension $\frac{\eta}{2}$ in the Polchinski's equation i.e.

\begin{align}\label{pol}
\nonumber &\frac{\p \DD S}{\p t}= \int_p \bigg \lbrace (-K'(p^2)) [\frac{\dd^2 \DD S}{\dd \phi(p)\dd \phi(-p)}-2\frac{\dd S}{\delta \phi(p)}\frac{ \DD S}{\delta \phi(-p)}]-2\frac{p^2K'}{K}\phi(p)\frac{\dd \DD S}{\delta \phi(p)}+[(1-\Dt)N_\phi+D-N_p]\DD S\\
& -\frac{\eta}{2} \frac{K(p^2)(1-K(p^2)}{p^2} [\frac{\delta^2 \Delta S}{\delta \phi(p)\delta \phi(-p)} -2\frac{\delta S}{\delta \phi(p)}\frac{\delta \Delta S}{\delta \phi(-p)}]+ \frac{-\eta}{2} \phi(p) \frac{\delta \Delta S}{\delta \phi(p)} \bigg \rbrace= d_m \DD S + \beta(\lambda) \frac{\p \Delta S}{\p \la}
\end{align}

The action S at $\mathcal{O}(\la^2)$  is given by\cite{Dutta},

\begin{align*}
S=& \int_p \bigg \lbrace \frac{(-F\lambda)}{2-\epsilon}-\frac{1}{2}\lambda^2 G(p)+\frac{1}{2}\frac{(-\lambda F^2)}{4}  h(p) \bigg \rbrace \phi(p)\phi(-p)\\
&+\frac{1}{4!} \int_{p_1,p_2,p_3} \bigg \lbrace \lambda  -\lambda^2 [\mathcal{F}(p_1+p_2)+\mathcal{F}(p_1+p_3)+\mathcal{F}(p_1+p_4)]+\frac{F\lambda^2}{2} \sum_{i=1}^4 h(p_i) \bigg \rbrace \phi(p_1)\phi(p_2)\phi(p_3)\phi(p_4)\\
&+\frac{1}{6!} \int_{p_1,p_2,p_3,p_4,p_5,p_6} (-\lambda^2) \sum_{10~perm~(i,j,k)} h(p_i+p_j+p_k) \phi(p_1)\phi(p_2)\phi(p_3)\phi(p_4)\phi(p_5)\phi(p_6)
\end{align*}

Where

 \begin{align*}
 G(p)= &\frac{1}{3}\int_{q,k} \frac{h(q)}{2}[ h(p+q+k)h(k)-h(k)h(k)]-\frac{1}{3} \int_q \frac{h(q)}{2}[h(q+k)h(k)-h(k)h(k)]\\
&+\frac{\eta p^2}{2\epsilon}-\frac{1}{2-2\epsilon} \bigg \lbrace \frac{2}{3}\beta^{(1)}v_2^{(1)}+ \int_q f(q)\mathcal{F}(q) \bigg \rbrace 
\end{align*}

\begin{align*}
 \beta^{(1)}=-\int_q f(q) h(q)\rightarrow_{\epsilon\rightarrow 0}-F; v_2^{(1)}=-\int_q f(q)h(q)\rightarrow_{\epsilon\rightarrow 0} -\frac{F}{2}
\end{align*}

\begin{align*}
\mathcal{F}(p_i+p_j)= \frac{1}{2} \int_q  \bigg \lbrace h(p_i+p_j+q)h(q)-h(q)h(q) \bigg \rbrace
\end{align*}

$\mathcal{F}(p)$ is defined by \eqref{F}.

\begin{align*}
 h(p)=\frac{1-K(p)}{p^2}; ~~f(q)=-2K^\prime(q)
\end{align*}

\[
\beta(\lambda)= \epsilon \lambda+\beta_1^{(1)}(\lambda); ~ \beta_1^{(1)}(\lambda)=-3F\lambda^2\]

The anomalous dimensions in subleading order are found to be as,

\begin{subequations}\label{dim_sublead}
\begin{empheq}[box=\fbox]{align}
& d_4= \frac{53}{3}\la^2 F^2\\
 & d_2= \frac{5}{6}\la^2 F^2
\end{empheq}
\end{subequations}

The corresponding eigenoperators are given by,

\begin{subequations}\label{irrelevant_sublead}
\begin{empheq}[box=\fbox]{align}
\nonumber & \mathcal{O}_4(0)= \int_{p_1,..p_7} \frac{\la^2}{8!}\phi(p_1)\phi(p_2)\phi(p_3)\phi(p_4)\phi(p_5) \phi(p_6)\phi(p_7)\phi(p_8)\\
& \nonumber \sum_{28~perm~(i,j,k)} \sum_{10~perm~(m,n)} 3 ~h(p_i+p_j+p_k)h(p_i+p_j+p_k+p_m+p_n)\\
\nonumber &+ \int_{p_1,..p_5} \frac{\la^2}{6!}\phi(p_1)\phi(p_2)\phi(p_3)\phi(p_4)\phi(p_5) \phi(p_6)\\
& \nonumber \bigg( 3  \sum_{10~perm~(i,j,k)}\sum_{3~perm~(\alpha,\beta)} \int_p \big \lbrace h(p_i+p_j+p_k) [ h(p_\alpha+p_\beta+p)h(p) - h(p)h(p)] \big \rbrace\\
\nonumber &+ \frac{-3 F}{2} \sum_{10~perm~(i,j,k)}h(p_i+p_j+p_k)h(p_i+p_j+p_k) + \frac{-3 F}{2}\sum_{l=1}^6 \sum_{10~perm~(i,j,k)} h(p_l)h(p_i+p_j+p_k)\\
\nonumber & +\frac{1}{2} \int_p \sum_{15~perm~(i,j)} \sum_{6~perm~(\alpha,\beta)} \bigg \lbrace h(p_i+p_j+p)h(p_i+p_j+p_\alpha+p_\beta+p)h(p)\bigg \rbrace \bigg )\\
& \nonumber + \int_{p_1,p_2,p_3} \frac{1}{4!} \phi(p_1)\phi(p_2)\phi(p_3)\phi(p_4)\\
& \nonumber \bigg( -\frac{6\la^2 F}{4}\sum_{l=1}^4 \big \lbrace h(p_l) \big \rbrace \sum_{3~perm~(i,j)} \mathcal{F}(p_i+p_j) -\frac{3 \la^2 F}{2}\sum_{3~perm~(i,j)} \bar{H}_3(p_i+p_j)\\
& \nonumber +\frac{3}{4} \la^2 F^2\sum_{l=1}^4 \big \lbrace h(p_l)h(p_l)\big \rbrace + \frac{3}{8}\frac{\lambda^2 F^2}{4!} \sum_{i \neq j} h(p_i) h(p_j) + \frac{F\lambda \epsilon}{2} \sum_{i=1}^4 h(p_i)- \frac{3F^2\lambda^2}{2} \sum_{i=1}^4 h(p_i)\\
& \nonumber +\frac{3\la^2}{4} \sum_{6~perm~(i,j)} \left \lbrace I_4(p_i+p_j;p_i)+ I_4(p_i+p_j;p_j) \right \rbrace \\
& \nonumber -12\la^2 \sum_{3~perm~(i,j)} \int_{p,q} \left \lbrace h(p_i+p_j+p+q)h(p+q)h(q)h(p)-h(q)h(p)h(p+q)h(p+q) \right \rbrace\\
& \nonumber +6\lambda^2 \sum_{3~perm~(i,j)}\int_{p,q} \left \lbrace h(p_i+p_j+q)h(p+q)h(q)h(p)-h(q)h(p+q)h(q)h(p) \right \rbrace\\
& \nonumber + 6\lambda^2 \sum_{3~perm~(i,j)} \int_{p,q} \left \lbrace h(p_i+p_j+p)h(p+q)h(q)h(p)-h(p)h(p+q)h(q)h(p) \right \rbrace\\
& \nonumber + \frac{\la^2}{2} \sum_{i=1}^4 h(p_i)F_3(p_i)+ 3 \la^2 \sum_{i=1}^4 h(p_i)\int_{q} f(q)\mathcal{F}(q)+ 3\la^2 \sum_{3~perm~(i,j)}  \mathcal{F}(p_i+p_j)\mathcal{F}(p_i+p_j)\\
& \nonumber + \sum_{i=1}^4 \frac{\eta}{2\epsilon}p_i^2 h(p_i)+ 9F \lambda^2 \sum_{3~perm~(i,j)} \int_\Lambda^\infty \int_{\bar{q}}\frac{d \Lambda^\prime}{\Lambda^\prime} \bigg \lbrace h\bigg( \frac{p_i}{\Lambda^\prime} + \frac{p_j}{\Lambda^\prime}+ \bar{q}\bigg)h\left(\bar{q}\right )-h \left(\bar{q}\right)h\left(\bar{q}\right) \bigg \rbrace \bigg)
\\
& +\hf \la \int_p \bigg( \frac{2F^2}{2-\epsilon}-\frac{2}{3} F_3(p)- \int_q f(q)h(q)-\frac{F^2}{2}h(p) \bigg)\phi(p)\phi(-p)
\end{empheq}
\end{subequations}

\begin{subequations}\label{relevant_sublead}
\begin{empheq}[box=\fbox]{align}
& \nonumber \mathcal{O}_2(0)= \frac{\la^2}{6!} \int_{p_1,..p_5} \phi(p_1)\phi(p_2)\phi(p_3)\phi(p_4)\phi(p_5) \phi(p_6)\\
& \nonumber +\bigg (\sum_{10~perm~(i,j,k)} h(p_i+p_j+p_k)h(p_i+p_j+p_k)+ \sum_{10~perm~(i,j,k)} h(p_i+p_j+p_k) \sum_{l=1}^6 h(p_l) \bigg)\\
& \nonumber +\frac{\la^2}{4!}\int_{p_1,p_2,p_3}\phi(p_1)\phi(p_2)\phi(p_3)\phi(p_4)\\
&\nonumber \bigg(\sum_{3~perm~(i,j)}\bar{H}_3(p_i+p_j)+ \sum_{l=1}^4 h(p_l) \sum_{3~perm~(i,j)} \mathcal{F}(p_i+p_j) -F \big \lbrace \frac{1}{2} \sum_{i \neq j} h(p_i) h(p_j) + \sum_{l=1}^4 h^2(p_l) \big \rbrace \bigg)\\
& +\nonumber \hf \int_{p} \phi(p) \phi(-p)\\
& \nonumber \bigg(-\frac{\la^2}{3} \int_{q,k} \big \lbrace  h(p+q+k)h(p)h(q)h(k)- h(q)h(q+k)h(k) \big \rbrace\\
& \nonumber -\frac{\la^2}{2} \int_{q,k} \big \lbrace h(p+q+k)h(q)h(q)h(k)-h(q+k)h(q)h(q)h(k) \big \rbrace\\
&  -\lambda^2 F^2 h(p)+\frac{\epsilon \lambda}{2} h(p)+ \frac{3}{4} F^2 \lambda^2 h^2(p)-\la^2 h(p)\int_q f(q) \mathcal{F}(q)+\frac{\eta}{\epsilon} p^2 h(p)
\end{empheq}
\end{subequations}

Where all the functions have been defined in Appendix \eqref{appF}. As mentioned in the previous section here also we did not find the 2-pt vertex to second order. First order 2-pt vertex is enough to find second order anomalous diemsnion. We have derived the above results in the next two subsections.

\subsection{The Irrelevant Operator, $\mathcal{O}_4^{(2)}(0)$}

The form of the irrelevant operator in the subleading order is given below. Note that at this order we need to include 8-pt vertex which is of $\mathcal{O}(\eps^2)$. We have just given the expressions of $\mathcal{O}_4(0)$ in this section. Equations to find them are given in Appendix \ref{appC}.

\begin{align*}
&\mathcal{O}_4(0)= \Delta S_2+\Delta S_4+\Delta S_6+\Delta S_8\\
&= \frac{1}{2!}\int_p \bigg \lbrace \frac{F}{d_4-2}+  A^{(1)}(p) \bigg \rbrace \phi(p)\phi(-p)\\
&+ \frac{1}{4!} \int_{p_1,p_2,p_3} \bigg \lbrace 1+ F\lambda \sum_{l=1}^4 h(p_l)-\lambda \int_{k} \sum_{3~perm~(i,j)}[ h(p_i+p_j+k)h(k)-h(k)h(k)]\\
& + B^{(2)}(p_1,p_2,p_3,p_4) \bigg \rbrace \phi(p_1)\phi(p_2)\phi(p_3)\phi(p_4)\\
&+ \frac{1}{6!}\int_{p_1,p_2,p_3,p_4,p_5} \bigg \lbrace -2\lambda\sum_{10~perm~(i,j,k)}^6 h(p_i+p_j+p_k)+ D^{(2)}(p_1,p_2,p_3,p_4,p_5,p_6) \bigg  \rbrace \phi(p_1)\phi(p_2)\phi(p_3)\phi(p_4)\phi(p_5) \phi(p_6)\\
&+ \frac{1}{8!} \int_{p_1,p_2,p_3,p_4,p_5,p_6,p_7}  E^{(2)}(p_1,p_2,p_3,p_4,p_5,p_6,p_7,p_8) \phi(p_1)\phi(p_2)\phi(p_3)\phi(p_4)\phi(p_5) \phi(p_6)\phi(p_7)\phi(p_8)
\end{align*}

with
\[
d_4= \eps-6F\la+ d_4^{(2)}+...\]

\subsubsection{$\phi^8$ equation-Determination of $\Delta S_8^{(2)}$}

The 8-pt vertex is found by solving the $\phi^8$ equation at $\mathcal{O}(\la^2)$. The $\phi^8$ equation is obtained as:

\begin{align} \label{phi8}
-& \nonumber 2 \int ( -K^\prime(p^2)) \bigg \lbrace \frac{\delta}{\delta \phi(p)} \frac{\lambda}{4!} \phi(p_1) \phi(p_2) \phi(p_3) \phi(p_4)\bigg \rbrace \times\\
&  \bigg \lbrace \frac{\delta}{\delta \phi(-p)} \sum_{10 perm (i,j,k)} \frac{D^{(1)}(p_i+p_j+p_k)}{6!} \phi(p_1)\phi(p_2)\phi(p_3)\phi(p_4)\phi(p_5)\phi(p_6) \bigg \rbrace \\ \nonumber
-&2 \int ( -K^\prime(p^2)) \bigg \lbrace \frac{\delta}{\delta\phi(p)} \sum_{i=1}^4 \frac{1}{4!} B^{(0)}(p_i) \phi(p_1)\phi(p_2)\phi(p_3)\phi(p_4) \bigg \rbrace \times\\
& \nonumber  \bigg \lbrace \frac{\delta}{\delta\phi(-p)} \sum_{10 perm (i,j,k)}\frac{V^{(2)}(p_i+p_j+p_k)}{6!} \phi(p_1)\phi(p_2)\phi(p_3)\phi(p_4)\phi(p_5)\phi(p_6) \bigg \rbrace\\ \nonumber
+&\frac{1}{8!} \bigg \lbrace  8-3D - \sum_{i=1}^8 p_i.\frac{d}{d p_i} \
\bigg \rbrace E^{(2)}(p_1,p_2,p_3,p_4,p_5,p_6,p_7,p_8)=0\\
\end{align}

The solution is given by:

\begin{align}\label{S8_irr}
E^{(2)}(p_1,p_2,p_3,p_4,p_5,p_6,p_7,p_8)= \sum_{28~perm~(i,j,k)} \sum_{10~perm~(m,n)} 3\la^2 ~h(p_i+p_j+p_k)h(p_i+p_j+p_k+p_m+p_n)
\end{align}

\begin{figure}[b]
   \centering
  \begin{tikzpicture}
  \begin{feynman}
    	\vertex (a) at (0,0);
    	\vertex (b) at (2,0);
    	\vertex (c) at (6,0);
    	\vertex (d) at (10,0);
    	\vertex (e) at (12,0);
    	\vertex (p) at (1,1);
    	\vertex (q) at (1,-1);
    	\vertex (m) at (5.5,1);
    	\vertex (n) at (6.5,1);
    	\vertex (r) at (11,1);
    	\vertex (s) at (11,-1);
    	\vertex (x) at (2,1);
      \diagram* {
        (a) -- [near start,edge label=$p_j$](b) -- [edge label=$h(p_i+p_j+p_k)$](c) --[edge label=$h(p_a+p_b+p_c)$] (d) --[near end,edge label=$p_b$](e),
        (p) -- [near start,edge label=$p_i$](b),
        (q) -- [near start,edge label=$p_k$](b),
        (m) -- [near start,edge label=$p_m$](c),
        (n) -- [near start, edge label=$p_n$](c),
        (r) -- [near start,edge label=$p_a$](d),
        (s) -- [near start,edge label=$p_c$](d),
        (x) -- [photon, near start, edge label=$\phi^4(0)$](b),
         };
    \end{feynman}
  \end{tikzpicture}
  \caption{The diagram for $E^{(2)}(p_1,p_2,p_3,p_4,p_5,p_6,p_7,p_8)$}
  \label{S_8}
  \end{figure}
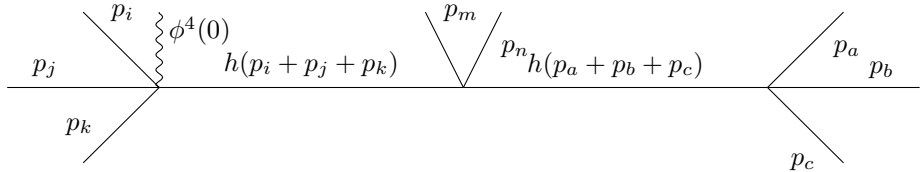

\subsubsection{$\phi^6$ equation - Determination of $\Delta S_6^{(2)}$}

Solving $\phi^6$ equation we get four kinds of solutions for 6-pt vertex at order $O(\la^2)$ based on their tensor structure (see \ref{solvphi6_irr} for details)

\begin{subequations}
\label{S6slead_irr}
\begin{align}
& D_I^{(2)}(p_1,p_2...,p_6)=3 \la^2 \sum_{10~perm~(i,j,k)}\sum_{3~perm~(\alpha,\beta)} \int_p \big \lbrace h(p_i+p_j+p_k) [ h(p_\alpha+p_\beta+p)h(p) - h(p)h(p)] \big \rbrace\\
& D_{II}^{(2)}(p_1,p_2,p_3,p_4,p_5,p_6)=\frac{-3\la^2 F}{2} \sum_{10~perm~(i,j,k)}h(p_i+p_j+p_k)h(p_i+p_j+p_k)\\
& D_{III}^{(2)}(p_1,p_2,p_3,p_4,p_5,p_6)=\frac{-3 \la^2 F}{2}\sum_{l=1}^6 \sum_{10~perm~(i,j,k)} h(p_l)h(p_i+p_j+p_k)\\
& D_{IV}^{(2)}(p_1,p_2,p_3,p_4,p_5,p_6)= \frac{\la^2}{2} \int_p \sum_{15~perm~(i,j)} \sum_{6~perm~(\alpha,\beta)} \bigg \lbrace h(p_i+p_j+p)h(p_i+p_j+p_\alpha+p_\beta+p)h(p)\bigg \rbrace
\end{align}
\end{subequations}

\subsubsection{$\phi^2$ equation at $\mathcal{O}(\epsilon)$: Determination of $A^{(1)}(p)$}
\label{A_phi2}

The $\phi^2$ equation at order $\lambda^1$ is given below (Note that we do not have to consider $\beta(\lambda) \frac{\p \Delta S}{\p \la}$ part because we want to find $A(p)$ at order $\eps^1$ or $\la^1$ only):

\begin{align}\label{phi2}
\nonumber &\int_q ( -K^\prime(q^2)) B_I^{(1)}(q)+ F B_I^{(1)}(p)+ \frac{1}{2} F B_{II}^{(1)}(0)- \frac{1}{2} \int_q K^\prime(q^2) [ B_{II}^{(1)}(p+q)+ B_{II}^{(1)}(p-q) ] -2 ( -K^\prime(q^2)) U_2^{(1)}(p) A^{(0)}(p)\\
&+ A^{(1)}(p)- \frac{1}{2} p.\frac{d}{dp} A^{(1)}(p)= d_m^{(1)} \frac{A^{(1)}(p)}{2}
\end{align}

Solving the $\phi^2$ equation we found the $A^{(0)}(p)$ and three kinds of 2-pt vertices based on their tensor structure.
\begin{subequations}
\label{S2slead_irr}
\begin{align}
& A_I^{(1)}(p)=\frac{2F^2\la}{2-\epsilon}\\
& A_{II}^{(1)}(p)=-\frac{2\la}{3} F_3(p)- \la\int_q f(q)h(q) \\
& A_{III}^{(1)}(p)=-\frac{F^2\la}{2}h(p)
\end{align}
\end{subequations}

From \eqref{A0} we get,

\begin{align}
A^{(0)}= -\frac{F}{2}-\frac{F\epsilon}{4}
\end{align}

Where  $\bar{F}_3(p)=\int_{q,k}h(p+q+k)h(q)h(k)$, $F_3(p)= \bar{F}_3(p)-\bar{F}_3(0)=\int_q 2h(q)\left[ \mathcal{F}(p+q)-\mathcal{F}(q) \right]$. They are defined by \eqref{F3_bar} and \eqref{F_3}.

\subsubsection{$\phi^4$ equation-Determination of $B^{(2)}(p_1,p_2,p_3,p_4)$}

Solving the $\phi^4$ equation we get total nine kinds of 4-pt vertices based on their tensor structure (see Appendix \ref{solvphi4_irr} for more details).
 
\begin{subequations}
\label{S4slead_irr}
\begin{align}
&\frac{1}{4!} B_{I}^{(2)}(p_1,p_2,p_3,p_4)=-\frac{6\la^2}{4} \frac{ F}{4!} \sum_{l=1}^4 \big \lbrace h(p_l) \big \rbrace \sum_{3~perm~(i,j)} \mathcal{F}(p_i+p_j)\\
&\frac{1}{4!} B_{II}^{(2)}(p_1,p_2,p_3,p_4)=-\frac{1}{4!}\frac{3 \la^2 F}{2}\sum_{3~perm~(i,j)} \bar{H}_3(p_i+p_j)\\
& \frac{1}{4!} B_{III}(p_1,p_2,p_3,p_4)= \frac{3}{4} \frac{ \la^2 F^2}{4!}\sum_{l=1}^4 \big \lbrace h(p_l)h(p_l)\big \rbrace + \frac{3}{8}\frac{\lambda^2 F^2}{4!} \sum_{i \neq j} h(p_i) h(p_j)\\
&B_{IV}^{(2)}(p_1,p_2,_3,p_4)= \frac{F\lambda \epsilon}{2} \sum_{i=1}^4 h(p_i)- \frac{3F^2\lambda^2}{2} \sum_{i=1}^4 h(p_i) \\
&\frac{1}{4!} B_{V}^{(2)}(p_1,p_2,p_3,p_4)=  \frac{1}{4!} \frac{3\la^2}{4} \sum_{6~perm~(i,j)} \left \lbrace I_4(p_i+p_j;p_i)+ I_4(p_i+p_j;p_j) \right \rbrace\\
&  \nonumber\frac{1}{4!} B_{VI}^{(2)}(p_1,p_2,p_3,p_4)\\
& =-\frac{\la^2}{2} \sum_{3~perm~(i,j)} \int_{p,q} \left \lbrace h(p_i+p_j+p+q)h(p+q)h(q)h(p)-h(q)h(p)h(p+q)h(p+q) \right \rbrace\\
&\nonumber +\frac{\lambda^2}{4} \sum_{3~perm~(i,j)}\int_{p,q} \left \lbrace h(p_i+p_j+q)h(p+q)h(q)h(p)-h(q)h(p+q)h(q)h(p) \right \rbrace\\
&\nonumber +\frac{\lambda^2}{4} \sum_{3~perm~(i,j)} \int_{p,q} \left \lbrace h(p_i+p_j+p)h(p+q)h(q)h(p)-h(p)h(p+q)h(q)h(p) \right \rbrace\\
&\nonumber \frac{1}{4!} B_{VII}(p_1,p_2,p_3,p_4)\\
& \nonumber = \frac{1}{4!} B_{VII}^{(2)}(p_1,p_2,p_3,p_4)|_1+ \frac{1}{4!} B_{VII}^{(2)}(p_1,p_2,p_3,p_4)|_2\\
& =\frac{1}{4!}\frac{\la^2}{2} \sum_{i=1}^4 h(p_i)F_3(p_i)+ \frac{1}{4!} 3 \la^2 \sum_{i=1}^4 h(p_i)\int_{q} f(q)\mathcal{F}(q)\\
&\frac{1}{4!}B_{VIII}^{(2)}(p_1,p_2,p_3,p_4)= \frac{3\la^2}{4!}\sum_{3~perm~(i,j)}  \mathcal{F}(p_i+p_j)\mathcal{F}(p_i+p_j)\\
&\frac{1}{4!} B_{IX}(p_1,p_2,p_3,p_4)=  \frac{1}{4!}\sum_{i=1}^4 \frac{\eta}{2\epsilon}p_i^2 h(p_i)\\
& \frac{1}{4!}B^{(2)}_{X}(\frac{p_1}{\Lambda},\frac{p_2}{\Lambda},\frac{p_3}{\Lambda},\frac{p_4}{\Lambda})= \frac{ 9F \lambda^2}{4!} \sum_{3~perm~(i,j)} \int_\Lambda^\infty \int_{\bar{q}}\frac{d \Lambda^\prime}{\Lambda^\prime} \bigg \lbrace h\bigg( \frac{p_i}{\Lambda^\prime} + \frac{p_j}{\Lambda^\prime}+ \bar{q}\bigg)h\left(\bar{q}\right )-h \left(\bar{q}\right)h\left(\bar{q}\right) \bigg \rbrace
\end{align}
\end{subequations}

Where  $\bar{F}_3(p)=\int_{q,k}h(p+q+k)h(q)h(k)$, $F_3(p)= \bar{F}_3(p)-\bar{F}_3(0)=\int_q 2h(q)\left[ \mathcal{F}(p+q)-\mathcal{F}(q) \right]$. They are defined by \eqref{F3_bar} and \eqref{F_3}.

Also

\[
\bar{H}_3(p)=\int_q h(p+q)h(q)h(q)
\]

and 

\begin{align}
 I_4(p_i+p_j;p_i)= &\nonumber \bar{I}_4(p_i+p_j;p_i)-\bar{I}_4(0;0)\\
 & =\sum_{6~perm~(i,j)}\int_{p,q}  \big \lbrace h(p_i+p_j+q)  h(p+q+p_i) h(p) h(q)- h(p+q)h(p)h(q)h(q) \big \rbrace
\end{align}

$\bar{H_3}(p)$ and $I_4(p_i+p_j;p_i)$ are defined by \eqref{H3_bar} and \eqref{I4} respectively.

\subsubsection*{Equation for $B_{IV}^{(2)}(p_1,p_2,p_3,p_4)$ and $B_V(p_1,p_2,p_3,p_4)$}

We will show one sample calculation here in order to explain how we have used Feynman diagram as a guide in the calculations.

Taking \eqref{Bfive1_i}, \eqref{Bfive2_i} and \eqref{Bfive3_i}, we get 

\begin{align*}
\nonumber & \frac{3\lambda^2}{4!}\int_{p,q} \big\lbrace-K^\prime(p^2)\big \rbrace  \sum_{6~perm~(i,j)} \big \lbrace h(p_i+p_j+p) [ h(p+q+p_j)+h(p+q+p_i)-2h(q)]h(q) \big \rbrace\\
\nonumber &+ \frac{2\lambda^2}{4!}\int_p \int_q  \big \lbrace -K^\prime(p^2) \big \rbrace\sum_{6~perm~(i,j)} \big \lbrace h(p_i+p_j+q)[ h(p+q+p_i)+h(p+q+p_j)]h(q)  \big \rbrace\\
\nonumber &+  \frac{\lambda^2}{4!} \int_q \int_p \big \lbrace -K^\prime(p^2) \big \rbrace \sum_{l=1}^4 \sum_{3~perm~(i,j)}\big \lbrace h(p_l+p+q)h(p_l+p_i+p_j+p+q)h(q) \big \rbrace\\
&+ \big \lbrace 4-D-\sum_{i=1}^4 p_i.\frac{d}{dp_i} \big \rbrace \frac{1}{4!} B_{IV}^{(2)}(p_1,p_2,p_3,p_4)=0
\end{align*}

\begin{figure}[b]
   \centering
  \begin{tikzpicture}
  \begin{feynman}
    	\vertex (a) at (0,0);
    	\vertex (b) at (2,0);
    	\vertex (c) at (4,0);
    	\vertex (d) at (6,0);
    	\vertex (m) at (1,1);
    	\vertex (x) at (3,1);
    	\vertex (y) at (2.5,1.5);
    	\vertex (z) at (3.5,1.5);
      \diagram* {
       (a) -- [edge label=$p_i$,insertion=1.0](b) --[edge label=$h(q)$] (c) --[edge label=$p_j$](d),
       (b) -- [quarter left, near end, edge label= $K^\prime(p)$](x) -- [edge label=$h(p_i+p_j+p)$,quarter left](c),
       (c) -- [half left, edge label= $h(p_i+p+q)$](b),
       (c) -- (d),
       (m) -- [photon, near end,edge label=$\phi^4(0)$](b),
       (y) -- [edge label=$p_i+p_j$](x),
       (z) -- (x),
       };
    \end{feynman}
  \end{tikzpicture}
  \hspace{0.2 in}
  \begin{tikzpicture}
  \begin{feynman}
    	\vertex (a) at (0,0);
    	\vertex (b) at (2,0);
    	\vertex (c) at (4,0);
    	\vertex (d) at (6,0);
    	\vertex (m) at (1,1);
    	\vertex (x) at (3,1);
    	\vertex (y) at (2.5,1.5);
    	\vertex (z) at (3.5,1.5);
      \diagram* {
       (a) -- [edge label=$p_i$,insertion=1.0](b) --[edge label=$K^\prime(p)$] (c) --[edge label=$p_j$](d),
       (b) -- [quarter left, near end, edge label= $h(q)$](x) -- [edge label=$h(p_i+p_j+q)$,quarter left](c),
       (c) -- [half left, edge label= $h(p_i+p+q)$](b),
       (c) -- (d),
       (m) -- [photon, near end,edge label=$\phi^4(0)$](b),
       (y) -- [edge label=$p_i+p_j$](x),
       (z) -- (x),
       };
    \end{feynman}
  \end{tikzpicture}
  \hspace{0.2 in}
  \begin{tikzpicture}
  \begin{feynman}
    	\vertex (a) at (0,0);
    	\vertex (b) at (2,0);
    	\vertex (c) at (4,0);
    	\vertex (d) at (6,0);
    	\vertex (m) at (1,1);
    	\vertex (x) at (3,1);
    	\vertex (y) at (2.5,1.5);
    	\vertex (z) at (3.5,1.5);
      \diagram* {
       (a) -- [edge label=$\frac{p_i}{\Lambda}$,insertion=1.0](b) --[edge label=$h(\frac{q}{\Lambda})$] (c) --[edge label=$\frac{p_j}{\Lambda}$](d),
       (b) -- [quarter left, near end, edge label= $h(\frac{p}{\Lambda})$](x) -- [edge label=$h(\frac{p_i}{\Lambda}+\frac{p_j}{\Lambda}+\frac{p}{\Lambda})$,quarter left](c),
       (c) -- [half left, edge label= $h(\frac{p_i}{\Lambda}+\frac{p}{\Lambda} + \frac{q}{\Lambda})$](b),
       (c) -- (d),
       (m) -- [photon, near end,edge label=$\phi^4(0)$](b),
       (y) -- [edge label=$\frac{p_i}{\Lambda}+\frac{p_j}{\Lambda}$](x),
       (z) -- (x),
       };
    \end{feynman}
  \end{tikzpicture}
  \caption{Application of $\Lambda \frac{d}{d \Lambda} $ on the diagram at the bottom gives the two diagrams at the top}
  \label{B_four}
\end{figure}
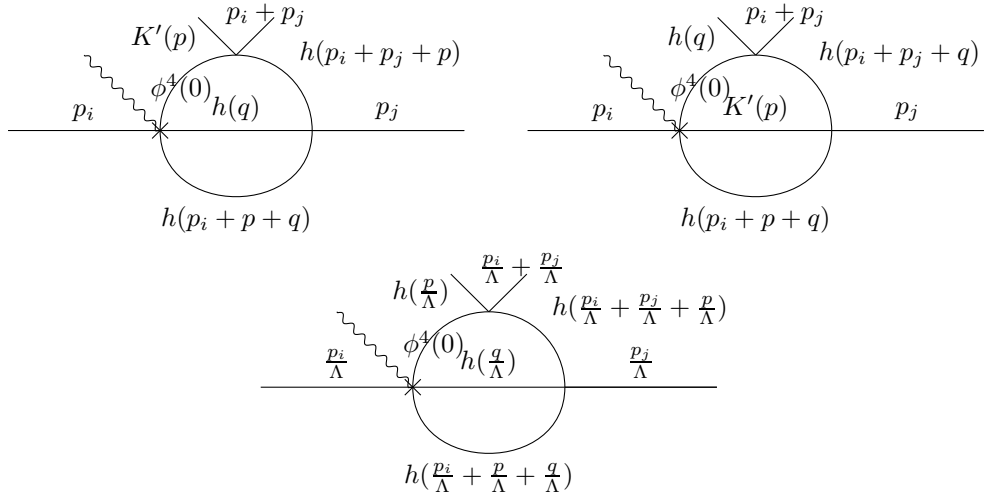

We aim to solve

\begin{align}\label{B_four}
\nonumber & \frac{3\lambda^2}{4!}\int_{p,q} \lbrace-K^\prime(p^2)\rbrace  \sum_{6~perm~(i,j)} h(p_i+p_j+p) \big \lbrace[ h(p+q+p_j)+h(p+q+p_i)]h(q)-2 h(p+q)h(q) \big \rbrace \\
\nonumber &+ \frac{2\lambda^2}{4!}\int_p \int_q  \big \lbrace -K^\prime(p^2) \big \rbrace\sum_{6~perm~(i,j)}h(p_i+p_j+q) \big \lbrace[ h(p+q+p_i)+h(p+q+p_j)] h(q)-2h(p+q)h(q) \big \rbrace \\
\nonumber &+  \frac{\lambda^2}{4!} \int_q \int_p \big \lbrace -K^\prime(p^2) \big \rbrace \sum_{l=1}^4 \sum_{3~perm~(i,j)}h(p_l+p+q)h(p_l+p_i+p_j+p+q)h(q)\\
\nonumber &+\frac{\lambda^2}{6} \int_{p,q} \big \lbrace K^\prime(p^2)\big \rbrace \sum_{3~perm~(i,j)}h(p+q)h(p_i+p_j+p+q)h(q)\\
&+ \big \lbrace-2( 4-D)-\sum_{i=1}^4 p_i.\frac{d}{dp_i} \big \rbrace \frac{1}{4!}  B_{IV}^{(2)}(p_1,p_2,p_3,p_4)=0
\end{align}

To solve this equation first note that the second and third term on the LHS are equal. The first and second  term is represented by the first and second diagram respectively on the top of Fig.\ref{B_four}. Now observe we are basically trying to find $\Delta S$ such that $-\Lambda\frac{d}{d \Lambda}\Delta S \propto \Delta S$, so if we write $\Lambda$ explicitly i.e. $p_i \rightarrow \frac{p_i}{\Lambda}$ we get,

\begin{align}
p_i. \frac{d}{d p_i}= -\Lambda \frac{d}{d \Lambda}
\end{align}

Now if we consider the third diagram at the bottom of Fig \ref{B_four} and apply $\Lambda \frac{d}{d \Lambda}$ we get back terms corresponding to the other two diagrams i.e.

\begin{align*}
&\Lambda \frac{d}{d \Lambda} \left[\int_{\frac{p}{\Lambda}, \frac{q}{\Lambda}} h\left(\frac{p_i+p_j+p}{\Lambda}\right)h\left(\frac{p}{\Lambda}\right)h\left(\frac{p_i+p+q}{\Lambda}\right)h\left(\frac{q}{\Lambda}\right)\right]\\
=& 4 \int_{\frac{p}{\Lambda}, \frac{q}{\Lambda}} K^\prime\left(\frac{p}{\Lambda}\right)h\left(\frac{p_i+p_j+p}{\Lambda}\right)h\left(\frac{q}{\Lambda}\right)h\left(\frac{p_i+p+q}{\Lambda}\right)+4 \int_{\frac{p}{\Lambda}, \frac{q}{\Lambda}} K^\prime\left(\frac{q}{\Lambda}\right)h\left(\frac{(p_i+p_j+p)}{\Lambda}\right)h\left(\frac{q}{\Lambda}\right)h\left(\frac{p_i+p+q}{\Lambda}\right)\\
\end{align*}

We can expect $B_{IV}(p_1,p_2,p_3,p_4)$ to be of the form $\int_{\frac{p}{\Lambda}, \frac{q}{\Lambda}} h\left(\frac{p_i+p_j+p}{\Lambda}\right)h\left(\frac{p}{\Lambda}\right)h\left(\frac{p_i+p+q}{\Lambda}\right)h\left(\frac{q}{\Lambda}\right)$. So we use \eqref{I4} and get the solution as:

\begin{align}
\nonumber  \frac{1}{4!} B_{IV}^{(2)}(p_1,p_2,p_3,p_4)& =  \frac{1}{4!}\frac{3}{4}\lambda^2  \int_{p,q} \sum_{6~perm~(i,j)} \big \lbrace h(p_i+p_j+q) \sum_{a=i,j} h(p+q+p_a) h(p) h(q)- 2h(p+q)h(p)h(q)h(q) \big \rbrace\\
& =\frac{1}{4!}\frac{3\la^2}{4} \sum_{6~perm~(i,j)}\left[ I_4(p_i+p_j;p_i)+I_4(p_i+p_j;p_j)\right]
\end{align}

In the L.H.S of \eqref{BA_irr} we are left with

\begin{align}
& \nonumber \frac{1}{4!} 3(4-D)B_{IV}^{(2)}(p_1,p_2,p_3,p_4)\\
& \nonumber +\frac{\lambda^2}{2} \big \lbrace -K^\prime(p^2) \big \rbrace \sum_{3~perm~(i,j)} h(p_i+p_j+p) \bigg \lbrace h(p+q)h(q)- h(q)h(q) \bigg \rbrace\\
& \nonumber +\frac{\lambda^2}{3} \big \lbrace -K^\prime(p^2) \big \rbrace \sum_{3~perm~(i,j)} h(p_i+p_j+q)h(p+q)h(q)\\
& +\frac{\lambda^2}{6} \big \lbrace -K^\prime(p^2) \big \rbrace \sum_{3~perm~(i,j)} h(p_i+p_j+p+q)h(p+q)h(q)
\end{align}

Ignoring $\mathcal{O}(\epsilon^3)$ term and aiming to solve the following equation from the left over terms:

\begin{align}
\nonumber &+\frac{\lambda^2}{2} \big \lbrace -K^\prime(p^2) \big \rbrace \sum_{3~perm~(i,j)}\left [ h(p_i+p_j+p) \big \lbrace h(p+q)h(q)- h(q)h(q) \big \rbrace - h(p)\big \lbrace h(p+q)h(q)-h(q)h(q) \big \rbrace  \right]\\
\nonumber & +\frac{\lambda^2}{3} \big \lbrace -K^\prime(p^2) \big \rbrace \sum_{3~perm~(i,j)}\left[ h(p_i+p_j+q)h(p+q)h(q)-h(q)h(p+q)h(q)\right]\\
\nonumber & +\frac{\lambda^2}{6} \big \lbrace -K^\prime(p^2) \big \rbrace \sum_{3~perm~(i,j)} \left[ h(p_i+p_j+p+q)h(p+q)h(q)-h(p+q)h(p+q)h(q) \right]\\
& + \left \lbrace  -2(4-D)- \sum_{i=1}^4 p_i.\frac{\partial}{\partial p_i} \right \rbrace \frac{1}{4!} B_{V}^{(2)}(p_1,p_2,p_3,p_4)=0
\end{align}

We can write a solution symmetric in variable $p$ and $q$.

\begin{align}
& \nonumber \frac{1}{4!} B_{V}^{(2)}(p_1,p_2,p_3,p_4)=-\frac{\la^2}{2} \sum_{3~perm~(i,j)} \left \lbrace h(p_i+p_j+p+q)h(p+q)h(q)h(p)-h(p)h(q)h(p+q)h(p+q) \right \rbrace\\
&\nonumber +\frac{\lambda^2}{4} \sum_{3~perm~(i,j)}\left \lbrace h(p_i+p_j+q)h(p+q)h(q)h(p)-h(q)h(p+q)h(q)h(p) \right \rbrace\\
& +\frac{\lambda^2}{4} \sum_{3~perm~(i,j)}\left \lbrace h(p_i+p_j+p)h(p+q)h(q)h(p)-h(p)h(p+q)h(q)h(p) \right \rbrace
\end{align}

And on LHS of \eqref{BA_irr} we are left with

\begin{align}\label{anom_irr}
& \nonumber \frac{1}{4!} 3(4-D) B_{V}^{(2)}(p_1,p_2,p_3,p_4)\\
& \nonumber  +\int_{p,q} \bigg \lbrace \frac{3\lambda^2}{2} \big \lbrace K^\prime(p^2)\big \rbrace h(p)\left[h(p+q)h(q)-h(q)h(q)\right] + \lambda^2 \big \lbrace K^\prime(p^2)\big \rbrace h(q)h(p+q)h(q)\\
&+\frac{\lambda^2}{2}\big \lbrace K^\prime(p^2)\big \rbrace h(p+q)h(p+q)h(q) \bigg \rbrace
\end{align}

Following this procedure we can solve all the equations given in \ref{solvphi4_irr} to get the 4-point composite operator vertices given above.

\subsubsection{Calculation of Anomalous Dimension}

To get the anomalous dimension we collect  the leftover terms which remain unused i.e. \eqref{anom_irr}, \eqref{anom_eta_i} in the LHS. All other left over terms are either cancelled or of $\mathcal{O}(\eps \la^2)$ or $O(\eps^3)$.

\begin{align}\label{final}
\nonumber &\int_{p,q} \bigg \lbrace \frac{3\lambda^2}{2} \big \lbrace -K^\prime(p^2)\big \rbrace h(p)\left[h(p+q)h(q)-h(q)h(q)\right] + \lambda^2 \big \lbrace- K^\prime(p^2)\big \rbrace h(q)h(p+q)h(q)\\
& +\frac{\lambda^2}{2}\big \lbrace- K^\prime(p^2)\big \rbrace h(p+q)h((p+q)h(q) \bigg \rbrace -\frac{4}{4!}\frac{\eta}{2}\sum_{i=1}^4 B^{(0)}(p_i) = \frac{d_m}{4!} \big \lbrace \sum_{i=1}^4 B^{(0)}(p_i) \big \rbrace
\end{align}

The first three terms on the LHS can be written as:

\begin{align}\label{integral}
 &\frac{3\lambda^2}{2} \left[\int_{p,q} \big \lbrace -K^\prime(p^2)\big \rbrace h(p) h(p+q)h(q)+\big \lbrace -K^\prime(p^2)\big \rbrace h(q)h(p+q)h(q)\right] - \frac{3\la^2}{2} \int_{p,q} \big \lbrace -K^\prime(p^2) \big \rbrace h(p)h(q)h(q)\\
& =-\frac{1}{4} \frac{3 \la^2}{2} \Lambda \frac{\partial}{\partial \Lambda} \int_{\frac{p}{\Lambda},\frac{q}{\Lambda}} h\left(\frac{p}{\Lambda}\right)h\left(\frac{p}{\Lambda}\right)h\left(\frac{p+q}{\Lambda}\right)h\left(\frac{q}{\Lambda}\right)+ \frac{3\la^2}{2} \int_{p,q} \big \lbrace K^\prime(p^2) \big \rbrace h(p)h(q)h(q)
\end{align}

Where in the second line we have rewritten the integral in terms of dimensionful momenta and written $\lm$ explicitly. This gives a convenient way of doing the integrals. It also reveal the relation with log divergences in Feynman diagrams.
 While evaluating the integral we have taken $h(p/\Lambda)$ as $\frac{K(p/\Lambda_0)-K(p/\Lambda)}{p^2/\Lambda^2}$ instead of $\frac{1-K(p/\Lambda)}{p^2/
\Lambda^2}$. 

We keep $\lo$ finite initially to make all the integrals finite and well defined and take $\lo \to \infty$ at the end.


Now we note the Feynman diagrams of the above terms. The first(second)term in the first line of \eqref{integral} represent the first(second) diagram at the top of Fig.\ref{B_four} (if we make all external momenta as zero). Similarly, the first term on the second line represents the diagram at the bottom of the same figure. As written above we will find this integral of the second line of \eqref{integral} and then apply $-\frac{1}{4} \Lambda \frac{\partial}{\partial \Lambda}$ to get our desrired integral (see Appendix \ref{appD}).

The value of the integrals in limit of $\Lambda_0 \rightarrow \infty$ is listed below.

\begin{align}\label{integral_main}
\nonumber a.~~~&\int_{p,q} \left[\big \lbrace -K^\prime(p^2)\big \rbrace h(p) h(p+q)h(q)+\big \lbrace -K^\prime(p^2)\big \rbrace h(q)h(p+q)h(q)\right]\\
&=F^2 \left(\hf-\log 2+\frac{1}{2}\log \frac{\Lambda_0^2}{\Lambda^2} \right)
\end{align}

and similarly one can calculate using method shown in Appendix \ref{appD}:

\begin{align}\label{integral_counter}
b.~~~\int_{p,q} \big \lbrace K^\prime(p^2) \big \rbrace h(p)h(q)h(q)=-F^2 \left\lbrace-\log 2+\frac{1}{2}\log \Lambda_0^2+\frac{1}{2}\log \Lambda^2 + \log \left(\frac{1}{\Lambda^2}+\frac{1}{\Lambda_0^2}\right)\right\rbrace
\end{align}

So in \eqref{final} we use $B^{(0)}(p_i)=\frac{1}{4}$, combine \eqref{integral_main} and \eqref{integral_counter} to get the anomalous dimension. Note that the logarithmic divergences gets exactly cancelled so the \eqref{integral_counter} is in fact originated from a counterterm.

\[
\tcboxmath{
 \frac{1}{4!} d_4=\frac{3}{4} \lambda^2 F^2 -\frac{4}{4!} \frac{\eta}{2}= \frac{1}{4!}\frac{53}{3}\la^2 F^2 }
\]

Where $\frac{\eta}{2}=\frac{\la^2 F^2}{12}$ at the fixed point and  $F=\frac{1}{16\pi^2}$. This value matches with \eqref{expect_irr}.

\subsection{The Relevant Operator, $\mathcal{O}_2^{(2)}(0)$}

The form of relevant composite operator $\mathcal{O}_2(0)$ in the subleading order is assumed as.

\begin{align*}  
&\mathcal{O}_2(0)= \Delta S_2+\Delta S_4+\Delta S_6\\
&= \frac{1}{2!}\int_p \bigg \lbrace 1 + A^{(1)}(p) \bigg \rbrace \phi(p)\phi(-p)\\
&+ \frac{1}{4!} \int_{p_1,p_2,p_3} \bigg \lbrace 1-\lambda \sum_{l=1}^4 h(p_l)+B^{(2)}(p_1,p_2,p_3,p_4) \bigg \rbrace \phi(p_1)\phi(p_2)\phi(p_3)\phi(p_4)\\
&+ \frac{1}{6!}\int_{p_1,p_2,p_3,p_4,p_5} \bigg \lbrace D^{(2)}(p_1,p_2,p_3,p_4,p_5,p_6) \bigg  \rbrace \phi(p_1)\phi(p_2)\phi(p_3)\phi(p_4)\phi(p_5) \phi(p_6)\\
\end{align*}

with \[ d_2= 2 -F \la + d_2^{(2)}\]

In this section we have written the final expressions of $\Delta S$. The details are given in Appendix \ref{appD}.

\subsubsection{Determination of $D^{(2)}(p_1,p_2,p_3,p_4,p_5,p_6)$ from $\phi^6$ equation}

There are two kinds of 6-pt vertices distinguised according to their tensor structure (see \ref{solvphi6_rel} for details).

\begin{subequations}
\label{S6slead_rel}
\begin{align}
& D_I^{(2)}(p_1,p_2,p_3,p_4,p_5,p_6)=  \la^2\sum_{10~perm~(i,j,k)} h(p_i+p_j+p_k)h(p_i+p_j+p_k)\\
& D_{II}^{(2)}(p_1,p_2,p_3,p_4,p_5,p_6)= \la^2\sum_{10~perm~(i,j,k)} h(p_i+p_j+p_k) \sum_{l=1}^6 h(p_l)
\end{align}
\end{subequations}

\subsubsection{Determination of $B^{(2)}(p_1,p_2,p_3,p_4)$ from $\phi^4$ equation}

Similarly there are 3 kinds of 4-pt vertices (see \ref{solvphi4_rel} for details).

\begin{subequations}
\label{S4slead_rel}
\begin{align}
& B_I^{(2)}(p_1,p_2,p_3,p_4)= \la^2\sum_{3~perm~(i,j)}\bar{H}_3(p_i+p_j)\\
& B_{II}^{(2)}(p_1,p_2,p_3,p_4)= \la^2 \sum_{l=1}^4 h(p_l) \sum_{3~perm~(i,j)} \mathcal{F}(p_i+p_j)\\
& B_{III}^{(2)}(p_1,p_2,p_3,p_4)= -F \lambda^2 \big \lbrace \frac{1}{2} \sum_{i \neq j} h(p_i) h(p_j) + \sum_{l=1}^4 h^2(p_l) \big \rbrace
\end{align}
\end{subequations}

$\bar{H}_3(p)$ and $\mathcal{F}(p)$ is defined in \eqref{H3_bar} and \eqref{F} respectively.

\subsubsection{Determination of $A^{(1)}(p)$ from $\phi^2$ equation}

This $\phi^2$ equation is solved by six kinds of $A^{(2)}$s according to different tensor structures (see \ref{solvphi2_rel} for details).

\begin{subequations}
\label{S2slead_rel}
\begin{align}
& A_I^{(2)}(p)= -\frac{\la^2}{3} \int_{q,k} \big \lbrace  h(p+q+k)h(p)h(q)h(k)- h(q)h(q+k)h(k) \big \rbrace\\
& A_{II}^{(2)}(p)= -\frac{\la^2}{2} \int_{q,k} \big \lbrace h(p+q+k)h(q)h(q)h(k)-h(q+k)h(q)h(q)h(k) \big \rbrace\\
& A_{III}^{(2)}(p)= -\lambda^2 F^2 h(p)+\frac{\epsilon \lambda}{2} h(p)\\
& A_{IV}^{(2)}(p)= \frac{3}{4} F^2 \lambda^2 h^2(p)\\
& A_V^{(2)}(p)= -\la^2 h(p)\int_q f(q) \mathcal{F}(q)\\
& A_{VI}^{(2)}(p)=\frac{\eta}{\epsilon} p^2 h(p)
\end{align}
\end{subequations}

\subsubsection{Anomalous dimension}

We collect the unused leftover terms like we did in the previous subsection to get the anomalous dimension:

\begin{align}\label{final_rel}
 \nonumber &\lambda^2 \int_{q,k} \big \lbrace -K^\prime(q) \big \rbrace h(q+k)h(k)h(k)+ \lambda^2 \int_{q,k} \big \lbrace -K^\prime(q)h(q) \big \rbrace \big \lbrace h(q+k)h(k)- h(k)h(k) \big \rbrace+ \frac{3}{2} \int_q K^\prime(q) h^2(q)\\
 &+ \lambda^2 \frac{1}{2} \int_k h(k)h(k)h(k)-\frac{\eta}{2}A^{(0)}(p)= d_2 \hf A^{(0)}(p)
\end{align}

Like we have seen in the calculation of anomlous dimension of the irrelevant operator here also the anomalous dimension is coming from a diagram as shown in Fig.\ref{relevant_sublead} which is logarithmically divergent but made finite by adding a counterterm.

\begin{figure}[b]
   \centering
  \begin{tikzpicture}
  \begin{feynman}
    	\vertex (a) at (0,0);
    	\vertex (b) at (2,0);
    	\vertex (c) at (4,0);
    	\vertex (d) at (6,0);
    	\vertex (x) at (3,1);
    	\vertex (y) at (3,2.0);
      \diagram* {
       (a) -- (b) --[edge label=$h(q/\Lambda)$] (c) --(d),
       (b) -- [quarter left, near end, edge label= $h(p/\Lambda)$](x) -- [edge label=$h(p/\Lambda)$,quarter left](c),
       (c) -- [half left, edge label= $h(p/\Lambda + q/\Lambda)$](b),
       (c) -- (d),
       (y) -- [photon, insertion= 1.0, edge label=$\phi^2(0)$](x),
       };
    \end{feynman}
  \end{tikzpicture}
  \caption{Diagram contributing to $d_2^{(2)}$ for the relevant operator}
  \label{relevant_sublead}
\end{figure}
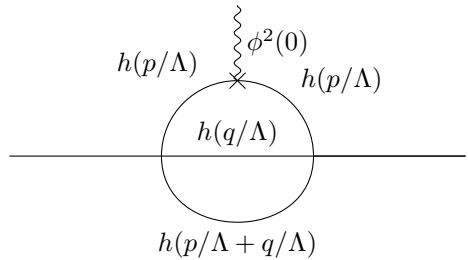

\vspace{0.2 in}

\textbf{Evaluation of Integrals}

\vspace{0.1 in}

\begin{align*}
a.~~\int_q K^\prime(q) h^2(q)= 2 \log 2- \log 3
\end{align*}

\begin{align*}
b.~~\int_k h(k)h(k)h(k)= 3 \log 3 -6 \log 2
\end{align*}

So the third and fourth term on LHS of \eqref{final_rel} cancels among each other. The rest of the integrals in the LHS we know from the previous subsection. So in the limit of $\Lambda_0 \rightarrow \infty $ we obtain the anomalous dimension as,

\[
\tcboxmath{
d_2=2 \left( \frac{\lambda^2 F^2}{2}-\frac{\eta}{2}\right)=\frac{5}{6} \la^2 F^2
}
\]

Where $F=\frac{1}{16\pi^2}$ and $\frac{\eta}{2}=\frac{\lambda^2 F^2}{12}$. This agrees with \eqref{expect_rel}.

\section{Conclusion}

In this paper two composite operators in the $\phi^4$ scalar field theory at the Wilson-Fisher fixed point in $D=4-\eps$ dimension have been constructed. The composite operators and their anomalous dimensions are listed in \eqref{dim_lead},\eqref{relevant_lead}, \eqref{irrelevant_lead}, \eqref{dim_sublead}, \eqref{irrelevant_sublead} and \eqref{relevant_sublead}.

These  operators are eigenfunctions of the  ERG evolution equation for linearized perturbations about the fixed point. Thus they have definite dimension. The dimension of the operators are also calculated to $O(\eps^2)$ and agree with known results. These operators reduce to $\phi^2$ and $\phi^4$ as the coupling constant goes to zero.
At the W-F fixed point this would mean $\eps\to 0$.

Dimension of an operator is a well defined concept only if the underlying theory is scale invariant (at least in some approximation).
The fixed point condition of the ERG equation is a condition for scale invariance of the action.  This was solved to  $O(\eps^2)$ in \cite{Dutta}.  The energy momentum tensor was also shown to be traceless, thus verifying that this theory is also conformal invariant - as expected on general grounds. Thus the operators constructed in this theory should correspond to primary operators of this CFT. However this need to be verified by checking the Conformal Ward Identities, which requires a {\em local} operator, i.e. ${\cal O}_2 (q), {\cal O}_4(q)$ with $q\neq 0$. We leave this for the future.

The main point of this paper (and also of \cite{Dutta}) is that the UV cutoff is kept finite throughout. Thus both the fixed point action constructed in \cite{Dutta} and the composite operators constructed here are valid at all length scales. In particular scale and conformal invariance of the action is not an approximate statement valid at energies $p<<\lm$ but is valid for all $p$. In the same way the expressions for the composite operators in terms of fundamental fields are valid when the internal momentum circulating is arbitrarily large. (Note that because of the analytic form of the cutoff function, loop momenta are not restricted to be less than $\lm$.)  

As mentioned in the introduction, CFT's and more generally field theories with a finite UV cutoff are conceptually interesting and generalize the notion of scale invariance in the presence of a UV cutoff. These could have applications in condensed matter physics and critical phenomena, because these systems always have an underlying short distance cutoff. 

The results of this paper and \cite{Dutta} are also relevant for a better understanding of holography. The bulk AdS dual of the O(N) model has been studied. The connection between ERG and Holographic RG has also been studied recently and in these approaches a finite  cutoff plays a crucial role \cite{Sathiapalan:2017frk,Sathiapalan:2019zex,Sathiapalan:2020cff}.


 There are several other open questions.  One is to understand the precise role of the irrelevant terms in the Wilson Action when constructing the bulk AdS-dual.
 It would also be interesting to have more examples of such constructions in other CFT's and in other dimensions where a Lagrangian description is available, for eg., Wess-Zumino-Witten models and $O(N)$ models in 3 dimensions.
 
  Finally and perhaps most important is the inclusion of gravity in these theories and the connection with string theory. If one were to speculate (as for instance in \cite{BSLV}) that underlying space time in string theory is not a continuum then it may also be necessary to understand properties of theories with finite cutoff where the underlying ``lattice" is dynamical.

{\bf Acknowledgements:} We thank Hidenori Sonoda for useful discussions. SD thanks Pavan Dharanipragada for suggestions in the calculations.

\begin{appendices}
\section{Local Operators}
Under a scale transformation
\be
\bar x=\la x~~~~,~~~\bp = \frac{p}{\la}
\ee
\[
\bphi (\bp) = \la^{-d_\phi^p}\phi(p)
\]
Here $d_O^x$ is the scaling dimension of any operator $O(x)$ and $d_O^p=d_O^x-D$ is the scaling dimension of $O(p)$.  Let $\la=e^{-t}$ and $\bp=pe^t$.
\[
\bphi(pe^t)=e^{d_\phi^pt}\phi(p)
\]
\[
e^{-d_\phi^pt}\bphi(pe^{t})=\phi(p)
\]
We hold $p$ fixed and change $t$:
\[
\frac{\p\phi(p)}{\p t}= (-d_\phi^p+\pdp)\phi(p)
\]
and more generally for any operator with mass scaling dimension $d_O^p$:
\be  \label{local}
\frac{\p O(p)}{\p t}= (-d_O^p+\pdp)O(p)
\ee 
One can also call $-d_O^p$ the length scaling dimension.

Let us consider operators of the form
\be
\DD S = \int_q B(q) O(q)
\ee
Then the change under scaling can be written as
\[
\frac{\p \DD S}{\p t} = \int_q B(q)(-d_O^q+\qdq ) O(q)
\]
\[=
\int_q [(-d_O^p-D -\qdq) B(q)]O(q)=\int_q [(-d_O^x -\qdq)B(q)] O(q)
\]
This gives the action on the coefficient functions in the composite operator. 

Thus if we have 
\[
O = \int_{p_1}\int_{p_2} A(p_1,p_2) \phi(p_1)\phi(p_2)
\]
Then
\be   \label{gdil}
\frac{\p O}{\p t} =\int_{p_1}\int_{p_2}[(-p_1\frac{d}{d p_1}-p_2\frac{d}{d p_2}-2d_\phi^x) A(p_1,p_2)] \phi(p_1)\phi(p_2)
\ee
The operator acting on the coefficient functions $A$ has been called ${\cal G}^c_{dil}$ in the literature. The superscript $c$  denotes that it is the contribution to scaling due to the classical or engineering dimensions. (see for eg.\cite{Bagnuls1,Bagnuls2}).
 
Let us consider some simple examples that will be used.

\begin{enumerate}
\item
\be
A(p_1,p_2)=\dd (p_1+p_2-q)
\ee
Then using
\be   \label{dfn}
(p_1\frac{d}{d p_1}+p_2\frac{d}{d p_2}+\qdq) \dd(p_1+p_2-q)=-D\dd(p_1+p_2-q)
\ee
we obtain
\[
\frac{\p O}{\p t}=\int_{p_1}\int_{p_2}(-2d_\phi^x +D +\qdq)\dd (p_1+p_2-q)\phi(p_1)\phi(p_2)
\]
\[
\frac{\p O}{\p t}=(-d_O^p +\qdq)\int_{p_1}\int_{p_2}\dd (p_1+p_2-q)\phi(p_1)\phi(p_2)
\]
as required.
\item
More generally
\be  \label{a68}
A(p_1,p_2)=\dd (p_1+p_2-q)B(p_1,p_2,q)
\ee
Then going through the same steps one obtains
\[
\frac{\p O}{\p t}=\Big((-2d_\phi^x+D +\qdq)\int_{p_1}\int_{p_2}\dd (p_1+p_2-q)B(p_1,p_2,q) \]
\[+\int_{p_1}\int_{p_2}\dd (p_1+p_2-q)(-p_1\frac{d}{d p_1}-p_2\frac{d}{d p_2}-\qdq)B(p_1,p_2,q)\Big) \phi(p_1)\phi(p_2)
\]
If $B(p_1,p_2,q)$ has a well defined scaling dimension it adds to $d_\phi^p$. For eg
if $B(p_1,p_2,q)=p_1.p_2$ the operator is just the kinetic term and we get $-2d_\phi^x+D-2=0$, which is the dimension of $\int_p(p.(q-p))\phi(p)\phi(q-p)$.

\end{enumerate}

\section{Exact Renormalization Group }\label{A}
The Polchinski's Exact Renormalization Group equation for the Wilson Action (see \cite{Igarashi,MorrisERG,Bagnuls2,Rosten:2010vm, Rosten1,Wetterich, Igarashi-gamma}) is the following:

\begin{align}\label{}
\nonumber &\frac{\partial  S}{\partial t}= \int_p  \bigg \lbrace (-K'(p^2)) [\frac{\delta^2  S}{\delta \phi(p)\delta \phi(-p)}-\frac{\delta S}{\delta \phi(p)}\frac{\delta  S}{\delta \phi(-p)}] -2\frac{p^2K'}{K}\phi(p)\frac{\delta  S}{\delta \phi(p)}+\\
 \nonumber &+ \frac{-\eta}{2} \frac{K(p^2)(1-K(p^2)}{p^2} [\frac{\delta^2  S}{\delta \phi(p)\delta \phi(-p)} -\frac{\delta S}{\delta \phi(p)}\frac{\delta  S}{\delta \phi(-p)}]+ \frac{-\eta}{2} \phi(p) \frac{\delta  S}{\delta \phi(p)} \bigg \rbrace\\
&+\underbrace{[(1-\frac{D}{2})N_\phi+D-N_p]}_{{\cal G}_{dil}} S
\end{align}

We have used some simplified notation: $N_\phi$ counts the number of $\phi$'s in any term. $N_p$ counts the powers of momenta in any expression. It can also be written as $\sum_i p_i\frac{\p}{\p p_i}$, where $p_i$ are all the inependent momenta.  ${\cal G}_{dil}$ thus counts the engineering dimension of a given term. This is an extra contribution that arises when one changes from dimensionful to dimensionless variables. $\frac{\eta}{2}$ is the anomalous dimension of $\phi$ associated with a fixed point. And the operator it multiplies is just the composite number operator $[\int_p \phi(p) \frac{\dd}{\dd \phi(p)}]$.

 A perturbation is made to the action: $S\to S+\DD S$. The linearized equation for $\DD S$ is given below and is satisfied by all composite operators by definition.

\begin{align}\label{appA}
\nonumber &\frac{\partial \Delta S}{\partial t}= \int_p  \bigg \lbrace (-K'(p^2)) [\frac{\delta^2 \Delta S}{\delta \phi(p)\delta \phi(-p)}-2\frac{\delta S}{\delta \phi(p)}\frac{\delta \Delta S}{\delta \phi(-p)}] -2\frac{p^2K'}{K}\phi(p)\frac{\delta \Delta S}{\delta \phi(p)}+\\
 \nonumber &+ \frac{-\eta}{2} \frac{K(p^2)(1-K(p^2)}{p^2} [\frac{\delta^2 \Delta S}{\delta \phi(p)\delta \phi(-p)} -2\frac{\delta S}{\delta \phi(p)}\frac{\delta \Delta S}{\delta \phi(-p)}]+ \frac{-\eta}{2} \phi(p) \frac{\delta \Delta S}{\delta \phi(p)} \bigg \rbrace\\
&+[(1-\frac{D}{2})N_\phi+D-N_p]\Delta S
\end{align}

This is the equation used in the paper along with the eigenvalue condition (see \eqref{local})
\be
\frac{\p \DD S(q)}{\p t} = [d_m +\qdq]\DD S(q)+ \beta(\lambda) \frac{\p \Delta S}{\p \la}
\ee
Here $d_m$ is the length scaling dimension.
For much of this paper we will in fact set $q=0$ for simplicity. This is equivalent to considering the  operator integrated over space.

\section{Composite operators at the leading order}
\label{appB}

In this appendix we have calculated different parts of \eqref{pol_lead} upto $\lambda^1$. Note that we have marked different parts as \textbf{(1)}, \textbf{(2)},\textbf{(3)} and \textbf{(4a)} respectively. As we have considered only the leading order terms we remove the superscript $(1)$ from 4 and 6-pt vertices $B_I$ , $B_{II}$ and $D$.

\textbf{(1)}

\[
\int_p \lbrace -K'(p^2) \rbrace
\frac{\dd^2 \DD S}{\dd \phi(p)\dd \phi(-p)}=
\int_p \lbrace -\kp \rbrace \hf\int_{p_1,p_2} \dd(p_1+p_2-q)\Big(B^{(0)} +(B_I(p_1,q)+B_I(p_2,q)+B_I(p,q)+B_I(-p,q))+
\]

\[+\hf [ B_{II}(p_1+p_2,q)+B_{II}(p_1+p,q)+B_{II}(p_1-p,q)+B_{II}(p_2+p,q)+B_{II}(p_2-p,q)+B_{II}(0,q)]\Big)\phi(p_1)\phi(p_2)
\]

\[+\frac{1}{4!}\int_p \lbrace -\kp \rbrace \hf\int_{p_1,p_2,p_3,p_4}\dd(p_1+p_2+p_3+p_4-q)[(D(p_1,q)+D(p_2,q)+D(p_3,q)+D(p_4,q)+
\]
\[+D(p_1+p_2+p_3,q)+D(p_1+p_2+p_4,q)+D(p_1+p_3+p_4,q)+D(p_2+p_3+p_4,q))+
\]
\[+(D(p_1+p_2+p,q)+D(p_1+p_3+p,q)+D(p_1+p_4+p,q)+D(p_3+p_2+p,q)+D(p_4+p_2+p,q)+D(p_3+p_4+p,q))+
\]
\[+(D(p_1+p_2-p,q)+D(p_1+p_3-p,q)+D(p_1+p_4-p,q)+D(p_3+p_2-p,q)+D(p_4+p_2-p,q)+D(p_3+p_4-p,q))]\]
\[\phi(p_1)\phi(p_2)\phi(p_3)\phi(p_4)\]

If we set $q=0$ in the above things simplify considerably:

\[\int_p \lbrace -K'(p^2) \rbrace \frac{\dd^2 \DD S}{\dd \phi(p)\dd \phi(-p)}=\int_p \lbrace -\kp \rbrace B_I(p)\int_q\phi(q)\phi(-q)+F\int_q B_I(q)\phi(q)\phi(-q)+\]\[\hf F B_{II}(0)\int_q \phi(q)\phi(-q)+ \hf \int_p \lbrace -\kp \rbrace  \int_q [B_{II}(p+q)+B_{II}(p-q)]\phi(q)\phi(-q)\]\[+\frac{1}{4!}\int_p \lbrace -\kp \rbrace \int_{p_1,p_2,p_3}[D(p_1)+D(p_2)+D(p_3)+D(p_4)\]\[+ D(p+p_3+p_4)+D(p+p_3+p_2)+D(p+p_3+p_1)\]\[+D(p-p_3-p_4)+D(p-p_3-p_2)+D(p-p_3-p_1)]\]\[\phi(p_1)...\phi(p_4)~~~~~p_4=-p_3-p_2-p_1\]

\textbf{(2)+(3)}
\[-2\int_p \lbrace -\kp \rbrace \frac{\dd S}{\delta \phi(p)}\frac{\dd \DD S}{\delta \phi(-p)}-\int_p2\frac{p^2K'}{K}\phi(p)\frac{\dd \DD S}{\delta \phi(p)}=\]
\[-\int_{p_1,p_2}\sum_i \lbrace -K'(p_i^2) \rbrace U_2(p_i)\dd(p_1+p_2-q)[A^{(0)}+  A^{(1)}(p_1,p_2,q)]\phi(p_1)\phi(p_2)+
\]
\[-\frac{2}{4!}\la \int_{p_1,p_2,p_3,p_4}\dd(p_1+p_2+p_3+p_4-q)\sum _{i=1}^4 \lbrace -K'((p_i-q)^2)\rbrace [A^{(0)}+ A^{(1)}(p_i,p_j+p_k+p_l)]\phi(p_1)\phi(p_2)\phi(p_3)\phi(p_4)
\]
\[
-\frac{2}{4!} \int_{p_1,p_2,p_3,p_4}\dd(p_1+p_2+p_3+p_4-q) \sum_i \lbrace -K'(p_i^2) \rbrace U_2(p_i) \Big( B^{(0)}+ (B_I(p_1)+B_I(p_2)+B_I(p_3)+B_I(p_4) \]
\[  
+B_{II}(p_1+p_2,q)+B_{II}(p_1+p_3,q)+B_{II}(p_1+p_4,q)+\]
\[+B_{II}(p_2+p_3,q)+B_{II}(p_2+p_4,q)+B_{II}(p_3+p_4,q))\Big)
 \phi(p_1)\phi(p_2)\phi(p_3)\phi(p_4)
\]

Once again if we set $q=0$ the result is simpler:

\[-2\int_p \lbrace -\kp \rbrace \frac{\dd S}{\delta \phi(p)}\frac{\dd \DD S}{\delta \phi(-p)}-\int_p2\frac{p^2K'}{K}\phi(p)\frac{\dd \DD S}{\delta \phi(p)}= -2\Big[\int_p \lbrace -\kp \rbrace U_2(p)A(p)\phi(p)\phi(-p)\]\[+\frac{1}{3!} \int_p \lbrace -\kp \rbrace A(p)\int_{p_2,p_3}(\la +U_4(p,p_2,p_3,p_4))\phi(-p)\phi(p_2)\phi(p_3)\phi(p_4)~~;~p=p_2+p_3+p_4\]\[+\frac{1}{3!}\int_p \lbrace -\kp \rbrace U_2(p)\int_{q_2,q_3}[B_I(p)+B_I(q_2)+B_I(q_3)+B_I(q_4)+B_{II}(p+q_2)+B_{II}(p+q_3)+B_{II}(p+q_4)]\]\[\phi(p)\phi(q_2)\phi(q_3)\phi(q_4)\Big]~~~;-p=q_2+q_3+q_4\]

Rename \(p->p_1\) and then symmetrize:

\[=-2\Big[\int_p \lbrace -\kp \rbrace U_2(p)A(p)\phi(p)\phi(-p)\]\[+\frac{1}{4!} \int_{p_1,p_2,p_3}(\sum_{i=1}^4 \lbrace -\kpi \rbrace A(p_i))(\la +U_4(p_1,p_2,p_3,p_4))\phi(p_1)\phi(p_2)\phi(p_3)\phi(p_4)~~\]\[+\frac{1}{4!}\int_{p_1,p_2,p_3}(\sum_{i=1}^4 \lbrace -\kpi \rbrace U_2(p_i))[B_I(p)+B_I(q_2)+B_I(q_3)+B_I(q_4)+B_{II}(p+q_2)+B_{II}(p+q_3)+B_{II}(p+q_4)]\]\[\phi(p)\phi(q_2)\phi(q_3)\phi(q_4)\Big]~~~;p_4=-(p_1+p_2+p_3)\]

We write the \(\phi^6\) terms separately (we set $q=0$ here since these terms are not required for the relevant operator at leading order):

\[-\frac{4}{6!}\int_{p_1,...p_5} \sum_{10~perm~i,j,k} \lbrace -K'((p_i+p_j+p_k)^2)\rbrace[\la+U_4(p,p_i,p_j,p_k)]\]\[ [ B_I(p)+B_I(p_a)+B_I(p_b)+B_I(p_c)+B_{II}(p+p_a)+B_{II}(p+p_b)+B_{II}(p+p_c)]\phi(p_1)....\phi(p_6)\]\[
+ \frac{2}{6!}\int_{p_1,...p_5}[\sum_i \lbrace \lbrace -\kpi \rbrace \rbrace U_2(p_i)][\sum_{10~perm~i,j,k}D(p_i+p_j+p_k)]\phi(p_1)....\phi(p_6)\]\[+\frac{2}{6!}\int_{p_1,...p_5}[\sum_i -\kpi A(p_i)][V_6(p_1,...p_6)]\phi(p_1)....\phi(p_6)\]
\[p=p_i+p_j+p_k=-(p_a+p_b+p_c)\]

\textbf{(4a)}

The general form of the action of ${\cal G}_{dil}^c$ is given by
\[
{\cal G}_{dil}^c  \dd(\sum p_i -q) X (p_1,..,p_N)=((1-\Dt)N -\sum p_i\frac{\p}{\p p_i})\dd(\sum p_i -q) X (p_1,..,p_N)
\]
\[ =
((1-\Dt)N +D +\qdq)\dd(\sum p_i -q) X (p_1,..,p_N)-
\]
\be \label{cgdil}
\dd(\sum p_i -q)(\sum p_i\frac{\p}{\p p_i}+\qdq)X(p_1,..,p_N)
\ee

When $q=0$ we get:

\[{\cal G}_{dil}^c\hf\int _p A(p) \phi(p)\phi(-p)=\int_p(A(p)-p^2 \frac{d}{dp^2}A(p))\phi(p)\phi(-p)\]\[
{\cal G}_{dil}^c\frac{1}{4!}\int _{p_1,p_2,p_3}[\sum_i \lbrace B^{(0)}(p_i)+B_I(p_i) \rbrace+ (B_{II}(p_1+p_2)+B_{II}(p_1+p_3)+B_{II}(p_1+p_4))]\phi(p_1)...\phi(p_4)=\]\[\frac{1}{4!}\int _{p_1,p_2,p_3}[(4-D- \sum_i p_i\frac{d}{dp_i})[\sum_i \lbrace B^{(0)}(p_i)+ B_I^{(1)}(p_i)\rbrace + (B_{II}(p_1+p_2)+B_{II}(p_1+p_3)+B_{II}(p_1+p_4))]\phi(p_1)...\phi(p_4)\]\[
{\cal G}_{dil}^c\frac{1}{6!}\int_{p_1,...p_5}\sum_{10~perm~i,j,k}D(p_i+p_j+p_k)\phi(p_1)...\phi(p_6)~~~\]\[=\frac{1}{6!}\int_{p_1,...p_5}(6-2D- \sum_i p_i\frac{d}{dp_i})\sum_{10~perm~i,j,k}D(p_i+p_j+p_k)\phi(p_1)...\phi(p_6)~~~\]\[~~~p_6=-p_1...-p_5\]\[\]

\section{Irrelevant Operator at subleading order}
\label{appC}

\subsection{ The $\phi^6$ equation}
\label{solvphi6_irr}

\begin{align}\label{phi6_irr}
& \nonumber \int_p \lbrace - K^\prime(p^2) \rbrace \frac{\delta^2 \Delta S_8^{(2)}(0)}{\delta \phi(p) \delta \phi(-p)}-\frac{4}{6!} \sum_{10~perm~(i,j,k)} \big \lbrace -K^\prime(p_i+p_j+p_k) \big \rbrace \big \lbrace \lambda+ U_4^{(2)}(p,p_i,p_j,p_k) \big \rbrace\\ \nonumber
&\big \lbrace B^{(0)}(p)+B^{(0)}(p_a)+ B^{(0)}(p_b)+ B^{(0)}(p_c)+B_I^{(1)}(p)+B_I^{(1)}(p_a)+ B_I^{(1)}(p_b)+B_I^{(1)}(p_c)\\
\nonumber &+B_{II}^{(1)}(p+p_a)+B_{II}^{(1)}(p+p_b)+ B_{II}^{(1)}(p+p_c)\big \rbrace\\ \nonumber
-& \frac{2}{6!} \big \lbrace \sum_{l=1}^6 (-K(p_l^2)) U_2^{(1)}(p_l) D^{(1)}(p_1,p_2,p_3,p_4,p_5,p_6) \big \rbrace - \frac{2}{6!} \big \lbrace \sum_{l=1}^6 (-K(p_l^2)) A^{(0)}(p_l) V_6^{(2)}(p_1,p_2,p_3,p_4,p_5,p_6) \big \rbrace\\ \nonumber
+& \frac{1}{6!} \big \lbrace 6-2D-\sum_{i=1}^6 p_i. \frac{d}{d p_i} \big \rbrace D^{(2)}(p_1,p_2,p_3,p_4,p_5,p_6)+ \frac{1}{6!} (2\epsilon) \big \lbrace D^{(1)}(p_1,p_2,p_3,p_4,p_5,p_6)\\
=& \frac{1}{6!}  \lbrace \epsilon-6F\la \rbrace D^{(1)}(p_1,p_2,p_3,p_4,p_5,p_6)+\frac{1}{6!} \lbrace  \epsilon \lambda+\beta_1^{(1)}(\lambda) \rbrace \frac{\partial}{\partial \lambda} D^{(1)}(p_1,p_2,p_3,p_4,p_5,p_6)
\end{align}

The last term on LHS comes from putting $D=4-\epsilon$ in $(6-2D)D^{(1)}$ term.

Where

\[\beta_1^{(1)}(\lambda)= -3F\lambda^2 \]

So the first and 3rd term combined in RHS cancels the last term in LHS.

\begin{align}\label{U4}
\nonumber & U_4^{(2)}(p,p_i,p_j,p_k)\\
&= \underbrace{-\lambda^2 \big \lbrace \mathcal{F}(p+p_i)+\mathcal{F}(p+p_j)+\mathcal{F}(p+p_k)\big \rbrace }_{U_4^I}+\underbrace{\frac{F\lambda^2}{2} \sum_{i=1}^4 h(p_i)}_{U_4^{II}}
\end{align}

Where $\mathcal{F}(p)=\frac{1}{2}\int_k \big \lbrace h(p+k)h(k)-h(k)h(k) \big \rbrace $.

\subsubsection*{$\frac{\delta^2 \Delta S_8^{(2)}(0)}{\delta \phi(p) \delta \phi(-p)}$ in $\phi^6$ equation}

\begin{align}\label{ddphiS8}
& \nonumber \frac{\delta^2 \Delta S_8^{(2)}(0)}{\delta \phi(p) \delta \phi(-p)}\\ \nonumber
=& \frac{3\lambda^2}{8!} \bigg \lbrace 28 \sum_{10~perm~(1,j,k)} \sum_{l=1}^6 h(p_l)h(p_i+p_j+p_k)+ 28\times 4 \sum_{10~perm~(i,j,k)}\sum_{3~perm~(\alpha,\beta)} h(p_i+p_j+p_k)h(p_\alpha+p_\beta+p)\\ \nonumber
+& 56 \sum_{10~perm~(i,j,k)}h(p_i+p_j+p_k)h(p_i+p_j+p_k)+ 28 \sum_{l=1}^6 \sum_{10~perm~(i,j,k)}h(p_l)h(p_i+p_j+p_k)\\ \nonumber
+& 112 \sum_{10~perm~(i,j,k)}h(p_i+p_j+p_k)\sum_{3~perm~(\alpha,\beta)} h(p_\alpha+p_\beta+p)\\ \nonumber
+& 56 \sum_{15~perm~(i,j)}\sum_{6~perm~(\alpha,\beta)}h(p_i+p_j+p)h(p_i+p_j+p_\alpha+p_\beta+p) \bigg \rbrace \phi(p_1)\phi(p_2)\phi(p_3)\phi(p_4)\phi(p_5)\phi(p_6)\\
\end{align}

\subsubsection*{Equation for $D_I^{(2)}(p_1,p_2,p_3,p_4,p_5,p_6)$}

We take 2nd and 5th term of R.H.S of \eqref{ddphiS8}, Note that the coefficients $\epsilon D^{(1)}$ terms cancel, now considering all terms in RHS we get: 

\begin{align}\label{D6_1}
\nonumber &\bigg \lbrace 6-2D -2\sum_{l=1}^6 p_i.\frac{d}{dp_i}  \bigg \rbrace D^{(2)}_{I}(p_1,p_2,p_3,p_4,p_5,p_6)\\ \nonumber
+& \frac{12 \lambda^2}{6!} \int_p \big( -K^\prime(p^2) \big)\sum_{10~perm~(i,j,k)}\sum_{3~perm~(\alpha,\beta)}h(p_i+p_j+p_k)\bigg \lbrace h(p_\alpha+p_\beta+p)-h(p)\bigg \rbrace\\ \nonumber
+& \frac{4}{6!} \sum_{10~perm~(i,j,k)}K^\prime(p_i+p_j+p_k) U_4^I(p,p_i,p_j,p_k)\sum_{l=1}^4 B^{(0)}(p_l)\\ \nonumber
+&\frac{4}{6!} \sum_{10~perm~(i,j,k)}\lambda K^\prime(p_i+p_j+p_k) \bigg \lbrace B_{II}^{(1)}(p+p_a)+B_{II}^{(1)}(p+p_b)+B_{II}^{(1)}(p+p_c) \bigg \rbrace=0\\
\end{align}

\subsubsection*{Equation for $D_{II}^{(2)}(p_1,p_2,p_3,p_4,p_5,p_6)$ and $D_{III}^{(2)}(p_1,p_2,p_3,p_4,p_5,p_6)$}

We take 1st, 3rd and 4th term from \eqref{ddphiS8} and remaining all terms in \eqref{phi6_irr}:

\begin{align}\label{D6_23}
&\nonumber\frac{4}{6!} \sum_{10~perm~(i,j,k)} \big \lbrace K^\prime(p_i+p_j+p_k) \big \rbrace \big \lbrace \lambda \big \rbrace \big \lbrace \underbrace{B_I^{(1)}(p)}_{1}+B_I^{(1)}(p_a)+B_I^{(1)}(p_b)+B_I^{(1)}(p_c) \big \rbrace\\ \nonumber
&+\frac{4}{6!} \sum_{10~perm~(i,j,k)} \big \lbrace K^\prime(p_i+p_j+p_k) \big \rbrace \big \lbrace B^{(0)}(p)+B^{(0)}(p_a)+B^{(0)}(p_b)+B^{(0)}(p_c) \big \rbrace \underbrace{\frac{F\lambda^2}{2} \bigg \lbrace \underbrace{ h(p)}_{1}+h(p_i)+h(p_j)+h(p_k) \bigg \rbrace}_{U_4^{II}}\\ \nonumber
&-\frac{2}{6!} \sum_{l=1}^6\bigg \lbrace -K^\prime(p_l^2) U_2^{(1)}(p_l) \bigg \rbrace \bigg \lbrace D^{(1)}(p_1,p_2,p_3,p_4,p_5,p_6) \bigg \rbrace-\frac{2}{6!} \sum_{l=1}^6\bigg \lbrace -K^\prime(p_l^2) A^{(0)}(p_l) \bigg \rbrace \bigg \lbrace V_6^{(2)}(p_1,p_2,p_3,p_4,p_5,p_6) \bigg \rbrace\\ \nonumber
&+\frac{3\lambda^2}{6!} \sum_{l=1}^6 \sum_{10~perm~(i,j,k)} Fh(p_l)h(p_i+p_j+p_k)+\int_p \big(-K^\prime(p^2)\big) \frac{3\lambda^2}{6!} \sum_{10~perm~(i,j,k)} \bigg \lbrace h(p_i+p_j+p_k)h(p_i+p_j+p_k) \bigg \rbrace\\ \nonumber
&+\frac{1}{6!} \bigg( 6-2D-\sum_i p_i.\frac{d}{dp_i} \bigg) \bigg \lbrace  \underbrace{D_{II}^{(2)}(p_1,p_2,p_3,p_4,p_5,p_6)}_{1}+ D_{III}^{(2)}(p_1,p_2,p_3,p_4,p_5,p_6)\bigg \rbrace=0\\
\end{align}

Let's take collect all terms marked with "1" marked and the 6 th term on LHS,

\begin{align}
\nonumber &\frac{4}{6!} \sum_{10~perm~(i,j,k)} \big \lbrace K^\prime(p_i+p_j+p_k) \big \rbrace \big \lbrace \lambda \big \rbrace B_I^{(1)}(p)\\
\nonumber +&\frac{4}{6!} \sum_{10~perm~(i,j,k)} \big \lbrace K^\prime(p_i+p_j+p_k) \big \rbrace \big \lbrace B^{(0)}(p)+B^{(0)}(p_a)+B^{(0)}(p_b)+B^{(0)}(p_c) \big \rbrace \frac{F\lambda^2}{2} h(p)\\
\nonumber +&\int_p \big(-K^\prime(p^2)\big) \frac{3\lambda^2}{6!} \sum_{10~perm~(i,j,k)} \bigg \lbrace h(p_i+p_j+p_k)h(p_i+p_j+p_k) \bigg \rbrace\\
\nonumber +&\frac{1}{6!} \bigg( 6-2D-\sum_i p_i.\frac{d}{dp_i} \bigg) D_{II}^{(2)}(p_1,p_2,p_3,p_4,p_5,p_6) =0
\end{align}

Collecting other terms in \eqref{D6_23} we get equation to solve $D_{III}^{(2)}(p_1,p_2,p_3,p_4,p_5,p_6)$.

\subsubsection*{Equation for $D_{IV}^{(2)}(p_1,p_2,p_3,p_4,p_5,p_6)$}

At last, only term remains in \eqref{ddphiS8} is the 6 th term. So the equation for $D_{IV}^{(2)}(p_1,p_2,p_3,p_4,p_5,p_6)$

\begin{align}\label{D6_4}
\nonumber &\frac{3\lambda^2}{6!}\sum_{15~perm~(i,j)}\sum_{6~perm~(\alpha,\beta)}\int_p \big \lbrace -K^\prime(p^2) \big \rbrace h(p_i+p_j+p)h(p_i+p_j+p_\alpha+p_\beta+p)\\
&+\big(6-2D- p_i.\frac{d}{dp_i} \big) \frac{1}{6!} D_{IV}^{(2)}(p_1,p_2,p_3,p_4,p_5,p_6)=0
\end{align}

\subsection{The $\phi^4$ equation to determine $B^{(2)}(p_1,p_2,p_3,p_4)$}
\label{solvphi4_irr}

Now we will write $\phi^4$ contribution in \eqref{pol}. We recall that while calculating 4-pt vertex of leading order there were two left over terms $(4-D)B_I^{(1)}(p_1,p_2,p_3,p_4)$ and $2(4-D)B_{II}^{(1)}(p_1,p_2,p_3,p_4)$. We have added those terms in LHS of the equation below.

\begin{align}\label{BA_irr}
\nonumber &\overbrace{\int_p K^\prime(p^2) \frac{\delta}{\delta \phi(p)}\frac{\delta}{\delta \phi(-p)} D^{(2)}(p_1,p_2,p_3,p_4,p_5,p_6)}^{A}-\overbrace{\frac{2}{4!}\sum_{i=1}^4\big \lbrace -K^\prime(p_i^2) \big \rbrace \big \lbrace A^{(0)}(p_i)+ A^{(1)}(p_i) \big \rbrace \big \lbrace \lambda+U_4^{(2)}(p_1,p_2,p_3,p_4) \big \rbrace}^{B}\\ \nonumber
&-\overbrace{\frac{2}{4!}\sum_{i=1}^4\big \lbrace -K^\prime(p_i^2) \big \rbrace \big \lbrace U_2^{(1)}(p_i)+U_2^{(2)}(p_i) \big \rbrace  \big \lbrace \sum_{i=1}^4 B^{(0)}(p_i)+\sum_{i=1}^4 B_I^{(1)}(p_i)+B_{II}^{(1)}(p_1,p_2,p_3,p_4) \big \rbrace}^{C}\\ \nonumber
&\overbrace{-\frac{\eta}{2} \bigg \lbrace \sum_{i=1}^4 \frac{4}{4!}  B^{(0)}(p_i) \bigg \rbrace +\frac{1}{4!} \eta \sum_{i=1}^4 p_i^2 h(p_i)}^D\\
& \nonumber +\overbrace{\frac{1}{4!} \big \lbrace (4-D)- p_i.\frac{d}{dp_i} \big \rbrace \big \lbrace  B^{(2)}(p_1,p_2,p_3,p_4) \big \rbrace +2(4-D) B_{II}^{(1)}(p_1,p_2,p_3,p_4)+(4-D) B_I^{(1)}(p_1,p_2,p_3,p_4)}^E \\
& \nonumber = \frac{\epsilon-6F\la}{4!} \big \lbrace  B_I^{(1)}(p_1,p_2,p_3,p_4)+ B_{II}^{(1)}(p_1,p_2,p_3,p_4) +B^{(2)}(p_1,p_2,p_3,p_4) \big \rbrace+ \frac{d_4^{(2)}}{4!} \big \lbrace \sum_{i=1}^4 B^{(0)}(p_i) \big \rbrace\\
&+ \frac{1}{4!}\lbrace \epsilon \lambda+ \beta_1^{(1)}(\lambda) \rbrace \frac{\partial }{\partial \lambda} \lbrace B_I^{(1)}(p_1,p_2,p_3,p_4) + B_{II}^{(1)}(p_1,p_2,p_3,p_4) \rbrace
\end{align}

Where $B_I^{(1)}(p_1,p_2,p_3,p_4)= \la\sum_{i=1}^4 h(p_i)$ and $B_{II}^{(1)}(p_1,p_2,p_3,p_4)=-2\la \sum_{3~perm~(i,j)} \mathcal{F}(p_i+p_j)$.

\[ \beta_1^{(1)}(\la)= -3F\la^2 \]

\begin{align*}
U_2^{1}(p)= -\frac{\la F}{2-\eps};~~U_2^{(2)}(p)= -\lambda^2 G(p)-\frac{\lambda^2 F^2}{4} h(p)
\end{align*}

Where

\begin{equation*}
G(p)= \frac{1}{3}\int_{q,k} \frac{h(q)}{2}[ h(p+q+k)h(k)-h(k)h(k)]-\frac{1}{3} \int_q \frac{h(q)}{2}[h(q+k)h(k)-h(k)h(k)]+\frac{\eta p^2}{2\epsilon}-\frac{1}{2-2\epsilon} \bigg \lbrace \frac{2}{3}\beta^{(1)}v_2^{(1)}+ \int_q f(q)\mathcal{F}(q) \bigg \rbrace 
\end{equation*}

\begin{align*}
\beta^{(1)}=-\int_q f(q) h(q)\rightarrow_{\epsilon\rightarrow 0}-F; v_2^{(1)}=-\int_q f(q)h(q)\rightarrow_{\epsilon\rightarrow 0} -\frac{F}{2}
\end{align*}

\subsubsection*{Different parts of \eqref{BA_irr}}

\textbf{In the LHS,}

\vspace{0.2 in}

\textbf{A}.~Calculation of $\frac{\delta^2 }{\delta \phi(p) \delta \phi(-p)}  D^{(2)}(p_1,p_2,p_3,p_4,p_5,p_6) $

1. 

\begin{align*}
&\int_p \big \lbrace -K^\prime(p^2) \big \rbrace \frac{\delta^2}{\delta \phi(p) \delta \phi(-p)}  D_I^{(2)}(p_1,p_2,p_3,p_4) \phi(p_1) \phi(p_2\phi(p_3)\phi(p_4)\phi(p_5)\phi(p_6)\\
&=\int_p \big \lbrace -K^\prime(p^2) \big \rbrace \frac{\delta^2}{\delta \phi(p) \delta \phi(-p)} \\
& \bigg \lbrace \frac{3\lambda^2}{6!}  \sum_{10~perm~(i,j,k)}\sum_{3~perm~(\alpha,\beta)} \int_q h(p_i+p_j+p_k) [h(p_\alpha+p_\beta+q)h(q) -h(q)h(q)] \bigg \rbrace \phi(p_1)\phi(p_2)\phi(p_3)\phi(p_4)\phi(p_5)\phi(p_6)
\end{align*}

\begin{subequations}
\begin{align}
&=\label{Bone1_i}\int_q \frac{1}{4!}\frac{3\lambda^2 F}{2} \bigg \lbrace \sum_{3~perm~(i,j)}\sum_{l=1}^4 h(p_l) [h(p_i+p_j+q)h(q)-h(q)h(q)]\bigg \rbrace \phi(p_1)\phi(p_2)\phi(p_3)\phi(p_4)\\
& + \label{Bfive1_i} \frac{3\lambda^2}{4!}\int_{p,q} \big\lbrace-K^\prime(p^2)\big \rbrace \bigg\lbrace \sum_{6~perm~(i,j)}h(p_i+p_j+p) [ h(p+q+p_j)+h(p+q+p_i)-2h(q)h(q)] \bigg \rbrace \phi(p_1)\phi(p_2)\phi(p_3)\phi(p_4)\\
&+\label{Bseven1_i}\int_{p,q} \frac{6\lambda^2 }{4!} \big\lbrace-K^\prime(p^2)\big \rbrace  \sum_{3~perm~(i,j)}  h(p_i+p_j+p) \big \lbrace h(p_i+p_j+q)h(q)-h(q)h(q) \big \rbrace \phi(p_1)\phi(p_2)\phi(p_3)\phi(p_4)\\
&+\label{Bsix1_i}\int_{p,q} \big \lbrace- K^\prime(p^2) \big \rbrace \frac{3\lambda^2}{4!} \sum_{i=1}^4 h(p_i) \big \lbrace h(p_i+p+q) h(q)-h(q)h(q) \big \rbrace \phi(p_1)\phi(p_2)\phi(p_3)\phi(p_4)
\end{align}
\end{subequations}

\vspace{0.2 in}

2.
\begin{align*}
&\int_p \big \lbrace -K^\prime(p^2 \big \rbrace \frac{\delta^2}{\delta \phi(p) \delta \phi(-p)} D_{II}^{(2)}(p_1,p_2,p_3,p_4) \phi(p_1) \phi(p_2\phi(p_3)\phi(p_4)\phi(p_5)\phi(p_6)\\
&= \int_p \big \lbrace -K^\prime(p^2 \big \rbrace \frac{\delta^2}{\delta \phi(p) \delta \phi(-p)} \frac{1}{6!}\frac{-3\lambda^2F}{2}\sum_{l=1}^6 \sum_{10~perm~(i,j,k)} h(p_l)h(p_i+p_j+p_k)\phi(p_1)\phi(p_2)...\phi(p_5)\phi(p_6)
\end{align*}
gives

\begin{subequations}
\begin{align}
&\label{Bfour1_i}+\frac{-3}{2} \frac{\lambda^2F}{4!} \int_p \big \lbrace -K^\prime(p^2) \big \rbrace \big \lbrace 2 h(p) \big \rbrace \sum_{l=1}^4 \big \lbrace h(p_l) \big \rbrace \phi(p_1)\phi(p_2)\phi(p_3)\phi(p_4)\\
&\label{Bthree1_i} +\frac{-3}{2} \frac{\lambda^2F}{4!} \int_p \big \lbrace -K^\prime(p^2) \big \rbrace \big \lbrace \sum_{l=1}^4 h(p_l)h(p_l)+ \sum_{i \neq j}h(p_i) h(p_j) \big \rbrace \phi(p_1)\phi(p_2)\phi(p_3)\phi(p_4)\\
& \label{Btwo1_i}+ \frac{-3\lambda^2 F}{4!}\int_p \big \lbrace -K^\prime(p^2) \big \rbrace \big \lbrace 2 h(p) \big \rbrace \sum_{3~perm~(i,j)} \big \lbrace h(p_i+p_j+p) \big \rbrace \phi(p_1)\phi(p_2)\phi(p_3)\phi(p_4)\\
&\label{Bone2_i}+ \frac{-3\lambda^2 F}{4!}\int_p \big \lbrace -K^\prime(p^2) \big \rbrace \big \lbrace \sum_{l=1}^4 h(p_l) \big \rbrace \sum_{3~perm~(i,j)} \big \lbrace h(p_i+p_j+p) \big \rbrace \phi(p_1)\phi(p_2)\phi(p_3)\phi(p_4)
\end{align}
\end{subequations}

\vspace{0.2 in}

3.
\begin{align*}
&\int_p \big \lbrace -K^\prime(p^2) \big \rbrace \frac{\delta^2}{\delta \phi(p) \delta \phi(-p)} D_{III}^{(2)}(p_1,p_2,p_3,p_4) \phi(p_1) \phi(p_2\phi(p_3)\phi(p_4)\phi(p_5)\phi(p_6)\\
&=\int_p \big \lbrace -K^\prime(p^2 \big \rbrace \frac{\delta^2}{\delta \phi(p) \delta \phi(-p)} \frac{1}{6!}\frac{-3\lambda^2 F}{2} \sum_{10~perm~(i,j,k)}h(p_i+p_j+p_k)h(p_i+p_j+p_k)\phi(p_1)\phi(p_2)\phi(p_3)\phi(p_4)\phi(p_5) \phi(p_6)
\end{align*}
gives

\begin{subequations}
\begin{align}
&\label{Bthree2_i}\frac{1}{4!}\frac{-3\lambda^2 F^2}{2} \sum_{l=1}^4 h(p_l)h(p_l) \phi(p_1)\phi(p_2)\phi(p_3)\phi(p_4)\\
&\label{Btwo2_i}+\int_p \big \lbrace -K^\prime(p^2) \big \rbrace  \frac{-3\lambda^2 F}{4!} \sum_{3~perm~(i,j)} h(p_i+p_j+p)h(p_i+p_j+p) \phi(p_1)\phi(p_2)\phi(p_3)\phi(p_4) 
\end{align}
\end{subequations}

4.

\begin{align*}
&\int_p \big \lbrace -K^\prime(p^2) \big \rbrace \frac{\delta^2}{\delta \phi(p) \delta \phi(-p)} D_{IV}^{(2)}(p_1,p_2,p_3,p_4) \phi(p_1) \phi(p_2)\phi(p_3)\phi(p_4)\phi(p_5)\phi(p_6)\\
&=\int_p \big \lbrace -K^\prime(p^2 \big \rbrace \frac{\delta^2}{\delta \phi(p) \delta \phi(-p)}\\
& \bigg \lbrace \frac{1}{6!}\frac{\lambda^2}{2} \int_q \sum_{15~perm~(i,j)} \sum_{6~perm~(\alpha,\beta)} [ h(p_i+p_j+q)h(p_i+p_j+p_\alpha+p_\beta+q)h(q)] \phi(p_1)\phi(p_2) \phi(p_3) \phi(p_4) \phi(p_5) \phi(p_6) \bigg \rbrace
\end{align*}

gives

\begin{subequations}
\begin{align}
&=\label{Btwo3_i}\frac{3\lambda^2 F}{4!}\sum_{3~perm~(i,j)}\int_q \big \lbrace h(p_i+p_j+q)h(q)h(q) \big \rbrace \phi(p_1)\phi(p_2)\phi(p_3)\phi(p_4)\\
& \label{Bfive2_i} +\frac{2\lambda^2}{4!}\int_p \int_q  \big \lbrace -K^\prime(p^2 \big \rbrace\sum_{6~perm~(i,j)} \big \lbrace h(p_i+p_j+q) [ h(p+q+p_i)+h(p+q+p_j)]h(q)\big \rbrace \phi(p_1)\phi(p_2)\phi(p_3)\phi(p_4)\\
&\label{Bfive3_i} + \frac{\lambda^2}{4!} \int_q \int_p \big \lbrace -K^\prime(p^2 \big \rbrace \sum_{l=1}^4 \sum_{3~perm~(i,j)} \big \lbrace h(p_l+p+q)h(p_l+p_i+p_j+p+q)h(q) \big \rbrace \phi(p_1)\phi(p_2)\phi(p_3)\phi(p_4)
\end{align}
\end{subequations}

\vspace{0.5 in}

\textbf{B} 

\begin{align*}
&-\frac{2}{4!}\sum_{i=1}^4\big \lbrace -K^\prime(p_i^2) \big \rbrace \big \lbrace A^{(0)}(p_i)+ A^{(1)}(p_i) \big \rbrace \big \lbrace \lambda+U_4^{(2)}(p_1,p_2,p_3,p_4) \big \rbrace\\
&=-\frac{2}{4!}\sum_{i=1}^4\big \lbrace -K^\prime(p_i^2) \big \rbrace \bigg( \bigg \lbrace -\frac{F}{2} \bigg \rbrace \bigg \lbrace -\lambda^2 \sum_{3~perm~(i,j)}\mathcal{F}(p_i+p_j)+ \frac{F\lambda^2}{2} \sum_{i=1}^4 h(p_i) \bigg \rbrace+[A^{(0)}+A^{(1)}(p_i)] \lambda \bigg)\\
\end{align*}

where $\mathcal{F}(p_1+p_2)= \frac{1}{2} \int_q \big \lbrace h(p_1+p_2+q)h(q)-h(q)h(q) \big \rbrace$

\begin{subequations}
\begin{align}
&\label{Bone3_i} =\frac{1}{4!}\frac{F\lambda^2}{2}\sum_{l=1}^4\big \lbrace K^\prime(p_l^2) \big \rbrace \sum_{3~perm~(i,j)}\int_q \big \lbrace h(p_i+p_j+q)h(q)-h(q)h(q) \big \rbrace\\ 
&\label{Bthree3_i} -\frac{1}{4!}\frac{F^2\lambda^2}{2}\sum_{i=1}^4\big \lbrace K^\prime(p_i^2) \big \rbrace \sum_{l=1}^4 h(p_l)\\
&\label{A_i}+\frac{2}{4!} \sum_{i=1}^4 \big \lbrace K^\prime(p_i^2) \big \rbrace \lambda \big \lbrace A^{(0)}+A^{(1)}(p_i)\big \rbrace
\end{align}
\end{subequations}

\vspace{0.5 in}

\textbf{C}
\begin{align*}
-\frac{2}{4!}\sum_{i=1}^4\big \lbrace -K^\prime(p_i^2) \big \rbrace \big \lbrace U_2^{(1)}(p_i)+U_2^{(2)}(p_i) \big \rbrace  \big \lbrace \sum_{i=1}^4 B^{(0)}(p_i)+ B_I^{(1)}(p_1,p_2,p_3,p_4)+B_{II}^{(1)}(p_1,p_2,p_3,p_4) \big \rbrace\\
\end{align*}

\begin{subequations}
\begin{align}
& =\label{Bfour2_i} \frac{2}{4!} \frac{-F\lambda \epsilon}{4} \sum_{i=1}^4 K^\prime(p_i^2)\\
& =\label{Bone4_i} \frac{\lambda^2 F}{4!}\sum_{l=1}^4 \big \lbrace K^\prime(p_l^2)\big \rbrace \big \lbrace \int_q \sum_{3~perm~(i,j)} \big( h(p_i+p_j+q)h(q)-h(q)h(q) \big) \big \rbrace \\ 
&\label{Bthree4_i} -\frac{F^2\lambda^2}{4!}\sum_{i=1}^4\big \lbrace K^\prime(p_i^2) \big \rbrace \big \lbrace \sum_{i=1}^4 h(p_i) \big \rbrace\\
&\label{Bsix2_i} -\frac{1}{3}\frac{2 \lambda^2}{4!}\sum_{i=1}^4\big \lbrace K^\prime(p_i^2) \big \rbrace \big \lbrace \int_{q,k} h(q)\left[ h(p_i+q+k)h(k)-h(k)h(k)\right]-\int_{q,k} h(q) \left[ h(q+k)h(k)-h(k)h(k) \right] \big \rbrace \\
&\label{Beight1_i} -\frac{2 \lambda^2}{4!}\sum_{i=1}^4\big \lbrace K^\prime(p_i^2) \big \rbrace \big \lbrace \frac{\eta p_i^2}{2 \epsilon} \big \rbrace\\
& \label{Bsix3_i}-\frac{\lambda^2}{4!} \sum_{i=1}^4 K^\prime(p_i^2) \int_q K^\prime(q^2) \big \lbrace h(q+k)h(k)-h(k)h(k) \big \rbrace\\
& \label{Bfour3_i}+ \frac{2 \lambda^2}{4!}\sum_{i=1}^4\big \lbrace K^\prime(p_i^2) \big \rbrace \big \lbrace  \frac{1}{2-2\epsilon} \bigg( \frac{2}{3} \beta^{(1)}v_2^{(1)}\bigg) \big \rbrace\\
& \label{Bthree5_i}-\frac{1}{4!}\frac{ F^2\lambda^2}{2}\sum_{i=1}^4\big \lbrace K^\prime(p_i^2) h(p_i) \big \rbrace
\end{align}
\end{subequations}

\vspace{0.2 in}

\textbf{D.} 

\begin{align}
& \label{anom_eta_i} -\frac{4}{4!}\frac{\eta}{2}\sum_{i=1}^4 B^{(0)}(p_i)\\
& \label{Beight2_i}+\frac{1}{4!} \eta \sum_{i=1}^4 p_i^2 h(p_i)
\end{align}

Where at the fixed point $F\lambda=\frac{\epsilon}{3}$ , $\frac{\eta}{2} \longrightarrow \frac{\epsilon^2}{108} $
 
 \vspace{0.2 in}
 
\textbf{E.}

\begin{subequations}
\begin{align}
&\label{Bfour4_i}\frac{1}{4!} (4-D) B_I^{(1)}(p_1,p_2,p_3,p_4)\\
&\label{Bnine1_i}+\frac{1}{4!}2(4-D) B_{II}^{(1)}(p_1,p_2,p_3,p_4)\\
&+ \frac{1}{4!}(4-D- \sum_{i=1}^4 p_i.\frac{d}{d p_i}) B^{(2)}(p_1,p_2,p_3,p_4)
\end{align}
\end{subequations}

\vspace{0.2 in}

\textbf{In the RHS}

\begin{subequations}
\begin{align}
&= \label{Bfour5_i}\frac{\epsilon-6F\la}{4!} \big \lbrace \sum_{i=1}^4 B_I^{(1)}(p_i) \rbrace \\
& \label{Bnine2_i}+\frac{\eps-6F\la}{4!} B_{II}^{(1)}(p_1,p_2,p_3,p_4) \\
&+\frac{\eps-6F\la}{4!} \big \lbrace B^{(2)}(p_1,p_2,p_3,p_4) \big \rbrace\\
&+ \label{leftover_3} \frac{d_4^{(2)}}{4!} \big \lbrace \sum_{i=1}^4 B^{(0)}(p_i) \big \rbrace\\
&+ \label{Bfour6_i} \epsilon \frac{1}{4!} B_I^{(1)}(p_1,p_2,p_3,p_4)\\
&+ \label{Bnine3_i}\epsilon \frac{1}{4!} B_{II}^{(2)}(p_1,p_2,p_3,p_4)\\
&+ \label{Bfour7_i}(-3F \lambda) \frac{1}{4!} B_I^{(1)}(p_1,p_2,p_3,p_4)\\
&+ \label{Bnine4_i}(-3F \lambda) \frac{1}{4!} B_{II}^{(1)}(p_1,p_2,p_3,p_4)
\end{align}
\end{subequations}

We know all necessary terms to find  $B^{(2)}(p_1,p_2,p_3,p_4)$.  We will reorganize the terms and will make suitable ansatz about  $B^{(2)}(p_1,p_2,p_3,p_4)$ so that \eqref{BA_irr} is satisfied and at the end we get some numerical term proportional to $\sum_{i=1}^4 B^{(0)}(p_i)$ in the LHS of \eqref{BA_irr} so that  we can equate that with $\frac{d_m^{(2)}}{4!} \big \lbrace \sum_{i=1}^4 B^{(0)}(p_i) \big \rbrace$ in RHS to get the anomalous dimension.

\subsubsection*{Equation for $B_I^{(2)}(p_1,p_2,p_3,p_4)$}

Taking \eqref{Bone1_i},\eqref{Bone2_i},\eqref{Bone3_i} and \eqref{Bone4_i} and adding suitable couterterm,

\begin{align}
\nonumber &\int_q \frac{1}{4!}\frac{3\lambda^2 F}{2}  \sum_{3~perm~(i,j)}\sum_{l=1}^4 \big \lbrace h(p_l) \big \rbrace \big \lbrace h(p_i+p_j+q)h(q)-h(q)h(q)\big \rbrace\\
\nonumber & + \frac{-3\lambda^2 F}{4!}\int_p \big \lbrace -K^\prime(p^2) \big \rbrace  \sum_{l=1}^4 \big \lbrace h(p_l) \big \rbrace \sum_{3~perm~(i,j)} \big \lbrace h(p_i+p_j+p) \big \rbrace +\frac{9}{2} \frac{\lambda^2 F^2}{4!}\sum_{l=1}^4 h(p_l) \\
\nonumber & +\frac{1}{4!}\frac{\lambda^2 F}{2}\sum_{l=1}^4 \big \lbrace K^\prime(p_l^2)\big \rbrace \big \lbrace \int_q \sum_{3~perm~(i,j)} \big( h(p_i+p_j+q)h(q)-h(q)h(q) \big) \big \rbrace\\
\nonumber  &+\frac{\lambda^2 F}{4!}\sum_{l=1}^4 \big \lbrace K^\prime(p_l^2)\big \rbrace \big \lbrace \int_q \sum_{3~perm~(i,j)} \big( h(p_i+p_j+q)h(q)-h(q)h(q) \big) \big \rbrace\\
 &+\big \lbrace -(4-D)-\sum_{i=1}^4 p_i.\frac{d}{dp_i} \frac{1}{4!} \big \rbrace B_I^{(2)}(p_1,p_2,p_3,p_4)=0
\end{align}

On LHS of \eqref{BA_irr} we are left with,

\begin{align}\label{leftoverB_1}
2(4-D) \frac{1}{4!} B_I^{(2)}(p_1,p_2,p_3,p_4)-\frac{1}{4!}\frac{9\lambda^2 F}{2} \sum_{l=1}^4 h(p_l)
\end{align}

\subsubsection*{Equation for $B_{II}^{(2)}(p_1,p_2,p_3,p_4)$}

Taking \eqref{Btwo1_i},\eqref{Btwo2_i} and \eqref{Btwo3_i} we get,

\begin{align}
\nonumber & \frac{-3\lambda^2 F}{4!}\int_p \big \lbrace -K^\prime(p^2) \big \rbrace \big \lbrace 2 h(p) \big \rbrace \sum_{3~perm~(i,j)} \big \lbrace h(p_i+p_j+p) \big \rbrace \\
\nonumber & +\int_p \big \lbrace -K^\prime(p^2) \big \rbrace  \frac{-3\lambda^2 F}{4!} \sum_{3~perm~(i,j)}h(p_i+p_j+p)h(p_i+p_j+p)\\
\nonumber & + \frac{3\lambda^2 F}{4!}\sum_{3~perm~(i,j)}\int_q h(p_i+p_j+q)h(q)h(q)\\
& + \big \lbrace -(4-D)- \sum_{i=1}^4 p_i.\frac{d}{d p_i} \big \rbrace \frac{1}{4!} B_{II}^{(2)}(p_1,p_2,p_3,p_4)=0
\end{align}

On LHS of \eqref{BA_irr} we are left with,
\begin{align}
+2(4-D)\frac{1}{4!} B_{II}^{(2)}(p_1,p_2,p_3,p_4)
\end{align}

\subsubsection*{Equation for $B_{III}^{(2)}(p_1,p_2,p_3,p_4)$}

Taking \eqref{Bthree1_i},\eqref{Bthree2_i},\eqref{Bthree3_i},\eqref{Bthree4_i}, \eqref{Bthree5_i} we get,(Note that we need to $A_{III}^{(1)}$ for the equation to be satisfied.

\begin{align}
\nonumber & \frac{-3}{2} \frac{\lambda^2F}{4!} \int_p \big \lbrace -K^\prime(p^2) \big \rbrace \big \lbrace \sum_{l=1}^4 h(p_l)h(p_l)+ \sum_{i \neq j}h(p_i) h(p_j) \big \rbrace \\
\nonumber &+ \frac{1}{4!}\frac{-3\lambda^2 F^2}{2} \sum_{l=1}^4 h(p_l)h(p_l) \\
\nonumber & -\frac{1}{4!}\frac{F^2\lambda^2}{2}\big \lbrace \sum_{i=1}^4 K^\prime(p_i^2) h(p_i)+ \sum_{i \neq j} K^\prime(p_i^2)h(p_j) \big \rbrace\\
\nonumber & -\frac{F^2\lambda^2}{4!}\sum_{i=1}^4 \big \lbrace \sum_{i=1}^4 K^\prime(p_i^2) h(p_i)+ \sum_{i \neq j} K^\prime(p_i^2)h(p_j) \big \rbrace\\
\nonumber & -\frac{1}{4!}\frac{ F^2\lambda^2}{2}\sum_{i=1}^4\big \lbrace K^\prime(p_i^2) h(p_i) \big \rbrace\\
& +\frac{2}{4!} \sum_{i=1}^4 \big \lbrace K^\prime(p_i^2) \big \rbrace \lambda A_{III}^{(1)}(p_i)-\sum_{i=1}^4 p_i.\frac{d}{dp_i} \frac{1}{4!} B_{III}^{(2)}(p_1,p_2,p_3,p_4)=0
\end{align}

where $A_{III}^{(1)}(p)=\frac{-\lambda F^2}{2} h(p)$.

\vspace{0.2 in}

On LHS of \eqref{BA_irr} we are left with

\begin{align}
&(4-D) \frac{1}{4!} B_{III}^{(2)}(p_1,p_2,p_3,p_4)
\end{align}

\subsubsection*{Equation for $B_{IV}^{(2)}(p_1,p_2,p_3,p_4)$}
Collecting \eqref{Bfour1_i},\eqref{Bfour2_i},\eqref{Bfour3_i},\eqref{Bfour4_i}, \eqref{Bfour5_i}, \eqref{Bfour6_i}, \eqref{Bfour7_i} and the second term of \eqref{leftoverB_1} we get,
\begin{align*}
& \frac{-3}{2} \frac{\lambda^2F}{4!} \int_p \big \lbrace -K^\prime(p^2) \big \rbrace \big \lbrace 2 h(p) \big \rbrace \sum_{l=1}^4 \big \lbrace h(p_l) \big \rbrace \\
& +\frac{2}{4!} \frac{-F\lambda \epsilon}{4} \sum_{i=1}^4 K^\prime(p_i^2)\\
& + \frac{2 \lambda^2}{4!}\sum_{i=1}^4\big \lbrace K^\prime(p_i^2) \big \rbrace \big \lbrace  \frac{1}{2-2\epsilon} \bigg(\frac{2}{3} \beta^{(1)}v_2^{(1)}\bigg) \big \rbrace\\
& +\frac{2}{4!}\sum_{i=1}^4 \big \lbrace K^\prime(p_i^2) \big \rbrace  \lambda \big \lbrace A^{(0)}+  A_{I} ^{(1)}(p_i) \big \rbrace+ \bigg \lbrace (4-D)-\sum_{i=1}^4 p_i.\frac{d}{dp_i} \bigg \rbrace \frac{1}{4!} B_{IV}^{(2)}(p_1,p_2,p_3,p_4)+\frac{4-D}{4!}\sum_{i=1}^4 \lambda F h(p_i)\\
&-\frac{1}{4!}\frac{9}{2} F^2 \lambda^2 \sum_{l=1}^4 h(p_l)\\
&= \frac{\eps-6F\la}{4!}\sum_{i=1}^4 \lambda F h(p_i)+(\epsilon \lambda-3F\lambda^2) F \sum_{i=1}^4 h(p_i)
\end{align*}

Where $\frac{1}{3}\beta^{(1)}= -\int_q f(q) h(q) \rightarrow_{\epsilon \rightarrow 0} -F$, $v_2^{(1)}= -\frac{1}{2-\epsilon} \int_p -K^\prime(p^2) \rightarrow_{\epsilon \rightarrow 0} \frac{-F}{2}$, $A^{(0)}=-\frac{F\epsilon}{4}$,$ A_{I}^{(1)}(p)=F^2 \lambda$.

\subsubsection*{Equation for $B_{VI}^{(2)}(p_1,p_2,p_3,p_4)$}

Taking \eqref{Bsix1_i}, \eqref{Bsix2_i} and \eqref{Bsix3_i}, we get

\begin{align}
\nonumber &\int_{p,q} \bigg \lbrace -K^\prime(p^2) \bigg \rbrace \frac{3\lambda^2}{4!} \sum_{4~perm~(i,j,k)} \big \lbrace h(p_i+p_j+p_k) h(p_\alpha+p+q) h(q)-h(p+q)h(q) \big \rbrace\\
\nonumber & \frac{1}{2} \frac{\lambda^2}{4!} \int_{p,q} K^\prime(p) h(p+q) h(q)\\
\nonumber &- \frac{1}{3}\frac{\lambda^2}{4!}\sum_{i=1}^4\big \lbrace K^\prime(p_i^2) \big \rbrace \big \lbrace \int_{q,k} h(q)\left[ h(p_i+q+k)h(k)- h(q+k)h(k) \right] \big \rbrace \\
\nonumber & +\frac{\lambda^2}{4!} \sum_{i=1}^4 K^\prime(p_i^2)\int_{p,q} \big \lbrace -K^\prime(q) \big \rbrace \big \lbrace h(q+k)h(k)-h(k)h(k) \big \rbrace \\
\nonumber & +\frac{3\lambda^2}{4!} \sum_{i=1}^4 h(p_i)\int_{p,q} \big \lbrace -K^\prime(q) \big \rbrace \big \lbrace h(q+k)h(k)-h(k)h(k) \big \rbrace \\
\nonumber &+\frac{2}{4!} \sum_{i=1}^4 K^\prime(p_i^2) A_{II} (p_i) \lambda \\
& + \big \lbrace -2(4-D)- \sum_{i=1}^4 p_i.\frac{d}{dp_i} \big \rbrace \frac{1}{4!} \bigg \lbrace B_{VI}^{(2)}(p_1,p_2,p_3,p_4)|_1 \bigg \rbrace + \big \lbrace - \sum_{i=1}^4 p_i.\frac{d}{dp_i} \big \rbrace \frac{1}{4!}\bigg \lbrace B_{VI}^{(2)}(p_1,p_2,p_3,p_4)|_2 \bigg \rbrace =0
\end{align}

$A_{II}^{(1)}(p_i)= -\frac{\lambda}{3}  \int_{p,q} [h(p_i+p+q)h(p)h(q)-h(p+q)h(p)h(q)]-  \lambda \int_{p} K^\prime(p)\lbrace h(p+q)h(q)-h(q)h(q) \rbrace $.

\vspace{0.1 in}

On LHS of \eqref{BA_irr} we are left with

\begin{align}
\nonumber & 3(4-D)\frac{1}{4!} B_{VI}^{(2)}(p_1,p_2,p_3,p_4)|_1 + (4-D) \frac{1}{4!} B_{VI}^{(2)}|_2(p_1,p_2,p_3,p_4)|_2\\
\end{align}

\subsubsection*{Equation for $B_{VII}^{(2)}(p_1,p_2,p_3,p_4)$}

Considering \eqref{Bseven1_i}, 

\begin{align}
\nonumber &\int_{p,q} \frac{6\lambda^2  }{4!} \big \lbrace -K^\prime(p) \big \rbrace \sum_{3~perm~(i,j)}h(p_i+p_j+p) \big \lbrace h(p_i+p_j+q)h(q)-h(q)h(q) \big \rbrace \\
& + \big \lbrace -2(4-D)-p_i.\frac{d}{d p_i} \big \rbrace \frac{1}{4!} B_{VII}^{(2)}(p_1,p_2,p_3,p_4)=0
\end{align}

On LHS of \eqref{BA_irr} we are left with 

\begin{align}\label{Bseven_leftover}
3(4-D)\frac{1}{4!} B_{VII}^{(2)}(p_1,p_2,p_3,p_4)
\end{align}

\subsubsection*{Equation for $B_{VIII}^{(2)}(p_1,p_2,p_3,p_4)$}

Collecting \eqref{Beight1_i} and \eqref{Beight2_i} we get (because of the expected structure of $B_{VIII}^{(2)}$ as $\frac{\eta}{\eps}$ we consider the term $\eps \la \frac{\p B_{IX}}{\p \la}$ from RHS of \eqref{BA_irr}),

\begin{align*}
&-\frac{2}{4!}\sum_{i=1}^4\big \lbrace K^\prime(p_i^2) \big \rbrace \big \lbrace \frac{\eta p_i^2}{2 \epsilon} \big \rbrace +\frac{1}{4!} \eta \sum_{i=1}^4 p_i^2 h(p_i)+ \bigg \lbrace (4-D)-\sum_{i=1}^4 p_i.\frac{\partial }{\partial p_i} \bigg \rbrace \frac{1}{4!}B_{IX}(p_1,p_2,p_3,p_4)\\
& = \lbrace \eps-6F\la \rbrace \frac{1}{4!} B_{IX}(p_1,p_2,p_3,p_4)+ \frac{1}{4!}\epsilon \lambda \frac{\partial B_{IX}(p_1,p_2,p_3,p_4)}{\partial \lambda}
\end{align*}

 We ignore $\lambda B_{IX}^{(2)}$ oe $\epsilon B_{IX}^{(2)}$ terms being higher order and get

\begin{align}
-\frac{2}{4!}\sum_{i=1}^4\big \lbrace K^\prime(p_i^2) \big \rbrace \big \lbrace \frac{\eta p_i^2}{2 \epsilon} \big \rbrace +\frac{1}{4!} \eta \sum_{i=1}^4 p_i^2 h(p_i)+ \bigg \lbrace-\sum_{i=1}^4 p_i.\frac{\partial }{\partial p_i} \bigg \rbrace \frac{1}{4!} B_{IX}= \frac{1}{4!}\epsilon \lambda \frac{\partial B_{IX}}{\partial \lambda}
\end{align}

And on LHS of \eqref{BA_irr} we are left with

\begin{align}
(\epsilon- 6\la F )\frac{1}{4!} B_{IX}(p_1,p_2,p_3,p_4)
\end{align}

\subsubsection*{Equation for $B_{IX}^{(2)}(p_1,p_2,p_3,p_4)$}

At last we collect the terms \eqref{Bnine1_i}, \eqref{Bnine2_i}, \eqref{Bnine3_i} and \eqref{Bnine4_i} to get,

\begin{align}
\nonumber & \lbrace 4-D-\sum_{i=1}^4 p_i.\frac{d}{dp_i} \rbrace \frac{1}{4!}B^{(2)}_X(p_1,p_2,p_3,p_4)+2(4-D)\frac{1}{4!}B_{II}(p_1,p_2,p_3,p_4)\\
& = \left \lbrace \eps-6F\la + \epsilon-3F\lambda^2 \right \rbrace \frac{1}{4!}B_{II}^{(1)}(p_1,p_2,p_3,p_4)
\end{align}

We ignore the term $\epsilon \frac{1}{4!} B_{X}^{(2)}(p_1,p_2,p_3,p_4)$ and get the following euqations:

\begin{align}
&-\sum_{i=1}^4 p_i.\frac{d}{dp_i} B^{(2)}_X(p_1,p_2,p_3,p_4)- 9 F\lambda^2 \sum_{3~perm~(i,j)} \int_q \big \lbrace h(p_i+p_j+q)h(q)-h(q)h(q) \big \rbrace =0
\end{align}

To solve this, we use $\bar{p}=\frac{p}{\Lambda}$. In this noatation $-\sum_{i=1}^4 \bar{p_i}. \frac{d}{dp_i} B^{(2)}_X(\bar{p}_1,\bar{p}_2,\bar{p}_3,\bar{p}_4)$ can be written as,

\begin{align*}
 -\sum_{i=1}^4 \bar{p}_i. \frac{d}{d\bar{p}_i}B^{(2)}_X\bigg(\frac{p_1}{\Lambda},\frac{p_2}{\Lambda},\frac{p_3}{\Lambda},\frac{p_4}{\Lambda}\bigg) = \Lambda. \frac{d}{d \Lambda} B^{(2)}_X \bigg(\frac{p_1}{\Lambda},\frac{p_2}{\Lambda},\frac{p_3}{\Lambda},\frac{p_4}{\Lambda}\bigg)
\end{align*}

So the solution is given by,

\begin{align}
& \frac{1}{4!} B^{(2)}_X(\frac{p_1}{\Lambda},\frac{p_2}{\Lambda},\frac{p_3}{\Lambda},\frac{p_4}{\Lambda})= \frac{9F\lambda^2}{4!} \sum_{3~perm~(i,j)} \int_\Lambda^\infty \int_{\bar{q}} \frac{d \Lambda^\prime}{\Lambda^\prime} \bigg \lbrace h\bigg( \frac{p_i}{\Lambda^\prime} + \frac{p_j}{\Lambda^\prime}+ \bar{q}\bigg)h\left(\bar{q}\right )-h \left(\bar{q}\right)h\left(\bar{q}\right) \bigg \rbrace
\end{align}

In LHS we are left with 

\begin{align}
\frac{1}{4!}(4-D)B_{X}^{(2)}(p_1,p_2,p_3,p_4)
\end{align}

\section{Relevant operator at sub-leading operator}
\label{appD}

\subsection{The $\phi^6$ equation to find $ D^{(2)}(p_1,p_2,p_3,p_4,p_5,p_6)$ }
\label{solvphi6_rel}

$\phi^6$ equation is given by ( we donot have to consider $\beta(\lambda) \frac{\p \Delta S}{\p \la}$ part because there is no $D^{(1)}(p_1,p_2,p_3,p_4)$ in this case).

\begin{align}\label{phi6_rel}
\nonumber & -\frac{4}{6!} \sum_{10~perm~(i,j,k)} \big \lbrace -K^\prime(p_i+p_j+p_k) \big \rbrace \big \lbrace \lambda \big \rbrace \big \lbrace \underbrace{B(p_i)+B(p_j)+ B(p_k)}_1 + \underbrace{B(p_i+p_j+p_k)}_2 \big \rbrace\\
\nonumber & -\frac{2}{6!} \sum_{i=1}^4 \big \lbrace -K^\prime(p_i^2) \big \rbrace \underbrace{\big \lbrace A(p_i) \big \rbrace V_6^{(2)}(p_1,p_2,p_3,p_4,p_5,p_6)}_3 \\
\nonumber & +\frac{1}{6!} \bigg( 6-2D- \sum_{i=1}^4 p_i.\frac{\partial}{\partial p_i} \bigg) D^{(2)}(p_1,p_2,p_3,p_4,p_5,p_6)\\
& = \frac{d_2^{(0)}}{6!} D^{(2)}(p_1,p_2,p_3,p_4,p_5,p_6)
\end{align}

\[ d_2^{(0)}=2\]

We collect the terms marked '2' to find first kind of $D^{(2)}(p_1,p_2,p_3,p_4,p_5,p_6)$.

\begin{align}\label{D6I_rel}
\nonumber & -\frac{4}{6!} \sum_{10~perm~(i,j,k)} \big \lbrace -K^\prime(p_i+p_j+p_k) \big \rbrace \big \lbrace \lambda \big \rbrace \big \lbrace B(p_i+p_j+p_k) \big \rbrace\\
\nonumber & +\frac{1}{6!} \bigg( 6-2D- \sum_{i=1}^4 p_i.\frac{\partial}{\partial p_i} \bigg) D^{(2)}(p_1,p_2,p_3,p_4,p_5,p_6)\\
& = \frac{d_2^{(0)}}{6!} D_I^{(2)}(p_1,p_2,p_3,p_4,p_5,p_6)
\end{align}

Similarly collecting  the terms marked as '1' and '3' we get the following equation,

\begin{align}\label{D6II_rel}
\nonumber & -\frac{4}{6!} \sum_{10~perm~(i,j,k)} \big \lbrace -K^\prime(p_i+p_j+p_k) \big \rbrace \big \lbrace \lambda \big \rbrace \big \lbrace B(p_i)+B(p_j)+ B(p_k) \big \rbrace\\
\nonumber & -\frac{2}{6!} \sum_{i=1}^4 \big \lbrace -K^\prime(p_i^2) \big \rbrace \big \lbrace A(p_i) \big \rbrace V_6^{(2)}(p_1,p_2,p_3,p_4,p_5,p_6)\\
& = \frac{2}{6!} D_{II}^{(2)}(p_1,p_2,p_3,p_4,p_5,p_6)
\end{align}

\subsection{The $\phi^4$ equation to determine $B^{(2)}(p_1,p_2,p_3,p_4)$}
\label{solvphi4_rel}

The $\phi^4$ equation is given by,

\begin{align}\label{phi4_rel}
\nonumber & \frac{1}{6!}\int_p \big \lbrace -K^\prime(p) \big \rbrace \frac{\delta}{\delta \phi(p)} \frac{\delta}{\delta \phi(-p)} D^{(2)}(p_1,p_2,p_3,p_4,p_5,p_6) -\frac{2}{4!} \sum_{i=1}^4 \big \lbrace -K^\prime(p_i) \big \rbrace \big \lbrace A^{(0)}(p)+A^{(1)}(p) \big \rbrace \big \lbrace \lambda+ U_4^{(2)}(p_1,p_2,p_3,p_4) \big \rbrace \\
\nonumber &  -\frac{2}{4!} \sum_{i=1}^4 \big \lbrace -K^\prime(p_i) \big \rbrace \big \lbrace U_2^{(1)}(p_i)+  U_2^{(2)}(p_i)\big \rbrace \big \lbrace \sum_{i=1}^4 B^{(1)}(p_i) \big \rbrace+ \frac{1}{4!} \bigg( 4-D- \sum_{i=1}^4 p_i. \frac{\partial}{\partial p_i} \bigg) B^{(2)}(p_1,p_2,p_3,p_4)\\
& \nonumber +(4-D) B^{(1)}(p_1,p_2,p_3,p_4)\\
 & = \frac{d_2^{(1)}}{4!} B^{(1)}(p_1,p_2,p_3,p_4)+\frac{d_2^{(0)}}{4!}B^{(2)}(p_1,p_2,p_3,p_4) + \frac{1}{4!}\lbrace \epsilon \lambda + \beta_1^{(1)}(\lambda) \rbrace  \frac{\partial}{\partial \la} B^{(1)}(p_1,p_2,p_3,p_4)
\end{align}

Where \[
U_4^{(2)}(p_1,p_2,p_3,p_4)=\underbrace{-\lambda^2\sum_{3~perm~(i,j)} \mathcal{F}(p_i+p_j)}_{U_4^{I}} + \underbrace{\frac{F\lambda^2}{2} \sum_{i=1}^4 h(p_i)}_{U_4^{II}}
\]

\[\mathcal{F}(p_i+p_j)= \frac{1}{2} \int_k \lbrace h(p_i+p_j+k)h(k)-h(k)h(k) \rbrace\]

\[ d_2^{(1)}=-F\lambda \]

\[B^{(1)}(p_1,p_2,p_3,p_4)= -\la \sum_{i=1}^4 h(p_i)\]

\subsubsection*{ Calculation of $\frac{\delta}{\delta \phi(p)} \frac{\delta}{\delta \phi(-p)} D^{(2)}(p_1,p_2,p_3,p_4,p_5,p_6)$}

a)

\begin{subequations}
\begin{align}
\nonumber & \frac{\delta^2}{\delta \phi(p) \delta \phi(-p)} D_I^{(2)}(p_1,p_2,p_3,p_4,p_5,p_6)\phi(p_1)\phi(p_2)\phi(p_3)\phi(p_4)\phi(p_5)\phi(p_6)\\
\nonumber & = \frac{\delta^2}{\delta \phi(p) \delta \phi(-p)} \big \lbrace \sum_{10~perm~(i,j,k)}h(p_i+p_j+p_k) \big \rbrace \phi(p_1)\phi(p_2)\phi(p_3)\phi(p_4)\phi(p_5)\phi(p_6)\\
& = \label{B_III_one}30 \times \big \lbrace \sum_{l=1}^4 h(p_l) h(p_l) \big \rbrace  \phi(p_1)\phi(p_2)\phi(p_3)\phi(p_4)\\
&\label{B_one_1} + 60 \times \sum_{3~perm~(i,j)} \big \lbrace h(p_i+p_j+p)h(p_i+p_j+p) \big \rbrace \phi(p_1)\phi(p_2)\phi(p_3)\phi(p_4)
\end{align}
\end{subequations}

b)

\begin{subequations}
\begin{align}
\nonumber & \frac{\delta^2}{\delta \phi(p) \delta \phi(-p)} D_{II}^{(2)}(p_1,p_2,p_3,p_4,p_5,p_6)\phi(p_1)\phi(p_2)\phi(p_3)\phi(p_4)\phi(p_5)\phi(p_6) \\
\nonumber & \frac{\delta^2}{\delta \phi(p) \delta \phi(-p)} \sum_{10~perm~(i,j,k)} h(p_i+p_j+p_k) \sum_{l=1}^6 h(p_l)\phi(p_1)\phi(p_2)\phi(p_3)\phi(p_4)\phi(p_5)\phi(p_6)\\
& = \label{left_rhs} 30 \times \sum_{i=1}^4 h(p_i) 2h(p)\phi(p_1)\phi(p_2)\phi(p_3)\phi(p_4)\\
& \label{B_III_two} + 30 \times \sum_{i=1}^4 h(p_i)\sum_{j=1}^4 h(p_j) \phi(p_1)\phi(p_2)\phi(p_3)\phi(p_4)\\
& \label{B_one_2} + 60 \times \sum_{3~perm~(i,j)} h(p_i+p_j+p) \big \lbrace 2h(p) \big \rbrace \phi(p_1)\phi(p_2)\phi(p_3)\phi(p_4)\\
& \label{B_two_1}+ 60 \times \sum_{3~perm~(i,j)} h(p_i+p_j+p)\sum_{k=1}^4 h(p_k) \phi(p_1)\phi(p_2)\phi(p_3)\phi(p_4)
\end{align}
\end{subequations}

\subsubsection*{Equation for $B_{I}^{(2)}(p_1,p_2,p_3,p_4)$}

Collecting \eqref{B_one_1} and \eqref{B_one_2}, we get the following equations:

\begin{align}
\nonumber \la^2\int_p \big \lbrace -K^\prime(p) \big \rbrace [ 2 \sum_{3~perm~(i,j)} h(p_i+p_j+p)h(p_i+p_j+p)+ 4 \sum_{3~perm~(i,j)} h(p_i+p_j+p)h(p)]\\
\bigg \lbrace (4-D)- p_i.\frac{d}{d p_i} \bigg \rbrace B_I^{(2)}(p_1,p_2,p_3,p_4)= d_2^{(0)} B_I^{(2)}(p_1,p_2,p_3,p_4)
\end{align}

\subsubsection*{Equation for $B_{II}^{(2)}(p_1,p_2,p_3,p_4)$}

We take $\beta_1^{(1)}(\la)\frac{\p}{\p \la} B^{(1)}$ term from RHS. Collecting \eqref{B_two_1} and the term with $ U_4^{I} $ in the second line of \eqref{phi4_rel} we get the following equation,

\begin{align*}
& 2 \la^2 \sum_{3~perm~(i,j)} h(p_i+p_j+p) \big \lbrace -K^\prime(p) \big \rbrace \sum_{l=1}^4 h(p_l)-3\la^2 F\sum_{l=1}^4 h(p_l)\\
& +2 \sum_{i=1}^4 K^\prime(p_i) A^{(0)}(p_i)\bigg \lbrace -\frac{\lambda^2}{2} \int_k \sum_{3~perm~(i,j)}[h(p_i+p_j+k)h(k)-h(k)h(k)] \bigg \rbrace\\
& + \bigg \lbrace (4-D)-p_i.\frac{d}{d p_i} \bigg \rbrace B_{II}^{(2)}(p_1,p_2,p_3,p_4)=d_2^{(0)} B_{II}^{(2)}(p_1,p_2,p_3,p_4)
\end{align*}

\subsubsection*{Equation for $B_{III}^{(2)}(p_1,p_2,p_3,p_4)$}

Collecting \eqref{B_III_one}, \eqref{B_III_two} and the term containing $U_4^{II}$  we get

\begin{align*}
& \int_p \big \lbrace -K^\prime(p) \big \rbrace  \big \lbrace \sum_{l=1}^4 h(p_l)h(p_l)+ \sum_{i=1}^4 h(p_i) \sum_{j=1}^4 h(p_j) \big \rbrace\\
& + 2 \sum_{i=1}^4 K^\prime(p_i^2) \big \lbrace A^{(0)}(p_i) U_4^{(2)}|_2+ \lambda A^{(1)}(p_i) \big \rbrace \\
& + 2 \sum_{i=1}^4 K^\prime(p_i) U_2^{(1)}(p_i) \sum_{j=1}^4 B^{(1)}(p_i)\\
& + \big (4-D-p_i.\frac{d}{d p_i} \big) B_{III}^{(2)}(p_1,p_2,p_3,p_4)= d_2^{(0)} B_{III}^{(0)}(p_1,p_2,p_3,p_4)
\end{align*}

We have,

\begin{align*}
& A^{(0)}(p)=1~; A^{(1)}(p)= F\lambda h(p)~; U_2^{(1)}(p)=-\frac{\lambda F}{2-\epsilon}~; B^{(1)}(p)=-\lambda h(p)\\
\end{align*}

So the equation for $B_{III}^{(2)}(p_1,p_2,p_3,p_4) $ becomes,

\begin{align}
\nonumber & F \big \lbrace \sum_{l=1}^4 h(p_l)h(p_l) + \sum_{i \neq j} h(p_i)h(p_j) \big \rbrace + F \sum_{i \neq j} K^\prime(p_i^2)h(p_j) \\
& 2F \sum_{i \neq j} K^\prime(p_i) h(p_j) + 4F \sum_{i=1}^4 K^\prime(p_i) h(p_i)+ \big \lbrace 4-D- p_i.\frac{d}{d p_i} \big \rbrace B_{III}^{(2)}(p_1,p_2,p_3,p_4)= d_2^{(0)} B_{III}^{(2)}(p_1,p_2,p_3,p_4)
\end{align}

\subsubsection{Cancellation}

Note that last term in LHS and third term in RHS of \eqref{phi4_rel} cancels. 
Also the term \eqref{left_rhs} cancels with the term $d_2^{(1)} B^{(1)}(p_1,p_2,p_3,p_4)$.

\subsection{The $\phi^2$ equation to determine $A^{(2)}(p)$}
\label{solvphi2_rel}

$\phi^2$ equation is given by,

\begin{align}\label{phi2_rel}
&\nonumber  \int (-K^\prime(q)) \frac{\delta^2 }{\delta \phi(q)\phi(-q)} \big \lbrace B^{(2)}(p_1,p_2,p_3,p_4) \phi(p_1)\phi(p_2)\phi(p_3)\phi(p_4) \big \rbrace\\
& \nonumber + \left[ A(p)-p^2 A^\prime(p)-2 \big \lbrace -K^\prime(p^2) \big \rbrace U_2(p)A(p) \right] \phi(p)\phi(-p)\\
& +\eta \frac{K(p^2) (1-K(p^2))}{p^2} \frac{p^2}{K(p^2)}\phi(p)\phi(-p) -\frac{\eta}{2} A^{(0)}(p) \phi(p)\phi(-p)= \frac{d_2}{2} A(p)+(\epsilon \lambda+ \beta(\lambda))\frac{\partial }{\partial \lambda} \frac{A^{(2)}(p)}{2}
\end{align}

\begin{align*}
U_2^{1}(p)= -\frac{\la F}{2-\eps};~~U_2^{(2)}(p)= -\lambda^2 G(p)-\frac{\lambda^2 F^2}{4} h(p)
\end{align*}

Where

\begin{equation*}
G(p)= \frac{1}{3}\int_{q,k} \frac{h(q)}{2}[ h(p+q+k)h(k)-h(k)h(k)]-\frac{1}{3} \int_q \frac{h(q)}{2}[h(q+k)h(k)-h(k)h(k)]+\frac{\eta p^2}{2\epsilon}-\frac{1}{2-2\epsilon} \bigg \lbrace \frac{2}{3}\beta^{(1)}v_2^{(1)}+ \int_q f(q)\mathcal{F}(q) \bigg \rbrace 
\end{equation*}

\begin{align*}
\beta^{(1)}=-\int_q f(q) h(q)\rightarrow_{\epsilon\rightarrow 0}-F; v_2^{(1)}=-\int_q f(q)h(q)\rightarrow_{\epsilon\rightarrow 0} -\frac{F}{2}
\end{align*}

A.

\vspace{0.1 in}

\begin{align}
1.~~~\nonumber & \frac{\delta^2}{\delta \phi(q) \delta \phi(-q)}B_I^{(2)}(p_1,p_2,p_3,p_4) \phi(p_1)\phi(p_2)\phi(p_3)\phi(p_4)\\
\nonumber &= \frac{\delta^2}{\delta \phi(q) \delta \phi(-q)} \frac{1}{4!}\sum_{3~perm~(i,j)}\int_k \big \lbrace h(p_i+p_j+k)h(k)h(k)\big \rbrace \phi(p_1)\phi(p_2)\phi(p_3)\phi(p_4)\\
=& \label{A_II_one}  \int_{q,k} \big \lbrace h(p+q+k)h(k)h(k)\\
&+ \label{left_1}\frac{1}{2} h(k)h(k)h(k)
\end{align}

\begin{align}
2.~~~\nonumber & \frac{\delta^2}{\delta \phi(q) \delta \phi(-q)} \big \lbrace B_{II}^{(2)}(p_1,p_2,p_3,p_4) \phi(p_1)\phi(p_2)\phi(p_3)\phi(p_4)\big \rbrace \\
\nonumber= &  \frac{\delta^2}{\delta \phi(q) \delta \phi(-q)} \frac{1}{4!}\frac{1}{2} \int_k \sum_{3~perm~(i,j)}\big \lbrace h(p_i+p_j+k)h(k) -h(k)h(k) \big \rbrace \sum_{l=1}^4 h(p_l) \phi(p_1)\phi(p_2)\phi(p_3)\phi(p_4)\\
 = & \label{A_I_one}  \int_{q,k} \big \lbrace h(p+q+k)h(k)- h(k)h(k) \big \rbrace \big \lbrace  h(p) \big \rbrace \\
 + & \label{A_II_two}\int_{q,k}\big \lbrace h(p+q+k)h(k)- h(k)h(k) \big \rbrace \big \lbrace  h(q) \big \rbrace 
\end{align}

\begin{align}
3.~~~&\nonumber \int_q \big \lbrace -K^\prime(q) \big \rbrace \frac{\delta^2}{\delta \phi(q)\delta \phi(-q)} B_{III}^{(2)}(p_1,p_2,p_3,p_4)\phi(p_1)\phi(p_2)\phi(p_3)\phi(p_4)\\
&=\nonumber \int_q \big \lbrace -K^\prime(q) \big \rbrace  \frac{\delta^2}{\delta \phi(q)\delta \phi(-q)} \frac{1}{4!} ( -F ) \big \lbrace \sum_{i \neq j} \frac{1}{2} h(p_i) h(p_j) + \sum_{l=1}^4 h^2(p_l) \big \rbrace \phi(p_1)\phi(p_2)\phi(p_3)\phi(p_4)\\
&\nonumber  =  \int_q \big \lbrace -K^\prime(q) \big \rbrace \frac{-F}{4!} \left[ 6\times 2 \bigg \lbrace \frac{2 h^2(p)+ 2 h^2(q)+ 8h(p)h(q)}{2} \bigg \rbrace + 6\times 2 \big \lbrace 2 h^2(p)+ 2h^2(q) \big \rbrace \right]\\
& = \label{left_4} \frac{3 \lambda^2 F}{2} \int K^\prime(q) h^2(q)\\
& \label{A_IV_three} -\frac{3}{2}\lambda^2 F^2 h^2(p)\\
& \label{A_III_three} -\lambda^2 F^2 h(p)
\end{align}

\vspace{0.1 in}

\begin{align}
B.~~~\nonumber & 2 K^\prime(p^2) \big \lbrace U_2^{(2)}(p) A^{(0)}(p)+ U_2^{(1)} A^{(1)}(p) \big \rbrace\\
&= \label{A_III_one}2 K^\prime(p^2)\frac{-\lambda F}{2-\epsilon}\\
&= \label{A_IV_one} 2 K^\prime(p^2)\frac{-\lambda^2 F^2}{4} h(p)\\
&= \label{A_I_two} -2 K^\prime (p^2)A^{(0)}(p) \frac{1}{3} \left[ \int_{q,k} \frac{h(q)}{2} \big \lbrace h(p+q+k)h(k)-h(q+k)h(k) \big \rbrace \right]  \\
& \label{A_VI}- 2 K^\prime(p^2)A^{(0)}(p) \frac{\eta p^2}{2 \epsilon}\\
& \label{A_III_two} + 2 K^\prime(p^2)A^{(0)}(p) \frac{1}{2-2\epsilon} \big \lbrace \frac{2}{3} \beta^{(1)} v_2^{(1)} \big \rbrace\\
& \label{A_V_one}+ 2 K^\prime(p^2)A^{(0)}(p) \frac{1}{2-2\epsilon} \big \lbrace \int_q f(q) \mathcal{F}(q) \big \rbrace\\
& + \label{A_IV_two}2 K^\prime(p^2) \big \lbrace -\frac{\lambda F}{2-\epsilon} \big \rbrace \big \lbrace \lambda F h(p) \big \rbrace 
\end{align}

\vspace{0.1 in}

\begin{align}
C.~~~& \nonumber - \eta\frac{K(p^2) (1-K(p^2))}{p^2} \frac{p^2}{K(p^2)}-\frac{\eta}{2} A^{(0)}(p) \\
&= -\eta~ p^2 h(p)\phi(p)\phi(-p)\\
& \label{left_eta}-\frac{\eta}{2} A^{(0)}(p)
\end{align}

\subsubsection*{Equation for $A_I^{(2)}(p)$}

 We collect \eqref{A_I_one} and \eqref{A_I_two} to write the following equation,

\begin{align}
\nonumber & \int_{q,k} \big \lbrace -K^\prime(q)\big \rbrace \big \lbrace h(p+q+k)h(k)- h(q+k)h(k) \big \rbrace \big \lbrace  h(p) \big \rbrace\\
 & \nonumber -2 K^\prime (p^2)A^{(0)}(p) \frac{1}{3} \left[ \int_{q,k} \frac{h(q)}{2} \big \lbrace h(p+q+k)h(k)-h(q+k)h(k) \big \rbrace \right] \\
& + A_I^{(2)}(p)-p^2 \frac{\partial A_I^{(2)}(p)}{\partial p^2}= d_2 \frac{A_I^{(2)}(p)}{2}
\end{align}

On LHS of \eqref{phi2_rel} we are left with,

\begin{align}\label{left_2}
\int_{q,k} \big \lbrace - K^\prime(q) \big \rbrace h(p)  \big \lbrace h(q+k)h(k)-h(k)h(k) \big \rbrace
\end{align}

\subsubsection*{Equation for $A_{II}^{(2)}(p)$}

We collect \eqref{A_II_one} and \eqref{A_II_two} to write

\begin{align}
&  \nonumber \int_{q,k} \big \lbrace -K^\prime(q) \big \rbrace \big \lbrace h(p+q+k)h(k)h(k)-h(q+k)h(k)h(k) \big \rbrace+\int_{q,k} \big \lbrace -K^\prime(q) \big \rbrace \big \lbrace h(p+q+k)h(k)- h(q+k)h(k) \big \rbrace \big \lbrace  h(q) \big \rbrace\\
 & + A_{II}^{(2)}(p)-p^2 \frac{\partial A_{II}^{(2)}(p)}{\partial p^2}= d_2^{(0)} \frac{A_{II}^{(2)}(p)}{2} 
\end{align}

On LHS of \eqref{phi2_rel} we are left with

\begin{align}\label{left_3}
 \int_{q,k} \big \lbrace -K^\prime(q)h(q+k)h(k)h(k) \big \rbrace+\int_{q,k} \big \lbrace -K^\prime(q)h(q) \big \rbrace \big \lbrace h(q+k)h(k)- h(k)h(k) \big \rbrace
\end{align}

\subsubsection*{Equation for $A_{III}^{(2)}(p)$}

 We collect \eqref{A_III_one}, \eqref{A_III_two} and \eqref{A_III_three} to get,

\begin{align}
&\nonumber -\lambda^2 F^2 h(p)+2K^\prime(p^2)\big \lbrace -\frac{\lambda F \epsilon}{4} \big \rbrace + 2 K^\prime(p^2)A^{(0)}(p) \frac{1}{2-2\epsilon} \big \lbrace \frac{2}{3} \beta^{(1)} v_2^{(1)} \big \rbrace+A_{III}^{(2)}(p)-p^2 \frac{\partial}{\partial p^2} A_{III}^{(2)}(p)\\
&= d_2^{(0)} \frac{A_{III}^{(2)}(p)}{2}+ ( d_2^{(1)})\frac{ A^{(1)}(p)}{2}+ \lbrace \epsilon \la+ \beta(\la) \rbrace \hf\frac{\p A}{\p \la}
\end{align}

Where $d_2^{(1)}=-\lambda F$ and $A^{(1)}(p)=\lambda F h(p)$.

\subsubsection*{Equation for $A_{IV}^{(2)}(p)$}

Collecting \eqref{A_IV_one}, \eqref{A_IV_two} and \eqref{A_IV_three} we get,

\begin{align}
-\frac{3}{2}\lambda^2 F^2 h^2(p)+2 K^\prime(p^2)\frac{-\lambda^2 F^2}{4} h(p)+ 2 K^\prime(p^2) \big \lbrace -\frac{\lambda F}{2-\epsilon} \big \rbrace \big \lbrace \lambda F h(p) \big \rbrace +A_{IV}^{(2)}(p) -p^2 \frac{\partial}{\partial p^2} A_{IV}^{(2)}(p)= d_2^{(0)} \frac{A_{IV}^{(2)}(p)}{2}
\end{align}

\subsubsection*{Equation for $A_{V}^{(2)}(p)$}

We collect \eqref{A_V_one} and \eqref{left_1} to get the following equation,

\begin{align}
 \nonumber &2 K^\prime(p^2)A^{(0)}(p) \frac{1}{2-2\epsilon} \big \lbrace \int_q f(q) \mathcal{F}(q) \big \rbrace+\int_{q,k} \big \lbrace - K^\prime(q) \big \rbrace h(p)  \big \lbrace h(q+k)h(k)-h(k)h(k) \big \rbrace \\
 &+A_V^{(2)}(p)-p^2\frac{\partial}{\partial p^2} A_{V}^{(2)}(p)= d_2^{(0)} \frac{A_{V}^{(2)}(p)}{2}
\end{align}

\subsubsection*{Equation for $A_{VI}^{(2)}(p)$}

We collect 5th term of \eqref{phi2}, \eqref{A_VI} to get the following equation

\begin{align}
\nonumber &\eta \frac{K(p^2) (1-K(p^2))}{p^2} \frac{p^2}{K(p^2)} - 2 K^\prime(p^2)A^{(0)}(p) \frac{\eta p^2}{2 \epsilon}+A_{VI}^{(2)}(p)-\frac{1}{2} p.\frac{\partial}{\partial p}A_{VI}^{(2)}(p)\\
&= d_2^{(0)}\frac{A_{VI}(p)}{2}+d_2^{(1)} \frac{A_{VI}^{(2)}(p)}{2}+ \epsilon\lambda \frac{\partial}{\partial \lambda}\frac{A_{VI}^{(2)}(p)}{2}
\end{align}

\section{Evaluation of Integrals}
\label{appE}

\begin{align*}
&\int_{p,q} h\left(\frac{p+q}{\Lambda}\right)h\left(\frac{q}{\Lambda}\right)h\left(\frac{p}{\Lambda}\right)h\left(\frac{p}{\Lambda}\right)\\
=& \int_{p,q} \frac{K\left( \frac{p+q}{\Lambda_0} \right)-K\left( \frac{p+q}{\Lambda} \right)}{(p+q)^2}~~\frac{K\left( \frac{q}{\Lambda_0} \right)-K\left(\frac{q}{\Lambda} \right)}{(q)^2}~~\frac{K\left(\frac{p}{\Lambda_0} \right)-K\left(\frac{p}{\Lambda} \right)}{(p)^2}~~\frac{K\left(\frac{p}{\Lambda_0} \right)-K\left(\frac{p}{\Lambda} \right)}{(p)^2}
\end{align*}

We evaluate the integral for $K(x)=e^{-x^2}$.

\begin{align*}
=\int_{p,q} \frac{e^{-\frac{(p+q)^2}{\Lambda_0^2}}-e^{- \frac{(p+q)^2}{\Lambda^2}}}{(p+q)^2}~~\frac{e^{-\frac{q^2}{\Lambda_0^2}}-e^{-\frac{q^2}{\Lambda^2} }}{(q)^2}~~\frac{e^{\frac{p^2}{\Lambda_0^2}}-e^{\frac{p^2}{\Lambda^2}}}{(p)^2}~~\frac{e^{\frac{p^2}{\Lambda_0^2}}-e^{\frac{p^2}{\Lambda^2}}}{(p)^2}
\end{align*}

Now we apply Schwinger parametrization.

\begin{align*}
\int_{p,q} \int_{x,y,u,v=\frac{1}{\Lambda^2}}^{\frac{1}{\Lambda_0^2}}e^{-(p+q)^2x}e^{-q^2y}e^{-p^2 u}e^{-p^2 v}
\end{align*}

Now we do q inetgral first. We Complete the square on q and change integration varibale q. After that we do p inetegral. Also we change $x, y$ as $x \rightarrow \frac{1}{x},~ y \rightarrow \frac{1}{y}$. At the end we take $\Lambda_0 \rightarrow \infty$.

\begin{align*}
&\int_{x,y,u,v} \int_{p,q}\frac{1}{(x+y)^2} e^{-q^2} e^{-p^2\left( \frac{xy}{x+y}+u+v \right)}\\
&=F^2\int_{x,y,u,v=\frac{1}{\Lambda^2}}^{\frac{1}{\Lambda_0^2}} \frac{1}{\big \lbrace 1+(p+q)(u+v)\big \rbrace^2}\\
& = F^2 \bigg (\frac{1}{2} \lbrace \log 2 \rbrace^2-\frac{1}{2}\lbrace \log\frac{2\Lambda_0^2}{\Lambda^2}\rbrace^2+ \lbrace \log \frac{\Lambda_0^2}{\Lambda^2} \rbrace^2 + 2\log2-6\log 3+ 2\log \frac{\Lambda_0^2}{\Lambda^2}\\
& +\frac{1}{4} \lbrace \log 4 \rbrace^2 -\frac{1}{2} \lbrace \log \frac{2\Lambda_0^2}{\Lambda^2} \rbrace^2+\frac{1}{4}\lbrace \log \frac{4\Lambda_0^2}{\Lambda^2} \rbrace^2-8\log2+5\log5-\log\frac{\Lambda_0^2}{\Lambda^2}\\
&+ \frac{1}{4} \lbrace \log \frac{4\Lambda^2}{\Lambda_0^2} \rbrace^2 -\frac{1}{2} \lbrace \log 2\rbrace^2+ \frac{1}{4} \lbrace \log 4 \rbrace^2 +4\log 2-6\log 3+5\log5 \bigg)
\end{align*}

So,

\begin{align}\label{integral_value}
\nonumber &\int_{p,q} \left[\big \lbrace -K^\prime(p^2)\big \rbrace h(p) h(p+q)h(q)+\big \lbrace -K^\prime(p^2)\big \rbrace h(q)h(p+q)h(q)\right]\\
&=F^2 \left(\frac{1}{2}-\log 2+\frac{1}{2}\log \frac{\Lambda_0^2}{\Lambda^2} \right)
\end{align}

Using same procedure we can find all other integrals of this kind.

\section{Useful Mathematical identities}
\label{appF}

In this section we give various mathematical identities about the functions $h(p)$, $\mathcal{F}(p)$, $F_3(p)$ etc which were used in the main text to find the composite operators.

\[ h(p)= \frac{1-K(p^2)}{p^2} ~~ f(p)=-2K^\prime(p)\]

\begin{align}\label{h}
-p.\frac{\partial}{\partial p} h(p)= -f(p)+2h(p)
\end{align}

\[ \mathcal{F}(p)= \hf \int_q \left \lbrace h(p+q)h(q)-h(q)h(q) \right \rbrace \]

\begin{align}\label{F}
\left( p.\frac{d}{dp}+\epsilon \right)\mathcal{F}(p)= \int_q f(q) \left \lbrace h(p+q)-h(q) \right \rbrace
\end{align}

\[\bar{F}_3(p)=\int_{q,k}h(p+q+k)h(q)h(k)\]
\[,~~ F_3(p)= \bar{F}_3(p)-\bar{F}_3(0)=\int_q 2h(q)\left[ F(p+q)-F(q) \right]\]

\begin{align}\label{F3_bar}
\left( -\frac{p}{2}.\frac{d}{dp}+1 \right) \bar{F_3}(p)=-6\int_{q,k} f(q)h(p+q+k)h(k)
\end{align}

\begin{align}\label{F_3}
\left( p.\frac{d}{dp} -2 + 2\epsilon \right) F_3(p)= 3 \int_{q,k} f(k)h(q) \left[ h(q+k+p)-h(q+k) \right]
\end{align}

\[\bar{H}_3(p)=\int_q h(p+q)h(q)h(q)\]

\begin{align}\label{H3_bar}
\left( p.\frac{\partial}{\partial p} +2+ \epsilon \right) \bar{H}_3(p)= \int_q f(p)h(p+q)^2+ 2\int_q f(q)h(q)h(p+q)
\end{align}

\[I_4(p_i+p_j;p_i)= \bar{I}_4(p_i+p_j;p_i)-\bar{I}_4(0;0)=\sum_{6~perm~(i,j)}\int_{p,q}  \big \lbrace h(p_i+p_j+q)  h(p+q+p_i) h(p) h(q)- h(p+q)h(p)h(q)h(q) \big \rbrace\]

\begin{align}\label{I4}
\left( -\sum_{l=1}^4 p_l.\frac{d}{dp_l}-2\epsilon\right)\bar{I}_4(p_i+p_j;p_i)= -2\int_{p,q} f(p)\left[ h(p_i+p_j+p)+ h(p_i+p_j+q) \right] h(p_i+p+q)h(q)
\end{align}

\end{appendices}

\end{document}